\pdfoutput=1
\documentclass[reprint, superscriptaddress, amsmath,pre,amssymb, aps, floatfix]{revtex4-2}

\makeatletter
\newsavebox{\@brx}
\newcommand{\llangle}[1][]{\savebox{\@brx}{\(\m@th{#1\langle}\)}%
  \mathopen{\copy\@brx\kern-0.5\wd\@brx\usebox{\@brx}}}
\newcommand{\rrangle}[1][]{\savebox{\@brx}{\(\m@th{#1\rangle}\)}%
  \mathclose{\copy\@brx\kern-0.5\wd\@brx\usebox{\@brx}}}
\usepackage{lipsum}
\usepackage{graphicx}% Include figure files
\usepackage{dcolumn}% Align table columns on decimal point
\usepackage{bm}% bold math
\usepackage{xcolor}
\usepackage{xspace}
\usepackage{comment}
\newcommand{\csch}{\operatorname{csch}}
\usepackage{multirow}
\usepackage{tikz}
\usetikzlibrary{arrows.meta}

\usepackage[colorlinks,allcolors=blue]{hyperref}
\usepackage[makeroom]{cancel}
\usepackage{float}
\usepackage[percent]{overpic}
\usepackage{siunitx}
\usepackage{color}
\definecolor{dgreen}{rgb}{0,0.7,0}
\def\bluew#1{{\color{red} #1}}

\def\bluew#1{{\color{black} #1}}

\def\bluepw#1{{\color{black} #1}}
\newcommand{\beq}{\begin{equation}}
\newcommand{\eeq}{\end{equation}}
\newcommand{\bea}{\begin{eqnarray}}
\newcommand{\eea}{\end{eqnarray}}

\newcommand{\MeijerG}[7]{G \begin{smallmatrix} #1 & #2 \\ #3 & #4 \end{smallmatrix} \left( \begin{smallmatrix} #5 \\ #6 \end{smallmatrix} \middle\vert #7 \right) }
\usepackage{subfigure}
\usepackage{booktabs} 
\usepackage{natbib}
\usepackage[normalem]{ulem}
\usepackage{empheq}

\begin{document}

%\title{A trapping perspective on the Sokoban random walk: turnover and robust stretched-exponential relaxation}
\title{The Sokoban Random Walk: A Trapping Perspective}
\author{Prashant Singh} 
\email{prashantsinghramitay@gmail.com}
\address{Department of Physics, Bar-Ilan University, Ramat Gan 52900, Israel}
\author{Eli Barkai} 
\address{Department of Physics, Bar-Ilan University, Ramat Gan 52900, Israel}
\address{Institute of Nanotechnology and Advanced Materials, Bar-Ilan University, Ramat Gan 52900, Israel}
\author{David A. Kessler}
\address{Department of Physics, Bar-Ilan University, Ramat Gan 52900, Israel}

%\ead{karel.proesmans@nbi.ku.dk}
%\ead{prashant.singh@icts.res.in, saikat.santra@icts.res.in}
%\ead{anupam.kundu@icts.res.in}
\vspace{10pt}

\begin{abstract}
We study caging/trapping in Sokoban-type models, featuring a random walker moving through a disordered medium of obstacles and capable of pushing some obstacles blocking its path. In one-dimension, we allow the walker to push up to an arbitrary $N_{\rm P}$ number of obstacles. For $N_{\rm P}\gg 1$, we use large-deviation theory to show that the survival probability to remain uncaged exhibits crossover from an exponential decay with time at intermediate times to a stretched-exponential decay at long times, with an exponent $1/3$ independent of $N_{\rm P}$. The long-time exponent matches the Balagurov--Vaks--Donsker--Varadhan (BVDV) theory of the classical trapping problem, while the exponential decay is qualitatively distinct from the Rosenstock's intermediate-time theory for classical trapping. Similarly, in two dimensions, numerical simulations reveal that both the Sokoban model and its generalized version exhibit long-time stretched-exponential relaxation with exponent $1/2$, again consistent with the BVDV theory. Finally, in two dimensions, we find that the mean trap size is nonmonotonic in $\rho$: it is small at both low and high densities, but reaches a peak at a characteristic density $\rho_*$. We estimate $\rho_* \approx 0.55$ for the Sokoban model and $\rho_* \approx 0.675$ for the generalized Sokoban model. 
%We demonstrate that the nonmonotonicity is related to the emergence of dynamical confinement within self-created traps at low densities.

%We study the $\textit{Sokoban model}$, in which a random walker moves through a disordered medium of obstacles and can push at most one obstacle blocking its path. The limited pushing ability induces a caging/trapping effect that suppresses long-range transport even at low obstacle density $\rho$. We examine the robustness of the trapping phenomenon for several variants of the Sokoban model. In one dimension, we introduce the $N_{\rm P}$-Sokoban model where the walker can push up to $N_{\rm P}$ obstacles. 

%Using numerical simulations, we also study a two-dimensional generalized Sokoban model with microscopic dynamical rules modified relative to the original model. In both models, the survival probability still exhibits a long-time stretched-exponential relaxation, albeit with a different exponent that we estimate to be $1/2$. 

%We study several variants of the $\textit{Sokoban model}$ introduced recently by Reuveni et al. \href{https://doi.org/10.1103/PhysRevResearch.5.L042015}{[Phys. Rev. Res. \textbf{5}, L042015 (2023)]}, featuring a random walker moving through a disordered medium of obstacles and capable of pushing at most one obstacle blocking its path. 

\end{abstract}

\maketitle

\section{Introduction}
Transport in a disordered medium is a classic problem in statistical physics that has been extensively studied in the context of various physical systems, such as percolation \cite{Broadbent_Hammersley_1957, Weiss-1,Havlin-2}, charge transport \cite{PhysRevB.4.2612, PhysRevB.27.2583}, motion in the porous medium \cite{RevModPhys.65.1393}, diffusion of single molecules in the cell environment \cite{Barkai2012StrangeKinetics, C4CP03465A} and active models in complex environment \cite{Bechinger2016, Singhkundu2020}. 
%The complexity of the environment through which a particle moves can significantly influence its transport properties. To grasp some basic features of the problem, it is often desirable to study simple models that are amenable to analytical and numerical treatment. 
A commonly studied model in percolation is the de Gennes' \textit{ant in a labyrinth} model \cite{degenne}; see also \cite{Havlin-2, Weiss-1} for reviews. In this model, a discrete-time random walker moves in a two-dimensional disordered lattice with randomly placed fixed obstacles, and in every move, it can jump with equal probability to one of its neighboring sites, provided that the target site does not contain an obstacle. The obstacles themselves are initially randomly distributed with some fixed density $\rho$ where $0 \leq \rho \leq 1$. 

The  ant in a labyrinth model (henceforth referred to as AIL) exhibits a percolation transition at the critical density $\rho _c$: for $\rho \leq \rho _c$, the walker can escape to infinity while for $\rho >\rho _c$, the escape is not possible. The value of $\rho _c$ itself depends on the dimensionality as well as on the lattice type. For example, $\rho _c \approx 0.407$ in a two-dimensional square lattice \cite{Ziff_1986, PhysRevE.57.230}.

%At high densities, the lattice is predominantly occupied by the obstacles, with only small, sparsely distributed clusters of vacant sites. Consequently, the caging effect is quite high in this regime and the motion of the walker is strongly inhibited. When $\rho$ is decreased, these clusters of vacancies start to merge together, and at sufficiently small density a spanning cluster of vacant sites emerges. This critical density, denoted by $\rho _c$, marks the onset of the formation of an infinite cluster, and for $\rho \leq \rho _c$, the walker can escape to infinity. Hence, the ant in a labyrinth model (henceforth referred to as AIL) exhibits a percolation transition at the critical density $\rho _c$. The value of $\rho _c$ itself depends on the dimensionality as well as on the lattice type: in a one-dimensional lattice, $\rho _c = 0$ whereas $\rho _c \approx 0.407$ in a two-dimensional square lattice \cite{Ziff_1986, PhysRevE.57.230}.

%Observe that the AIL model represents a quenched disorder case, \emph{i.e.,} the obstacle configuration does not change with time for a given realization, but it varies from realization to realization. 
Interestingly, recent studies by Reuveni and coworkers reveal that allowing the walker to even minimally modify its local environment leads to the loss of the percolation transition in two dimensions \cite{Shlomi-1, Shlomi-2}. These studies introduced a new type of random walk, named \textit{Sokoban}, the model having been inspired by a video game created by Hiroyuki Imabayashi in 1981 \cite{SokobanWiki}. The Sokoban walker, unlike an AIL walker, has the ability to push some of the obstacles that block its way. This pushing ability, however, is limited in the sense that the walker can displace only a finite number of obstacles (explained precisely later). On physical grounds, such a finiteness can naturally arise whenever the obstacles are smaller than or comparable in size to the random walker, such as in experiment involving bristle robot moving in an arena of movable obstacles \cite{expt-1} or energized active particle pushing the surrounding particles \cite{expt-2, expt-3}. The loss of percolation for the Sokoban model clearly signals a fundamental difference between the scenarios where the environment is fixed versus those where it can be shaped to small extent by the walker.

\begin{figure*}[]
\centering

\begin{minipage}[t]{0.42\textwidth}
\centering
\textbf{(a) Sokoban model}\\[2mm]
\begin{overpic}[scale=0.25]{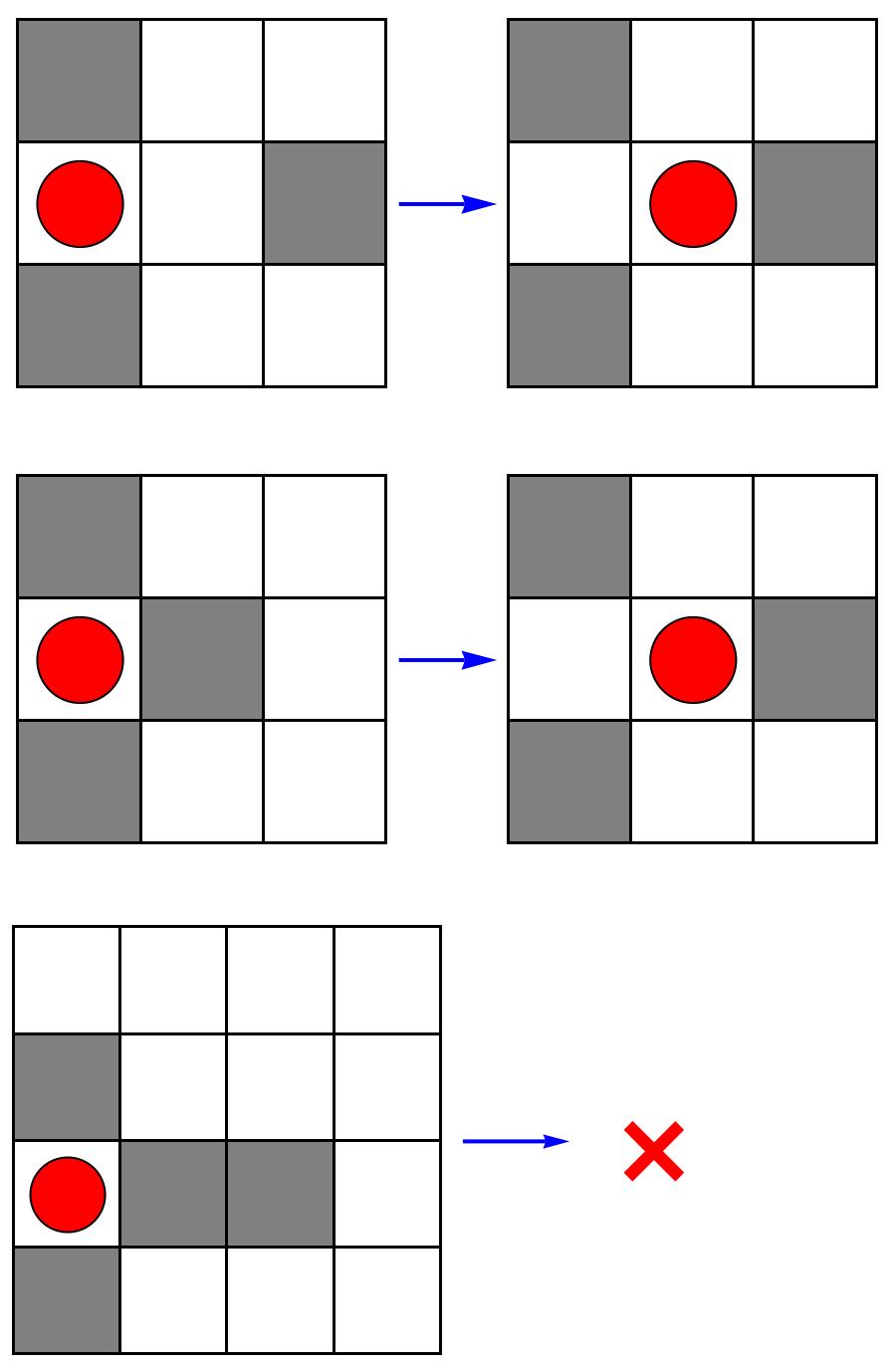}
\put(-13,83){\bfseries (i)}
\put(-13,50){\bfseries (ii)}
\put(-13,15){\bfseries (iii)}
\end{overpic}
\end{minipage}
\hspace{-0.5cm}
\begin{minipage}[t]{0.42\textwidth}
\centering
\textbf{(b) G-Sokoban model}\\[2mm]
\begin{overpic}[scale=0.25]{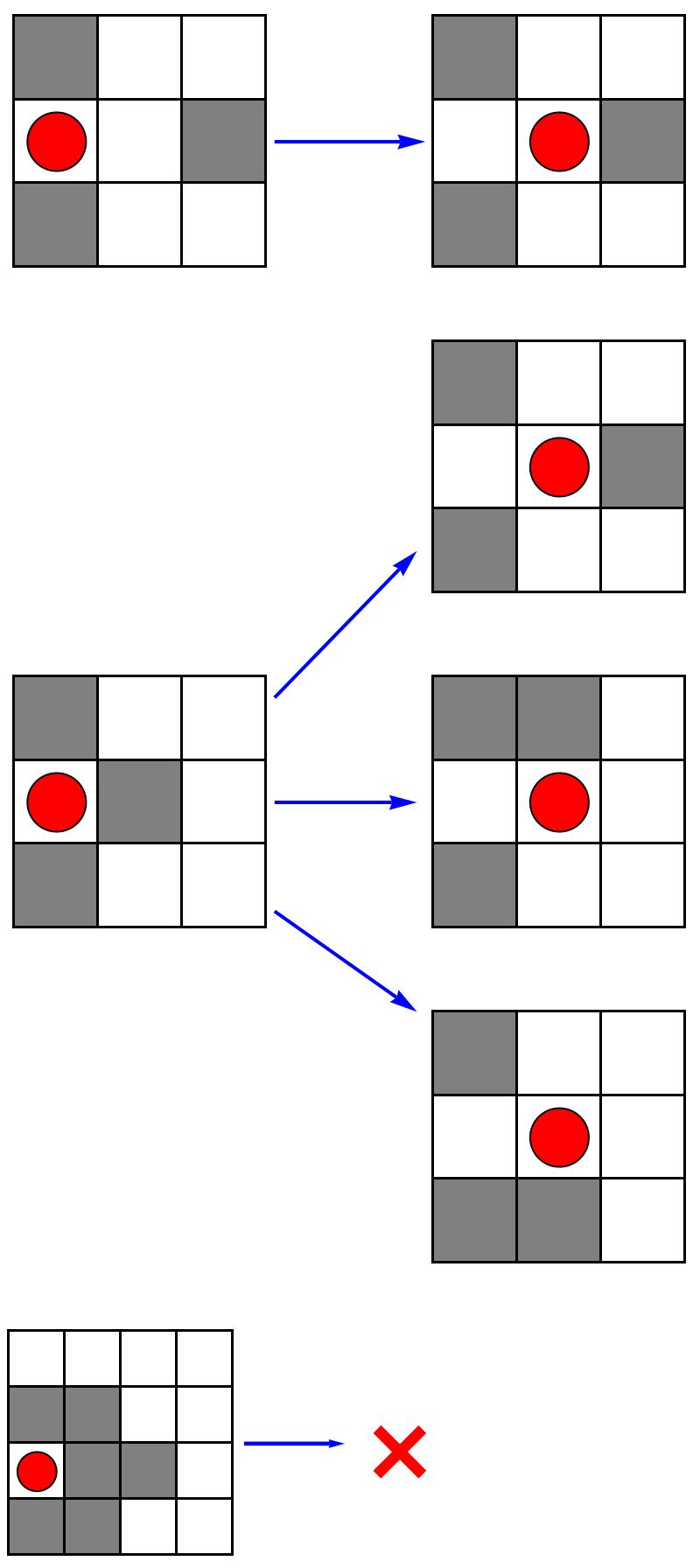}
\put(-10,88){\bfseries (i)}
\put(-10,48){\bfseries (ii)}
\put(-10,5){\bfseries (iii)}
\put(20,64){$\textcolor{blue}{\frac{1}{3}}$}
\put(20,51){$\textcolor{blue}{\frac{1}{3}}$}
\put(20,33){$\textcolor{blue}{\frac{1}{3}}$}
\end{overpic}
\end{minipage}
\hspace{-0.5cm}

\caption{Dynamical rules governing the time evolution of (a) the Sokoban walker and (b) the G-Sokoban walker. In both panels, the walker is shown in red, obstacles in gray, and vacant sites in white. The walker is attempting to move to the right. Let us focus on the Sokoban. In panel~(a-i), the Sokoban moves to its nearest-neighbor site since this target site is vacant before jump.  If the target site is occupied, as shown in panel~(a-ii), the Sokoban can still move by pushing the obstacle to the next-nearest site in the direction of its motion, provided that this site is vacant. If this condition is not satisfied, as shown in panel~(a-iii), the attempted jump is unsuccessful. Let us next turn to the G-Sokoban dynamics. As in the Sokoban model, the G-Sokoban walker can jump to a nearest-neighbor site if that site is vacant; see panel~(b-i). If the target site is occupied by an obstacle, as illustrated in panel~(b-ii), the walker can still move by pushing the obstacle. Unlike the Sokoban model, however, the obstacle-pushing is not just restricted in the direction of the walker's motion. Instead, the obstacle can be pushed to any one of its three neighboring sites (excluding the site occupied by the walker) with an equal probability $1/3$, provided that the chosen neighboring-site is vacant. If they are not vacant, as shown in panel~(b-iii), the jump is not successful. Note that, in the situation depicted in panel~(a-iii), a G-Sokoban walker can still move, since the obstacle can be pushed aside.}
\label{fig-update}
\end{figure*}

Although the percolation transition is lost in the two-dimensional Sokoban model, we showed in a recent Letter that it exhibits a dynamical crossover in the underlying trapping mechanisms \cite{SinghLetter}. As rigorously defined later, ``trapping" here means that the walker becomes caged to a finite, closed domain in an irreversible manner such that the long-range transport is impossible.  At high obstacle density, traps are essentially predetermined by the initial obstacle configuration, whereas at low density a self-trapping mechanism emerges in which the Sokoban dynamically generates its own trap.
In both cases, the walker cannot open the enclosing cage and is therefore permanently confined to a finite region of space. To characterize the trapping phenomena, we examine the following observables: (i) the disorder-averaged survival probability that the walker remains uncaged/untrapped until time $n$, (ii) the average trapping time $\langle n_{\rm T} \rangle $ and, (iii) the probability of trap sizes and its average $\langle A_{\rm T} \rangle $. Here $\langle n_{\rm T}\rangle$ and $\langle A_{\rm T}\rangle$ denote averages over the disorder realizations as well as the walker's stochastic dynamics.

The purpose of this paper is to investigate the trapping aspects of the Sokoban random walk in greater detail. We also study the variants of the model in order to test the universality of the emergent trapping features. To this end, we introduce the one-dimensional $N_{\rm P}$-Sokoban model, in which the walker can push up to an arbitrary $N_{\rm P}~(\geq 0)$ number of obstacles; $N_{\rm P}=1$ corresponds to the model of Reuveni and coworkers. For $N_{\rm P} \gg 1$, a large-deviation analysis reveals that the Sokoban model belongs to the Balagurov-Vaks-Donsker-Varadhan (BVDV) universality class for the classical trapping problem (see Eq.~\eqref{original-eqn} later) \cite{Havlin-2, Weiss-1}. In other words, the disorder-averaged survival probability to remain uncaged exhibits a stretched-exponential decay at late times, with stretch exponent $1/3$ that is independent of $N_{\rm P}$.

%calculation reveals that the disorder-averaged survival probability $S(n)$ crosses over from an exponential decay in time for $n \ll (N_{\rm P})^3$ to a universal stretched-exponential form for $n \gg (N_{\rm P})^3$, with stretch exponent $1/3$ that is independent of $N_{\rm P}$ and matches the Balagurov-Vaks-Donsker-Varadhan (BVDV) theory for the classical trapping problem (see Eq.~\eqref{original-eqn} later) \cite{Havlin-2, Weiss-1}.

In two dimensions, we perform numerical simulations of the Sokoban model and a generalized variant with modified microscopic rules introduced in Sec.~\ref{sec-model} and in Fig.~\ref{fig-update}. For both models, we find long-time stretched-exponential relaxation for the survival probability, with exponent $1/2$ again consistent with the BVDV theory. At the same time, the Sokoban model displays notable differences from the classical trapping problem: (i) modified prefactors accompanying the stretched-exponential decay, (ii) a distinct intermediate-time regime that deviates from the Rosenstock approximation for classical trapping (see Eq.~\eqref{original-eqn-2}), and (iii) geometry of the created trap in two-dimensions which bears signature of the dynamical crossover between the two trapping mechanisms discussed above.

The rest of our paper is structured as follows: Sec.~\ref{sec-model} introduces the models, defines precisely the observables of interest, and announces the main results of our paper. In Sec.~\ref{sec-1d-1}, we will use a renewal framework to derive the statistics of the trapping time and trap size for the $N_{\rm P}$-Sokoban model for a given initial obstacle configuration. We then perform the disorder-averaging of these results in Sec.~\ref{sec-1d-2} to calculate the survival probability. Building on the insights gained from these sections, we study the two-dimensional Sokoban model and its variant in Sec.~\ref{sec-2d-1} using numerical simulations. This is then followed by our conclusions and open directions in Sec.~\ref{sec-conclusion}.
 
%one-dimensional models. This framework is then applied to the Sokoban model with $N_{\rm P}=0$ in Sec.~\ref{sec-1d-2} and $N_{\rm P}=1$ in Sec.~\ref{sec-1d-3}. 

\section{Model, preliminaries and summary of the results}
\label{sec-model}
\textit{Sokoban model:} We first describe the Sokoban model introduced by Reuveni and co-workers \cite{Shlomi-1}. Consider a square $d$-dimensional lattice system of size $\mathcal{N}^d$, where initially every site can accommodate an obstacle with probability $\rho$ and remain vacant with the complementary probability $(1-\rho)$. Taking $\mathcal{N}$ to be odd, we place our discrete-time random walker initially at the center and consider the limit $\mathcal{N} \to \infty$. In every move, the walker chooses one of its $z_{d}$ neighboring sites with an equal probability $1/z_d$ and attempts a jump to that site. One has $z_d=2$ in one dimension, $z_d=4$ in two dimensions, and so on. 

%\begin{figure*}[]
%	\centering
%	\includegraphics[scale=0.2]{scheme-trap-2.jpeg}
%	\hspace{0.5 cm}
%	\includegraphics[scale=0.2]{scheme-trap-1.jpeg}
%	\hspace{0.5 cm}
%	\includegraphics[scale=0.2]{scheme-Omega.jpeg}
%	\caption{Caged trapping of a Sokoban random walker in two dimensions. The left panel shows the initial configuration of a part of the infinite system, with walker shown in red and located at the origin. The obstacles are shown in gray and green. They are identical, but distinguished here to illustrate that the green ones will be pushed by the walker. The middle panel presents the trapped scenario where, the green obstacles have been pushed by the walker and it is no longer able to visit any new lattice site. The trap size $A_{\rm T} = 21$ is demonstrated in blue. In the right panel, we have plotted  number of distinct sites visited $\Omega(n)$ as a function of time $n$. This number is saturated for all $ n \geq n _{\rm T} $ with $n _{\rm T} = 430$ for the considered example.}
%\label{fig-trap-size}
%\end{figure*}

\begin{figure}[t]
    \centering
    % Raise the first image by 1em (adjust as needed)
   \includegraphics[scale=0.3]{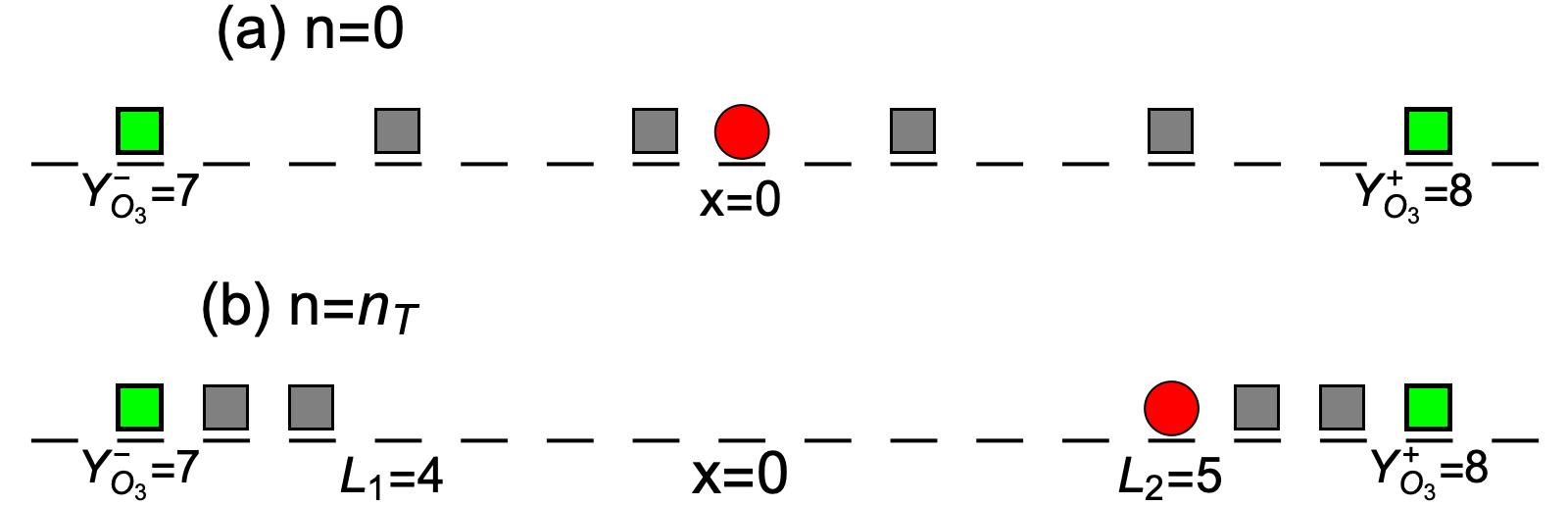}
    \caption{\bluew{Schematics of the one-dimensional $N_{\rm P}$-Sokoban model with $N_{\rm P} = 2$. Panel~(a) shows a realization of the walker and obstacles at the initial time. The walker, located at the origin, is shown in red and obstacles are  shown in gray and green.  They are identical, but distinguished here to illustrate that
the gray ones will be pushed by the walker. As time evolves, the gray obstacles on both sides of the origin are pushed until time $n=n_{\rm T}$ in panel~(b). Here, the pushed (gray) obstacles on each side lie on consecutive sites in front of the green obstacle. Beyond this point, the walker cannot push the obstacles due to its finite $(N_{\rm p}=2)$ pushing capacity.  Consequently, in every realization, the Sokoban walker is confined inside a finite interval $[-L_1, L_2]$ throughout its time evolution with $L_1$ and $L_2$ determined by the positions of the $(N_{\rm P}+1)$-th obstacles on the $x<0$ and $x>0$ sides, respectively, see Eq.~\eqref{hagqo}. These obstacles are shown in green in this figure. For the chosen realization in this figure, we have $L_1=4$ and $L_2=5$ determined by the positions of the green obstacles, $-Y_{O_{3}}^{-}=-7$ and $Y_{O_{3}}^{+}=8$, in either direction using Eq.~\eqref{hagqo}.}}
    \label{fig-new-1d-scheme}
\end{figure}

\begin{figure*}[t]
\centering

\begin{tikzpicture}[>=Stealth]

\node[anchor=north west] (A) at (0,0) {%
  \begin{minipage}[t]{0.5\linewidth}
    \vspace{0pt}\centering
    \includegraphics[width=1.1\linewidth]{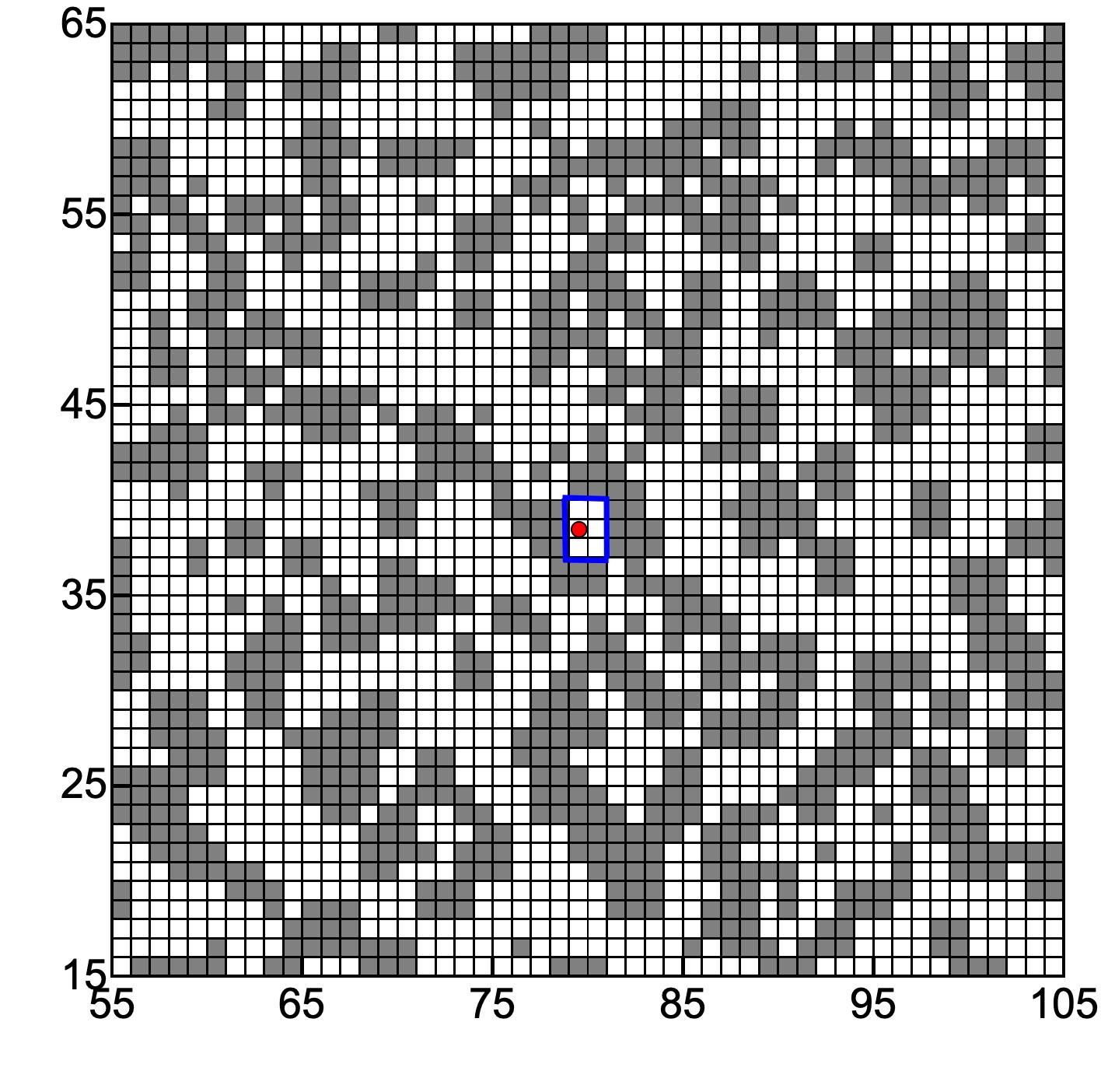}\\[-1mm]
    {\Large\textbf{(a)}}
  \end{minipage}
};

\node[anchor=north west] (B) at (12.5,0) {
  \begin{minipage}[t]{0.20\linewidth}
    \vspace{0pt}\centering
    \includegraphics[width=0.9\linewidth]{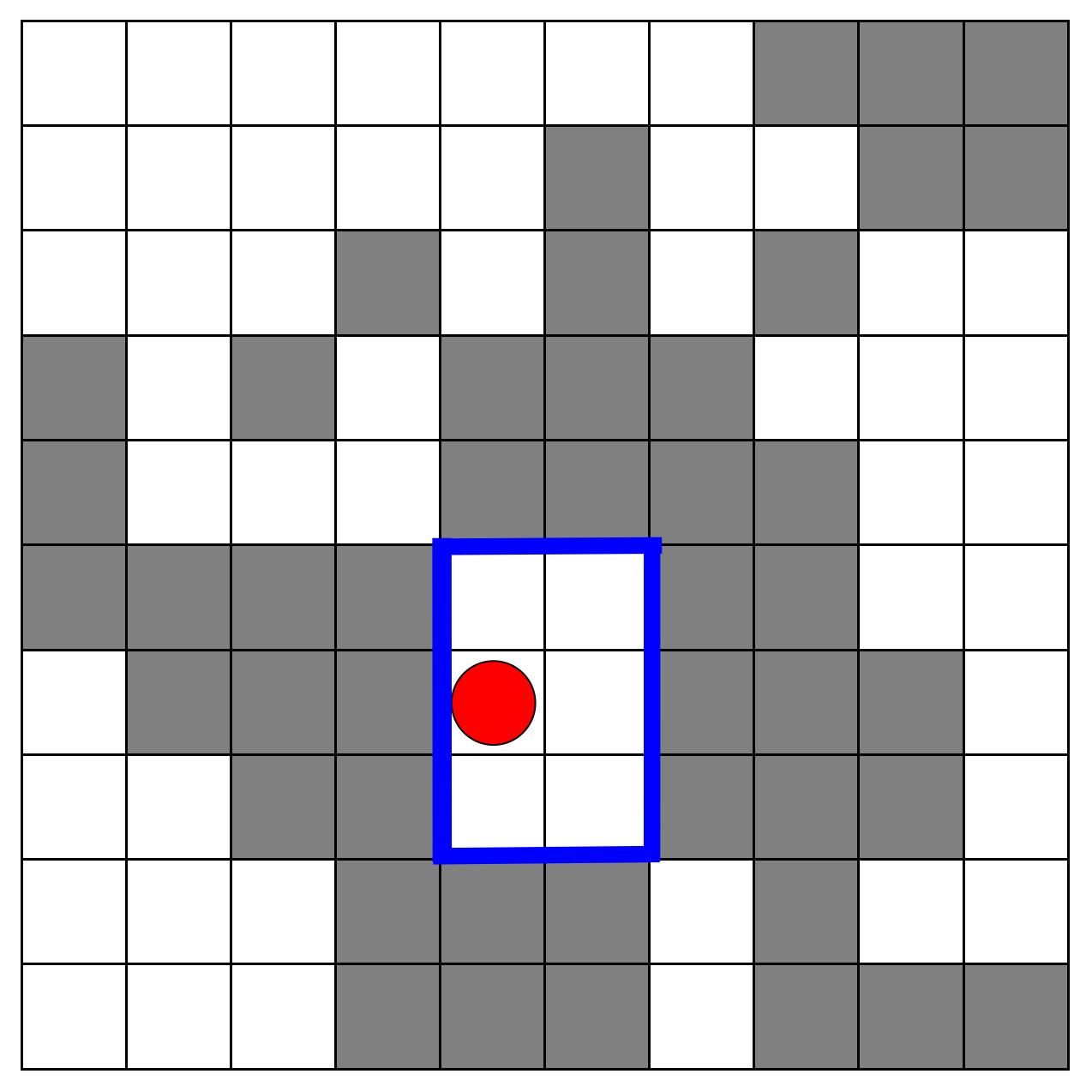}\\[-1mm]
    {\Large\textbf{(b)}}\\[1mm]
    \includegraphics[width=0.9\linewidth]{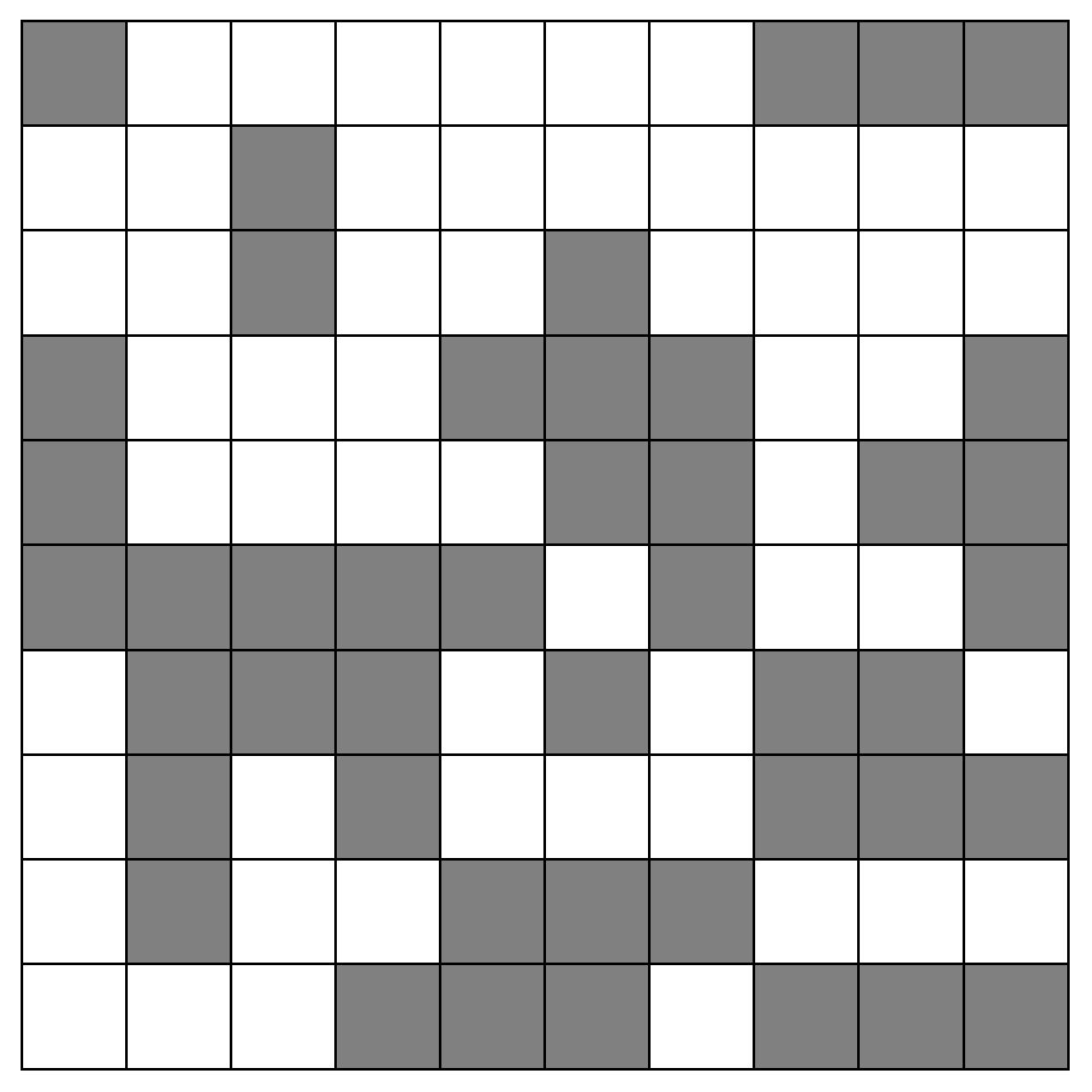}\\[-1mm]
    {\Large\textbf{(c)}}\\[1mm]
    \includegraphics[width=1\linewidth]{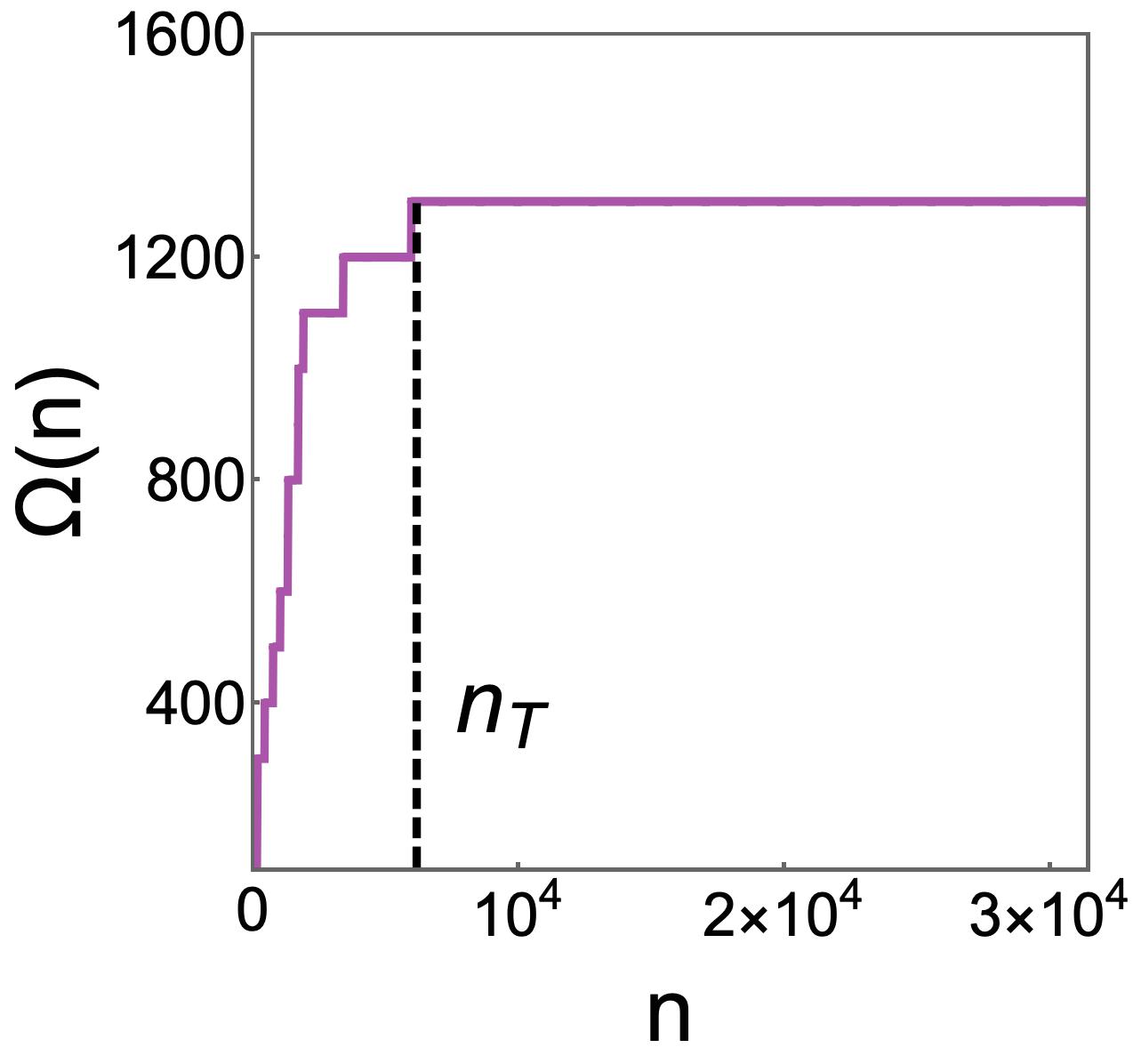}\\[-1mm]
   {\Large\textbf{(d)}}
  \end{minipage}
};

\draw[->,  line width=1.0pt, color=cyan!90!black]
 ([xshift=-35mm, yshift=6mm]A.east|-A.center)
  -- ([xshift=-2mm,yshift=-6mm]B.north west);

\draw[->,  line width=1.0pt, color=cyan!90!black]
  ([xshift=-35mm, yshift=6mm]A.east|-A.center)
  -- ([xshift=-2mm,yshift=-30mm]B.north west);
  \end{tikzpicture}

\caption{Trapping by caging of a Sokoban walker in two dimensions, starting from the origin. Panel~(a) displays the final configuration in a $50\times 50$ window, covering the ranges $55–105$ along the $x$-axis and $15–65$ along the $y$-axis, from a representative realization at $\rho=0.4$. The Sokoban walker (shown in red) is trapped inside a cage (shown in blue) formed by surrounding obstacles (shown in gray), and its position in the figure is $(80,40)$ (the walker initially is at the origin, outside the displayed window). Panel~(b) shows a magnified view of a part of the configuration responsible for trapping, with the trap size $A_{\rm T}=6$. In panel~(c), we show the same part of the configuration at the initial time and see that the trap did not exist initially. The Sokoban walker at some later time enters this region and modifies it into a trap in panel~(b) through successive interactions with the obstacles. Observe that in panel~(b) the allowed Sokoban moves cannot open the cage and therefore the walker is permanently confined inside this cage. As explained in the paper, this check is important for confirming that the caging is irreversible rather than a long-lived transient. Finally, in panel~(d), we have plotted the number of distinct visited sites, $\Omega(n)$, as a function of time $n$. This number is saturated for all $ n \geq n _{\rm T} $ with $n _{\rm T} = 6 \times 10^3$.}
\label{fig-trap-size}
\end{figure*}

The jump is successful if the target site is vacant; see Fig.~\ref{fig-update}(a-i). On the other hand, if the target site is occupied by an obstacle, the walker can still move one step by pushing the obstacle in the same direction. The pushing is allowed provided the next site beyond the obstacle is vacant; see panel~(a-ii) in the figure. If this condition is met, both the walker and the obstacle move one step in that direction. Otherwise, they do not move, as shown in panel~(a-iii). Hence, the Sokoban walker is capable of pushing the obstacles and it can at most push one obstacle in a given move in the direction of the attempted jump. In this paper, we focus on $d=1,2$.

\textit{$N_{\rm P}$-Sokoban model:} To test the universality of the trapping phenomenon, we also analyze variants of the Sokoban model with pushing rules modified relative to the one described above. As pointed out before, in one dimension, we consider the most general case where the Sokoban is capable of pushing up to an arbitrary $N_{\rm P}$ number of obstacles (``P" for ``pushing"). We will refer to this model as the $N_{\rm P}$-Sokoban walk. A schematic illustration of this model with $N_{\rm P}=2$ is shown in Fig.~\ref{fig-new-1d-scheme}. In this case, once $(N_{\rm P}+1)$ obstacles accumulate consecutively on one side of the walker, as shown in Fig.~\ref{fig-new-1d-scheme}(b), further pushing in that direction becomes impossible since the walker can push at most $N_{\rm P}$ obstacles only.

\textit{G-Sokoban model:} We also mentioned above a generalized two-dimensional variant of the Sokoban dynamics. Recall that in the Sokoban model, the obstacle in a given move can be pushed only in the direction of the attempted jump of the walker, provided that the site immediately beyond the obstacle in that direction is vacant; see Fig.~\ref{fig-update}(a-ii). Thus, the displacement of the obstacle is completely determined by the direction chosen by the walker for its motion. In the generalized model, this directional constraint is relaxed. As illustrated in Fig.~\ref{fig-update}(b-ii), when the walker attempts to move to a site containing an obstacle, the obstacle can be pushed to any one of its three neighboring sites with an equal probability $1/3$, provided that the chosen neighboring-site for the obstacle motion is vacant. Notice that the fourth neighboring-site of the obtacle contains the walker, and therefore the obstacle cannot move to that site. Moreover, if the other three neighboring-sites of the obstacle are not vacant, as shown in Fig.~\ref{fig-update}(b-iii), then the walker and the obstacle do not move.

We refer to this variant as the G-Sokoban walk (``G'' for ``generalized''). Compared to the Sokoban model, the G-Sokoban model enhances the obstacle mobility by allowing it to move to different available neighboring sites. At the same time, the basic ingredient of the walker having a limited pushing ability is kept intact since even the G-Sokoban walker can at most push only one obstacle. Our aim is to investigate whether the trapping phenomena observed in the Sokoban model are specific to its microscopic pushing rule or remain robust when the pushing dynamics is modified.

%In this extension, the obstacle can be pushed to any of its three neighbouring sites (except the site containing the walker) with equal probability $1/3$, provided that this target site is vacant. This is different than the two-dimensional Sokoban model introduced before, where the obstacle can be pushed only in the direction of the walker's attempted motion. We will refer to this variant as the G-Sokoban walk (``G" for ``generalized"). Our aim is to investigate whether the properties of the Sokoban model are model-specific or if they are common to other models with different pushing capabilities. 

\subsection{Reactive trapping: a brief survey of known results}
Before we proceed with our analysis, it is important to precisely explain what we mean by trapping. In the literature, a commonly studied form of trapping is reactive trapping \cite{Rosenstock, Balagurov1974, Donsker, Anlauf1984}. To understand it, consider a random walker moving through a medium of randomly placed obstacles and it is trapped immediately upon encountering an obstacle. The dynamics is terminated upon first encounter with an obstacle, and this mimics a perfect reaction center trapping the walker. 

In reactive trapping, the walker does not reshape its environment, and at first sight this setting may appear very different from the Sokoban model. However, the Sokoban model shares striking similarities with reactive trapping, while also exhibiting important differences that we analyze in later sections. For this reason, we briefly survey the relevant results for the reactive trapping problem.

%By contrast, in the caging case, the walker gets caged whenever the number of distinct sites visited by the walker saturates to a finite value. Once this saturation is achieved, the walker does not visit any new sites further and its motion is confined in a finite domain. This is referred to as trapping by caging, see the middle panel in Fig.~\ref{fig-trap-size}.

%In the literature, the reactive trapping turns out to be extensively studied and a quantity of keen interest is the survival probability $\phi (n)$ that the walker has not yet encountered any obstacle up to some time $n$ \cite{Balagurov1974, Rosenstock, Donsker, Weiss1980,  Grassberger1982, Redner-1983,Kayser1983,zumofen1983, Blumen1983-2, Anlauf1984, SHavlin-numerical-1984,Lubensky1984,SHavlin_PRL_1984,Weiss1d1984, Nieuwenhuizen1989, Gallos2001, Beijren2001}. Balagurov and Vaks \cite{Balagurov1974} and Donsker and Varadhan \cite{Donsker} proved that  this probability in $d$-dimension decays in the long-time limit as

A widely studied quantity in this context is the survival probability $\phi (n)$ that the walker has not yet encountered any obstacle up to some time $n$ \cite{Havlin-2, Weiss-1}. Balagurov and Vaks \cite{Balagurov1974} and Donsker and Varadhan \cite{Donsker} proved that this survival probability decays in the long-time limit, in $d$ dimensions, as
\begin{align}
\phi (n) \sim \exp \left( - \beta _{d} \lambda ^{\frac{2}{d+2}} n^{\frac{d}{d+2}} \right),~~~~ \text{with }n \lambda ^{2/d} \gg 1,
\label{original-eqn}
\end{align}
with $\lambda = |\ln(1-\rho)|$ and $\beta _ 1 =3\pi^{2/3}/2 $ and $ \beta _2 \approx 3.4$ in one and two dimensions \cite{Donsker, Rednerbook}. We refer to Eq.~\eqref{original-eqn} as the Balagurov-Vaks-Donsker-Varadhan (BVDV) theory, and it characterizes the asymptotic decay of the reactive-trapping-based survival probability at late times.

At intermediate times, an approximation due to Rosenstock \cite{Rosenstock, Anlauf1984} works well, especially at small densities:
\begin{align}
1- \phi (n)  \sim 
\begin{cases}
~~\sqrt{n \rho ^2}, & ~~~\text{in }1d , \\[4pt]
~~n \rho / \log n&~~~\text{in }2d. 
\end{cases}
\label{original-eqn-2}
\end{align} 
%However $\phi(n)$ does not characterize whether the walker is caged or not. For instance Eq.~\eqref{original-eqn} is valid even below the critical density in two dimensions \cite{SHavlin_PRL_1984, Gallos2001, Beijren2001}, when the walker percolates (\emph{i.e.,} escapes to infinity) at sufficiently long time. Moreover, the same expression for $\phi (n)$ is applicable even for the Sokoban model. From a physical standpoint, $\phi(n)$ depends only on the walker's first contact with a single obstacle. On the other hand, to know whether a walker is caged or not, one typically needs to look at its interaction with multiple obstacles. Consequently, $\phi(n)$ cannot be used to distinguish the caging/trapping properties for our system.

\noindent
%However, both of these results in Eqs.~\eqref{original-eqn} and \eqref{original-eqn-2} are related to the reactive trapping phenomenon. 
%Notably, for reactive trapping, the percolation transition becomes irrelevant, since the trapping occurs on the first encounter of the walker with an ``obstacle" (namely the obstacles are perfect reactive spots). If the walker is allowed to continue beyond this initial encounter, it may still visit hitherto unexplored regions, with the possibility of percolation depending on the obstacle density.

\noindent
One of our main findings is that the Sokoban model and its variants in one and two dimensions fall into the same universality class as reactive trapping, but with important differences that we demonstrate later.

\subsection{Trapping by caging}
%This brings us to the second form of trapping where the walker gets caged from all directions by the obstacles; see Figs.~\ref{fig-trap-size}(a,b,c). Absence of this form of caging means that the walker percolates to infinity. 
We now introduce cage-trapping, the process which will describe the Sokoban model. Here, trapping is induced by physical rearrangements of the environment, rather than by perfect chemical reaction between the walker and the obstacle. To characterize this form, we look at the number of distinct lattice sites $\Omega(n)$ visited by the Sokoban walker \cite{Numb-3, Numb-5, Numb-6}. \bluepw{For a given initial configuration of obstacles, $\Omega(n)$ is a nondecreasing function of $n$, and at each step it can either increase by one or remain unchanged. It saturates indefinitely only when the walker becomes caged from all directions by the obstacles and the Sokoban moves cannot reopen the cage. For example, in Fig.~\ref{fig-trap-size}(a,b,c), we show a schematic of the final configuration in which the walker has been caged. Fig.~\ref{fig-trap-size}(b) particularly illustrates that the Sokoban cannot reopen this cage and is therefore irreversibly confined inside the cage. As a result, $\Omega(n)$ saturates indefinitely to a finite value, as shown in panel~(d).} Therefore, by monitoring whether $\Omega \left(n \right)$ has indefinitely saturated or not, we can say whether the walker is cage-trapped or not for that trajectory.

%\bluew{In simulation, we cannot reach $n \to \infty$. So we adopt the following operational caging criterion. For each trajectory, we identify the first time at which $\Omega(n)$ appears to saturate, and we continue the simulation for an additional time window of length $\sim 100$ times this saturation time. If $\Omega(n)$ remains unchanged throughout this entire window, we classify the walker as caged.}

%Moreover, using $\Omega(n)$ in our description allows us to infer the caging without explicitly inspecting the environment, which can be computationally expensive. In this paper, we focus mainly on trapping by caging and we will, for simplicity, refer to this form as the trapping phenomenon. 

In this paper, we focus mainly on this trapping by caging and we will, for simplicity, refer to this form as the trapping phenomenon. Consider now the trapped scenario, i.e., $\Omega(n)$ has saturated indefinitely to a finite value for a given initial configuration and a given realization of the walker's stochastic dynamics. Let $n_{\rm T}$ be the time when $\Omega(n)$ attains its saturation value for the first time. This is the time when the walker completes its exploration of new distinct sites and it is clearly a random quantity. We will refer to $n_{\rm T}$ as the trapping time. Averaging it over initial obstacle configurations and the walker's stochastic dynamics, we obtain the average $ \langle n_{\rm T} \rangle $.

To infer trapping by caging, we have to monitor $\Omega(n \to \infty)$. However, in simulations, we cannot evolve the system for an infinite time and we therefore introduce a simulation stopping time $\tau_{\rm sim}$, which sets the duration for which we run each realization. In our work, we choose $\tau_{\rm sim} \sim 10^{3}-10^8$ depending on obstacle densities. Operationally, for each realization we record the final value $\Omega(\tau_{\rm sim})$ and then scan the trajectory backward to identify the earliest time $n_{\rm T}$ such that $\Omega(n_{\rm T})=\Omega(\tau_{\rm sim})$. For obstacle densities considered in this work, the chosen value of the simulation time is at least $\tau_{\rm sim} \sim 100 \langle n_{\rm T} \rangle $. This ensures that $\Omega(n)$ has reached its final value well before $\tau_{\rm sim}$ and remains unchanged at least over a time window $100 \langle n_{\rm T} \rangle$. \bluepw{In two dimensions, we further validate caging by inspecting the walker's environment at time $\tau _{\rm sim}$ and verifying that the Sokoban dynamics cannot open the cage, as explained before in Fig.~\ref{fig-trap-size}(b). This check confirms that the caging is irreversible. Otherwise, the plateau in $\Omega(n)$ could simply correspond to a long transient during which the walker takes a very long time to discover a new site. Given sufficient time, the walker could then escape the reversible cage, visit a new site, and cause $\Omega(n)$ to increase again. However, inspection of the environment rules out this possibility and ensures that the observed saturation of $\Omega(n)$ corresponds to irreversible caging and the motion is permanently confined inside this cage.}

%Importantly, since the value of$\Omega(n)$ may only increase by one at each move, its saturation in the given time window can be detected unambiguously for each realization; see Fig.~\ref{fig-trap-size}(d). Further details on the numerical simulations is provided in Appendix \ref{appen-simulation}.

%The chosen value of $\tau_{\rm sim}$ is sufficiently long that, for a given trajectory, once $\Omega(n)$ first saturates at $n = n_{\rm T}$ it remains unchanged over an extended window, at least up to $n \sim  100 n_{\rm T}$. 

Apart from $\langle n_{\rm T} \rangle $, we are also interested in the trap size $A_{\rm T}$, the number of vacant sites inside the trap. Once again, we have schematically illustrated $A_{\rm T}$ in Fig.~\ref{fig-trap-size} and the procedure of generating it in numerical simulations is provided in Appendix \ref{appen-simulation}. Averaging the trap size over initial obstacle configurations and the walker's stochastic dynamics, we obtain the average $\langle A_{\rm T} \rangle$.

In addition, we will also study the probability that $\Omega(n)$ has not attained its saturation value till time $n$. Physically, this represents the disorder-averaged survival probability $S(n)$ that the walker has not been caged up to time $n$. This is different from $\phi(n)$ introduced before in the reactive-trapping context. Remember that $\phi(n)$ depends only on the walker's first encounter with an obstacle. On the other hand, $S(n)$ depends on the walker's interaction with multiple obstacles. We study $S(n)$, $\langle n_{\rm T} \rangle $ and $ \langle A_{\rm T} \rangle $ for the Sokoban model and its variants introduced above. Let us now summarize the main results of our paper.

\subsection{Summary of results}
\begin{itemize}
\item \textit{Stretched-exponential relaxation:} We show that the survival probability, $S(n)$, exhibits a stretched-exponential decay at late times
\begin{align}
S(n) \sim \exp \left[ -f_d (\rho) n ^{\mu_d} \right],  \label{main-result-eq-2}
\end{align}
with exponents $\mu _1 = 1/3$ and $\mu _2 = 1/2$ in one and two dimensions, respectively. We analytically prove that $f_1 (\rho) = 3 \pi ^{\frac{2}{3}} |\ln(1-\rho) |^{\frac{2}{3}} / 2^{\frac{5}{3}} $ in one dimension. In two dimensions, we use numerical simulations to demonstrate Eq.~\eqref{main-result-eq-2} and extract an estimate of $f_2(\rho)$. Interestingly, both $\mu_1$ and $\mu _2$ exponents match with the BVDV formula in Eq.~\eqref{original-eqn}.

\item \textit{Moderate-time behavior:} Eq.~\eqref{main-result-eq-2} captures the long-time behavior of $S(n)$. For small to moderate values of $n$, we show 
\begin{align}
1- S (n)  \sim 
\begin{cases}
~~n^2 \rho^4  &~~~\text{for }1d \text{ Sokoban} , \\[4pt]
~~n ^{\epsilon(\rho)}, & ~~~\text{for }2d \text{ Sokoban}, 
\end{cases}
\end{align}
where the exponent $\epsilon(\rho)$ depends on the density. This behavior is qualitatively different from the moderate-time Rosenstock approximation for $\phi(n)$ in Eq.~\eqref{original-eqn-2}. This highlights how, despite similarities in long-time scaling, the short to moderate time dynamics of the two survival probabilities, $\phi(n)$ and $S(n)$, are completely different.

\item \textit{Generalized Sokoban models:} For the $N_{\rm P}$-Sokoban model, we prove that the survival probability $S(n)$ has a large-deviation form in the joint limit $N_{\rm P} \gg 1$ and $n \gg 1$ while keeping the ratio $\omega=\frac{n}{\left( N_{\rm P} \right)^3} $ fixed
\begin{align}
 \lim_{\substack{N_{\rm P} \gg 1 , n \gg 1 \\    \omega=n/ \left( N_{\rm P} \right)^3  \text{ fixed}         }} \frac{-\ln S(n)}{ N_{\rm P}}  =   \mathcal{I}(\omega)  . \label{main-result-eq-3}
\end{align}
The corresponding rate function $\mathcal{I}(\omega)$ is given by Eqs.~\eqref{gen-LDF-eq-8} and\eqref{gen-LDF-eq-9}. It has the asymptotic behaviors
\begin{align}
\mathcal{I}( \omega) \simeq
\begin{cases}
\frac{\pi^2 \rho ^2}{32(1-\rho)^2} \omega , & ~~\text{for small }  \omega, \\[4pt]
\frac{3 \pi ^{2/3} \lambda ^{2/3}}{2^{5/3}}~ \omega^{1/3} &~~\text{for large }  \omega,
\end{cases}
\label{main-result-eq-4}
\end{align} 
where the precise meaning of the large and small $\omega$ limits will be explained later in Eqs.~(\ref{gen-LDF-eq-10s}-\ref{gen-LDF-eq-10}). Eq.~\eqref{main-result-eq-4} implies that the survival probability changes from an exponential decay at $n \ll \left( N_{\rm P} \right)^3$ to the same $1/3$-stretched-exponential decay as in Eq.~\eqref{main-result-eq-2} at $n \gg \left( N_{\rm P} \right)^3 $. Thus, the long-time stretched-exponential decay of $S(n)$ is completely universal, independent of $N_{\rm P}$.
%Therefore, the long-time stretched-exponential decay of $S(n)$ is completely universal in one dimension and belongs to the BVDV trapping universality class, as described by Eq.~\eqref{original-eqn}. 

\quad For the two-dimensional G-Sokoban model, the survival probability, $S(n)$, is also characterized by the exponent $\mu_2 =1/2$, same as in Eq.~\eqref{main-result-eq-2}. Thus, modifying the microscopic pushing rules does not qualitatively alter the long-time stretched-exponential exponent even in two dimensions.
%Hence, in both one and two dimensions, the BVDV exponents are completely robust against modifications in the pushing rules of the walker.
%exhibits the same long-time exponent $\mu_2 =1/2$ (same as BVDV formula), albeit with a different $f_2(\rho)$.

%Therefore, the long-time stretched-exponential decay of $S(n)$ is completely universal in one dimension and it belongs to the BVDV trapping universality class in Eq.~\eqref{original-eqn}. Similarly, for the G-Sokoban model, a variant of the Sokoban model in two dimensions as introduced before, the survival probability $S(n)$ again has the same long-time decay as the BVDV formula.

\item \textit{Emergence of a trapping crossover:} For the two-dimensional Sokoban model, our simulations also suggest that the percolation transition is lost \cite{Shlomi-1}. However, our study goes further to reveal that the Sokoban model nonetheless exhibits a dynamical crossover in the underlying trapping mechanisms. We characterize the crossover in terms of the average trap size $\langle A_{\rm T} \rangle$ being a nonmonotonic function of the obstacle density $\rho$: $\langle A_{\rm T} \rangle$ initially increases as $\rho$ decreases, reaches a maximum at $\rho_*$, and then diminishes below this threshold; see Fig.~\ref{fig-2d-area-neww} later. For $\rho > \rho_*$, trapping is dominated by pre-existing cages in the initial obstacle configuration, whereas for $\rho < \rho_*$ it is dominated by self-created cages generated dynamically by the walker through interactions with obstacles; also see Fig.~\ref{fig-trap-size}(b,c). Using numerical simulations, we estimate $\rho _{*} \approx 0.55$ for the Sokoban model and $\rho _{*} \approx 0.675$ for the G-Sokoban model. 
%The emergent self-trapping mechanism cages the Sokoban and G-Sokoban walkers at low densities.
 %On the other hand, for the AIL model, the plot of $\langle A_{\rm T} \rangle$ versus $\rho$ is monotonic, as shown in the right panel of Fig.~\ref{turnover-fig}. 

%\quad
%In these models, the density $\rho_*$ represents a threshold that separates two distinct trapping mechanisms. Above this threshold, a pre-existing trapping regime dominates, where confinement arises due to the initial arrangement of obstacles. Below $\rho _*$, a self-trapping regime emerges, in which the walker becomes dynamically localized within a trap that it actively forms by reorganizing its local environment. This dynamical localization prevents the walker from percolating to infinity at low densities.

%While the dependence is monotonic in the AIL model, the mean trap size exhibits a nonmonotonic behavior with $\rho$ in the Sokoban and G-Sokoban models, see Fig.~\ref{turnover-fig}. It attains a maximum value at a characteristic density $\rho_*$. We identify two competing mechanisms that dominate the trapping behavior at densities below and above this threshold $\rho_*$. Using numerical simulations, we estimate $\rho _{*} \approx 0.55$ for the Sokoban model and $\rho _{*} \approx 0.675$ for the G-Sokoban model.
\end{itemize}

%in the mechanism by which confining cages are formed. Owing to this transition, the average trap size $\langle A_{\rm T} \rangle $ possesses a nonmonotonic dependence on $\rho$, see Fig.~\ref{turnover-fig}. The maximum average trap size occurs at some characteristic density $\rho_*$. Based on the numerical simulations, we estimate $\rho _{*} \approx 0.55$ for the Sokoban model and $\rho _{*} \approx 0.675$ for the G-Sokoban model. We identify two competing mechanisms that dominate trapping behavior at densities below and above $\rho_*$, respectively. Such nonmonotonicity , however, does not arise for $1d$ models (with or without pushing) or for the $2d$ AIL model, see the right panel of the above figure. 
\noindent
In what follows, we will derive all of the above results analytically in one dimensions. Building on the insights gained, we will then look at the two-dimensional models using extensive numerical simulations.

%\begin{figure}[]
%	\centering
%	\includegraphics[scale=0.2]{turnover-scheme.jpeg}
%	\hspace{0.3 cm}
%	\includegraphics[scale=0.2]{no-turnover-scheme.jpeg}
%	\caption{Schematic illustration of the mean trap size $\langle A_{\rm T} \rangle $ as a function of $\rho$, highlighting the presence or absence of nonmonotonic behavior. The presence of a maximum average trap size is observed only for the two-dimensional Sokoban and G-Sokoban models at densities $\rho _{*} \approx 0.55$ and $\rho _{*} \approx 0.675$ respectively (see Fig.~\ref{fig-2d-area-neww}). For the AIL models,  we find a monotonic dependence of $\langle A_{\rm T} \rangle $ on $\rho$. The density $\rho _*$ marks a crossover from the low-density self-trapping to the pre-existing traps a high density.	}
%\label{turnover-fig}
%\end{figure}

\section{$N_{\rm P}$-Sokoban model in one dimension}
\label{sec-1d-1}
\bluew{For the Sokoban walker moving in one dimension (say $x$ axis) and placed initially at the origin, the total space available for the motion, for a fixed \textit{initial} obstacle configuration, is determined by the number of obstacles it can push. For every initial realization of obstacle positions, the walker can push up to $N_{\rm P}$ obstacles on the positive $x$-axis and up to $N_{\rm P}$ obstacles on the negative $x$-axis. In each direction ($+x$ or $-x$), the obstacles can be pushed up to the point where the first $\left( N_{\rm P} +1\right)$ obstacles in that direction are situated on consecutive sites. Beyond this point, the walker cannot push obstacles further in that direction for that particular realization. For illustration, Fig.~\ref{fig-new-1d-scheme} shows the case $N_{\rm P}=2$. Panel~(a) shows the configuration at the initial time with walker at the origin. Panel~(b) shows the configuration at a later time where obstacles have been maximally pushed and cannot be pushed further. At this point, the first three obstacles in both $+x$ and $-x$ directions are located on consecutive sites. In the figure, observe that the third obstacle on each side (shown in green) remains fixed throughout the evolution.}

Likewise, for general $N_{\rm P}$, in each direction the $(N_{\rm P}+1)$-th obstacle remains immobile, while only the first $N_{\rm P}$ obstacles can be pushed until the first $(N_{\rm P}+1)$ obstacles in that direction occupy consecutive sites, as mentioned before. For a given initial realization of obstacle configuration, we denote the position of the $(N_{\rm P}+1)$-th obstacle in the $-x$ direction by $-Y_{O_{N_{\rm P}+1}}^{-}$, and the position of the $(N_{\rm P}+1)$-th obstacle in the $+x$ direction by $Y_{O_{N_{\rm P}+1}}^{+}$. Then, in this realization the walker is confined inside the interval $[-L_1, L_2]$ throughout its evolution such that
\begin{align}
&  \begin{matrix}
L_1 = \left(Y_{O_{N_{\rm P}+1}}^{-} -N_{\rm P}-1\right), \\
L_2 =  \left(Y_{O_{N_{\rm P}+1}}^{+} -N_{\rm P}-1\right).
\end{matrix}
\label{hagqo}
\end{align}
and reflecting conditions at the two ends; also see  Fig.~\ref{fig-new-1d-scheme}. Since the obstacles are initially distributed randomly, both $L_1$ and $L_2$ are also random quantities. Nonetheless, they are bounded from below as $L_1 \geq 0,~~L_2 \geq 0$. The idea now is to fix the values of $L_1$ and $L_2$ and derive the survival probability, trap size, and trapping time conditioned on these values. These conditional quantities will then be averaged over the probabilities of $L_1$ and $L_2$ to obtain the disorder-averaged observables.

\begin{figure*}[]
	\centering
	\includegraphics[scale=0.27]{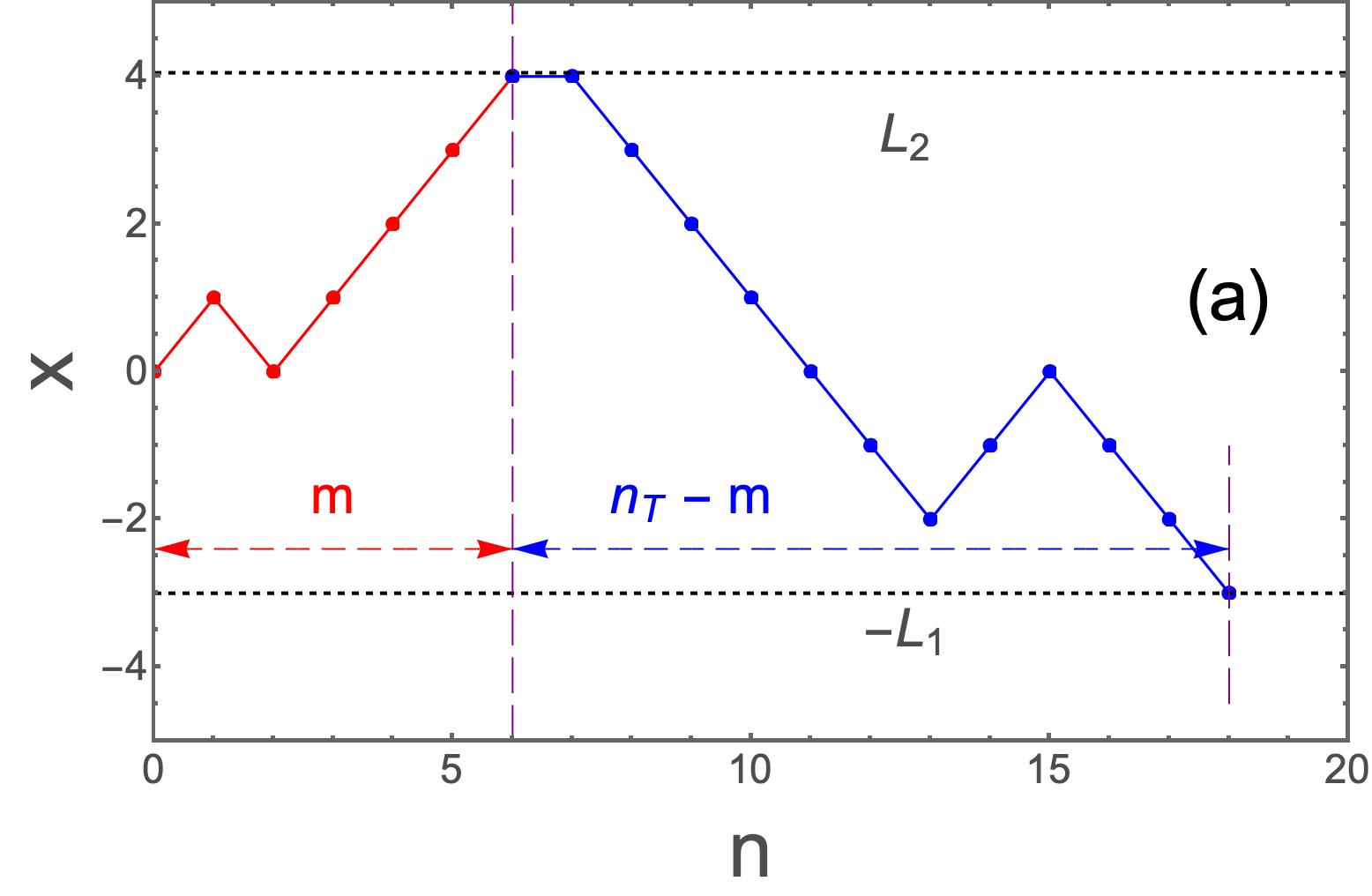}
	\includegraphics[scale=0.27]{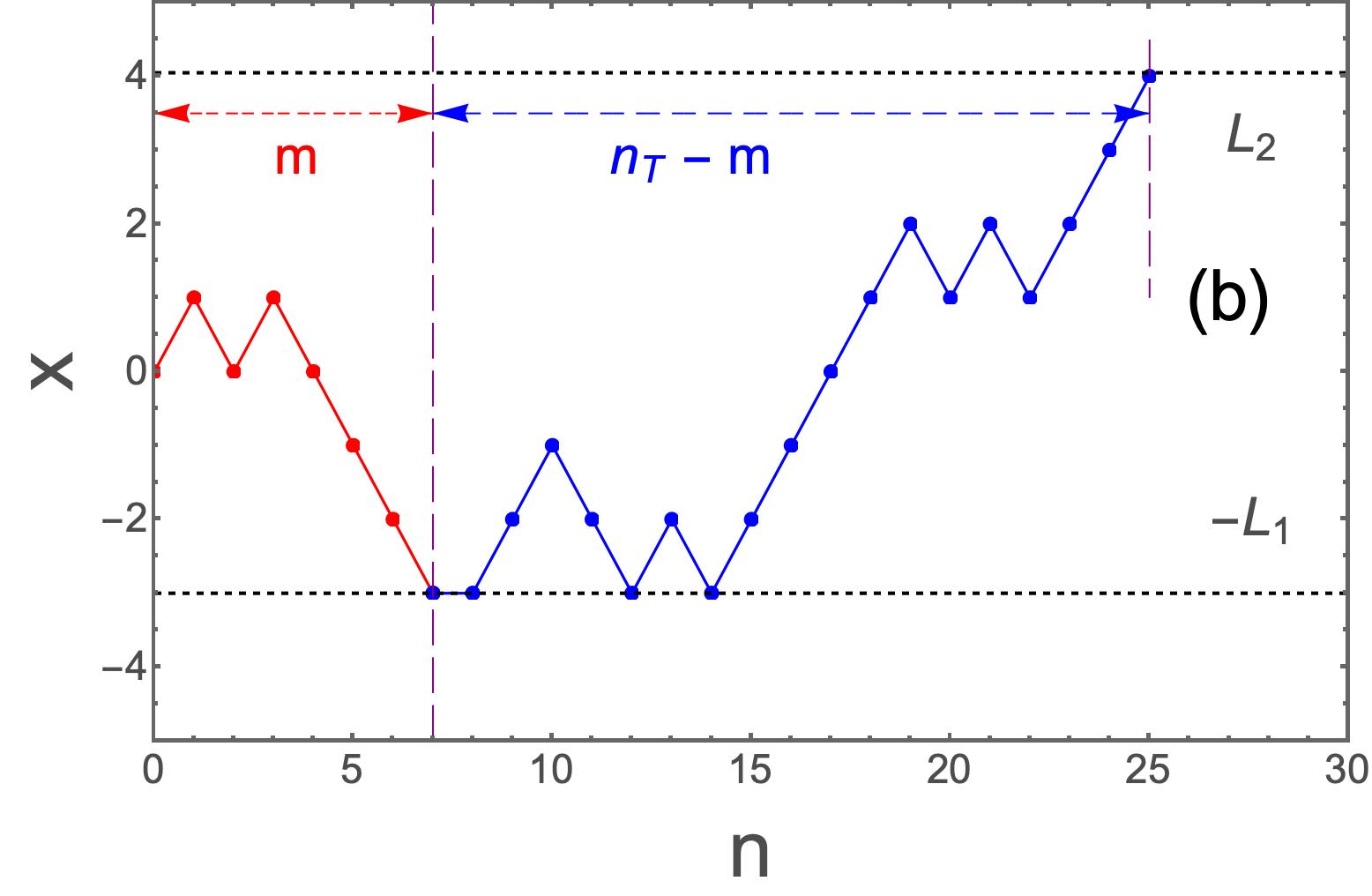}
	\caption{(a) A schematic illustration of the trajectory of the random walker, where it reaches \( x = -L_1 \) at the \( m \)-th time step and subsequently reaches \( x = L_2 \) at a later time step \( n_{\rm T} \). For the shown trajectory, $L_1=3$, $L_2=4$, $m=7$ and $n_{\rm T} = 25$. We have broken the trajectory in two parts: from $0$ to $m$ shown in red and from $m$ to $n_{\rm T}$ shown in blue. (b) Here we have shown an example of the second type of trajectories where the walker first reaches $x = L_2$ at the $m$-th time step and then reaches $x = -L_1$ at $n_{\rm T}$ time step. For this trajectory, $L_1=3$, $L_2=4$, $m=6$ and $n_{\rm T}=18$.	}
	\label{traj-scheme-1}
\end{figure*}

\subsection{Conditional survival dynamics for fixed $L_1$ and $L_2$}
For given $L_1$ and $L_2$, the walker will be trapped when the number of distinct sites it has visited saturates and no longer increases with time. As illustrated in Fig.~\ref{fig-new-1d-scheme}(b), this will happen whenever the walker has visited both the boundary sites, $-L_1$ and $L_2$, at least once. Let $n_{\rm T}$ be the first time it hits one of the interval boundaries given that it has previously hit the other one at least once. We denote this probability by $P \left( n_{\rm T} | L_1, L_2 \right)$ for given values of $L_1$ and $L_2$. As mentioned before, our eventual goal in this section is to calculate the disorder-averaged probability $\langle P \left( n_{\rm T} | L_1, L_2 \right) \rangle _{L_1, L_2}$. Throughout this section, we will use the notation $\langle \cdots \rangle _{L_1, L_2}$ to denote the expectation value with respect to the probabilities of $L_1$ and $L_2$.

To proceed, we will calculate various terms that contribute to $P \left( n_{\rm T} | L_1, L_2 \right)$. We begin by considering the simplest scenario where $L_1=0,~L_2 = 0$ (or equivalently $L=(L_1+L_2) = 0$). In this case, the walker is immediately trapped at its initial position. Consequently, one has
\begin{align}
P \left( n_{\rm T} | L_1=0, L_2=0 \right) = \delta _{n_{\rm T},0}.  \label{abxiqbz}
\end{align}
Here $ \delta _{n_{\rm T},0} $ stands for the Kronecker delta. Next, we consider the other case when $L \neq 0$, \emph{i.e.}, at least one of $L_1$ or $L_2$ is non-zero. Two possible scenarios can occur. The first is when the walker hits $x=-L_1$ first at least once and then reaches $x=L_2$ at some later time $n_{\rm T}$, see Fig.~\ref{traj-scheme-1}(a). The second scenario is the complementary one, where the walker first reaches $x=L_2$ and then hits $x=-L_1$ at time $n_{\rm T}$, see Fig.~\ref{traj-scheme-1}(b). We begin with the first case and denote by $m$ the first time the walker reaches $x = -L_1$. As shown in panel~(a), we now divide the whole trajectory into two parts: (i) the $[0,m]$ part shown in red and (ii) the $[m,n_{\rm T}]$ part shown in blue. In the red part, the walker starts from the origin and makes its first visit to $-L_1$ at time $m$ but with the constraint that it has not reached $L_2$ before. Thus the statistical weight of this part is equal to the first-passage probability $\mathcal{F}_1 \left( m|x_0 = 0, \{ x_{j} \} \big|_{j=1}^{m-1} < L_2  \right)$ to $-L_1$ such that the positions $x_1, x_2,\cdots, x_{m-1}$ of the random walk at all intermediate times are smaller than $L_2$. For the blue part, the walker starts at $x_0 = -L_1$ and reaches $L_2$ for the first time in the remaining time interval $ \left( n_{\rm T}-m \right)$, see Fig.~\ref{traj-scheme-1}(a). Hence, the statistical weight of this segment is equal to the first-passage probability $\mathcal{F}_2 \left(n_{\rm T}-m |x_0 = -L_1 \right)$ to $L_2$ without any constraint at intermediate times.

One can now express the conditional probability of observing a trapping time $n_{\rm T}$ for this class of trajectories as
\begin{align}
P_{\mathcal{C}_1} \left( n_{\rm T} | L_1, L_2 \right) =& \sum _{m=0}^{n_{\rm T}}  \mathcal{F}_1 \left( m|x_0 = 0, \{ x_{j} \} \big|_{j=1}^{m-1} < L_2  \right)  \nonumber \\ 
&~~~~~ \times  \mathcal{F}_2 \left(n_{\rm T}-m |x_0 = -L_1 \right).
\label{cont-eq-1}
\end{align}
Here the subscript `$\mathcal{C}_1$' in $P_{\mathcal{C}_1} \left( n_{\rm T} | L_1, L_2 \right) $ is used to indicate that we are looking at the probability conditioned on the fact that the walker first hits $x = -L_1$ and then reaches $x=L_2$ at some later time. Moreover, we have utilized the property that the two segments (red and blue) of the trajectory are statistically independent, allowing us to write the total probability as a sum of the product of the individual weights from the two segments.

We next consider the other case, in which the walker hits $x=L_2$ first and then reaches $x=-L_1$ at some later time. As shown in Fig.~\ref{traj-scheme-1}(b), one can again apply the same reasoning as before, dividing the trajectory into two segments and calculating the weights from each. The conditional probability from this class of trajectories is then given by
\begin{align}
P_{\mathcal{C}_2} \left( n_{\rm T} | L_1, L_2 \right) & =  \sum _{m=0}^{n_{\rm T}}  \mathcal{F}_2 \left( m|x_0 = 0, \{ x_{j} \} \big|_{j=1}^{m-1} >- L_1 \right) \nonumber \\
 & ~~~ \times \mathcal{F}_1 \left(n_{\rm T}-m |x_0 = L_2 \right),
\label{cont-eq-2}
\end{align}
where `${\mathcal{C}_2}$', as before, emphasizes the conditional nature of the probability. Summing Eqs.~(\ref{abxiqbz}-\ref{cont-eq-2}), we obtain the total probability $P \left( n_{\rm T} | L_1, L_2 \right)$ as 
\bluepw{\begin{equation}
\scalebox{0.85}{$
\begin{split}
& P \left( n_{\rm T} | L_1, L_2 \right) =  \delta _{n_{\rm T},0}\delta _{L_1+L_2,0}  + \sum _{m=0}^{n_{\rm T}} \Big[ \mathcal{F}_2 \left(n_{\rm T}-m |x_0 = -L_1 \right) \Big. \\
&\Big. \times \mathcal{F}_1 \left( m|x_0 = 0, \{ x_{j} \} \big|_{j=1}^{m-1} < L_2 \right) +\mathcal{F}_1 \left(n_{\rm T}-m |x_0 = L_2 \right) \Big.   \\
&~~\Big. \times  \mathcal{F}_2 \left( m|x_0 = 0, \{ x_{j} \} \big|_{j=1}^{m-1} >- L_1 \right)  \Big] \left(1-\delta_{L_1+L_2,0} \right). 
\end{split}$} \label{sok-surv-eq-1}
\end{equation}}
%Here $\Theta (z)$ stands for the Heaviside theta function, which takes the value $1$ if $z>0$ and $0$ otherwise.
Eq.~\eqref{sok-surv-eq-1} indicates that the problem of finding the trapping time probability has now been reduced to calculating various first-passage probabilities. For the random walk model, it turns out to be possible to calculate these probabilities exactly, and the details of this calculation are provided in Appendices~\ref{appen-FPT-1d},~\ref{appen-FPT-1d-2},~\ref{app:filtration}. Their final expressions read as
\begin{widetext}
%\twocolumngrid
\begin{align}
  & \mathcal{F}_1 \left( m|x_0 = 0, \{ x_{j} \} \big|_{j=1}^{m-1} < L_2 \right)    =  \frac{\left(1-\delta_{L_1,0} \right) \left(1-\delta_{m,0} \right)}{L}~\sum _{\theta = 1}^{L} \cos ^{m-1} \left( \frac{\pi \theta}{L} \right)~ \sin \left( \frac{\pi \theta L_1}{L} \right)  ~\sin \left(  \frac{ \pi \theta }{L} \right)+\delta _{L_1,0} \delta _{m,0},  \label{sok-surv-eq-2} \\
  & \mathcal{F}_2 \left( m|x_0 = 0, \{ x_{j} \} \big|_{j=1}^{m-1} >- L_1 \right)   =  \frac{\left(1-\delta_{L_2,0} \right) \left(1-\delta_{m,0} \right)}{L}~\sum _{\theta = 1}^{L} \cos ^{m-1} \left( \frac{\pi \theta}{L} \right)~ \sin \left( \frac{\pi \theta L_2}{L} \right)  ~\sin \left(  \frac{ \pi \theta }{L} \right)+\delta _{L_2,0} \delta _{m,0},   \label{sok-surv-eq-3} \\
  & \mathcal{F}_2 \left( m|x_0 = -L_1 \right)   = \frac{\left(1-\delta_{L,0} \right) \left(1-\delta_{m,0} \right)}{ \left( L + \frac{1}{2} \right)} ~\sum _{\theta = 0 }^{L}      (-1)^{\theta} \cos^{m-1} \left( \frac{\pi (2 \theta +1)}{2 \left( L + \frac{1}{2} \right)  }  \right)~\cos \left( \frac{\pi (2 \theta +1)}{4 \left( L + \frac{1}{2} \right)  }  \right) ~ \sin \left( \frac{\pi (2 \theta +1)}{2 \left( L + \frac{1}{2} \right)  }  \right),  \label{sok-surv-eq-4}
\end{align}
with $L = (L_1+L_2$). Also, by symmetry $\mathcal{F}_2 \left( m|x_0 = -L_1 \right) = \mathcal{F}_1 \left( m|x_0 = L_2 \right)$.
Plugging these expressions in Eq.~\eqref{sok-surv-eq-1}, we obtain the exact form of the trapping time probability $P \left( n_{\rm T} | L_1, L_2 \right)$. To check that this probability is correctly normalized, we first sum Eq.~\eqref{sok-surv-eq-1} over all possible values of $n_{\rm T}$ as
\begin{equation}
\scalebox{0.95}{$
\begin{split}
 \sum _{n_{\rm T}  = 0}^{\infty} P \left( n_{\rm T} | L_1, L_2 \right)  & = \delta _{L_1+L_2,0}+  \left(1-\delta_{L_1+L_2,0} \right) \sum _{m=0}^{\infty}   \mathcal{F}_1 \left( m|x_0 = 0, \{ x_{j} \} \big|_{j=1}^{m-1} < L_2 \right)  ~ \sum _{n=0}^{\infty} \mathcal{F}_2 \left(n |x_0 = -L_1 \right) \\
& +  \left(1-\delta_{L_1+L_2,0} \right)  \sum _{m=0}^{\infty} \mathcal{F}_2 \left( m|x_0 = 0, \{ x_{j} \} \big|_{j=1}^{m-1} >- L_1 \right)  ~\sum _{n=0}^{\infty} \mathcal{F}_1 \left(n |x_0 = L_2 \right).
\end{split}$}  \nonumber
\end{equation}
We now use the expressions in Eqs.~(\ref{sok-surv-eq-2}-\ref{sok-surv-eq-4}) to yield
\begingroup
\small
\begin{align}
& \sum _{m=0}^{\infty}   \mathcal{F}_1 \left( m|x_0 = 0, \{ x_{j} \} \big|_{j=1}^{m-1} < L_2 \right)   = \frac{L_2}{L},~~ \sum _{m=0}^{\infty} \mathcal{F}_2 \left( m|x_0 = 0, \{ x_{j} \} \big|_{j=1}^{m-1} >- L_1 \right)  = \frac{L_1}{L}, \nonumber \\
& ~~~~~~~~~~~~~~~~~~~~~~~~~~~\sum _{m=0}^{\infty} \mathcal{F}_2 \left(m |x_0 = -L_1 \right) = \sum _{m=0}^{\infty} \mathcal{F}_1 \left(m |x_0 = L_2 \right) =1. \label{abkmao1a}
\end{align}
\endgroup
\end{widetext}
%\twocolumngrid
The proof of these relations is relegated to Appendices~\ref{appen-FPT-1d},~\ref{appen-FPT-1d-2}. Using them, it now follows
\begin{align}
\sum _{n_{\rm T}  = 0}^{\infty} P \left( n_{\rm T} | L_1, L_2 \right) =  1.
\end{align}
Thus, our theoretical expression in Eq.~\eqref{sok-surv-eq-1} for the trapping time probability for fixed $L_1$ and $L_2$ is correctly normalized. Having obtained this probability, we can now look at the conditional survival probability that the Sokoban walker has not been trapped till time $n$
\begin{align}
Q \left( n|L_1, L_2 \right) = 1-\sum _{n_{\rm T}=1}^{n} P \left( n_{\rm T} | L_1, L_2 \right). \label{new-pahy1}
\end{align}
Given that we have an exact expression for $P \left( n_{\rm T} | L_1, L_2 \right)$, it follows that $Q \left( n|L_1, L_2 \right)$ is also exactly known. One can now utilize this exact expression to derive various asymptotic behaviors of $Q \left( n|L_1, L_2 \right)$. In particular, 
we analyze $Q \left( n|L_1, L_2 \right)$ in the joint limit of large $n$ and large $L$. The intuition behind this limit is that the walker can avoid being trapped over a long time interval only if the spatial region over which it moves is also large. This leads us to consider the diffusive scaling where both $n$ and $L$ are large but the ratio $n/L^2$ is finite. We show in Appendix \ref{appen-surv-longn} that one obtains different expressions depending on whether the scaled variable $n/L^2$ is large or small. For the case of large $n/L^2$, we find
\begin{equation}
\scalebox{0.9}{$
\begin{split}
Q \left( n|L_1, L_2 \right) \simeq \frac{4}{\pi}~\left[ \sin \left( \frac{\pi  L_1}{2L} \right) + \sin \left( \frac{\pi  L_2}{2L} \right)  \right]~\exp \left(  -\frac{ \pi ^2 n}{8L^2} \right),
\end{split}$} \label{sok-surv-eq-18}
\end{equation}
%\begin{align}
%Q \left( n|L_1, L_2 \right) \simeq & ~\frac{4\Theta(L)}{\pi}~\left[ \sin \left( \frac{\pi  L_1}{2L} \right) + \sin \left( \frac{\pi  L_2}{2L} \right)  \right] \nonumber \\
%& \times \exp \left(  -\frac{ \pi ^2 n}{8L^2} \right),~~~\text{for }n/L^2 \gg 1 \label{sok-surv-eq-18}
%\end{align}
while for small $n/L^2$, we get
\begin{equation}
\scalebox{0.9}{$
\begin{split}
Q \left( n|L_1, L_2 \right) \simeq &1- 2 \left[  \text{erfc} \left( \frac{L_1+2 L_2}{\sqrt{2n}}  \right) +\text{erfc} \left( \frac{2L_1+ L_2}{\sqrt{2n}}  \right)  \right],
\end{split}$}\label{sok-surv-eq-18kj08h1}
\end{equation}
%\begin{align}
%Q \left( n|L_1, L_2 \right) \simeq &1- 2 \left[  \text{erfc} \left( \frac{L_1+2 L_2}{\sqrt{2n}}  \right) +\text{erfc} \left( \frac{2L_1+ L_2}{\sqrt{2n}}  \right)  \right],\label{sok-surv-eq-18kj08h1}
%\end{align}
where $\text{erfc(z)}$ represents the complementary error function. Eqs.~(\ref{sok-surv-eq-18}-\ref{sok-surv-eq-18kj08h1}) give the survival probabilities conditioned on the fixed values of $L_1$ and $L_2$.

Let us next turn to the conditional trap size and trapping time. Recall that we define trap size $A_{\rm T}$ as the number of vacant sites available for the walker to move when it is trapped.
As seen in Fig.~\ref{fig-new-1d-scheme}(b), this will be equal to 
\begin{align}
A_{\rm T} (L_1,L_2) =( L_1+L_2+1). \label{eq-trapLL}
\end{align}
On the other hand, due to the stochastic motion of the walker, the trapping time, $n_{\rm T} (L_1,L_2)$ can take any integer value. In Appendix~\ref{appen-surv-longn}, we show that the conditional average $\langle n _{\rm T} \mid L_1, L_2 \rangle$ for both $L_1, L_2 \gg 1$ is given by
\begin{align}
\langle n _{\rm T} \mid L_1, L_2 \rangle \simeq  L_1^2+L_2^2+3 L_1 L_2. \label{new-pssR-eq-4}
\end{align}
As expected for a diffusive process, the mean time exhibits a quadratic dependence on $L_1$ and $L_2$, with an additional term that is linear in the product $L_1 L_2$. This expression, as shown later, will be useful for small densities of obstacles. Contrarily, for high-density values, the leading behavior will be described by $L_1$ and $L_2$ of unit order and here, we get 
\begin{align}
\langle n _{\rm T} \mid L_1, L_2 \rangle \simeq 2 \left( \delta_{L_1,0} \delta _{L_2,1}+ \delta_{L_1,1} \delta _{L_2,0} \right). \label{new-pssR-eq-4-high}
\end{align}
Once again we emphasize Eqs.~(\ref{eq-trapLL}-\ref{new-pssR-eq-4-high}) are all derived for fixed $L_1$ and $L_2$. 
 
\begin{figure*}[t]
 %\noindent\hspace*{3cm} $\longrightarrow $ decreasing density\\[2ex]
	\centering
	\includegraphics[scale=0.3]{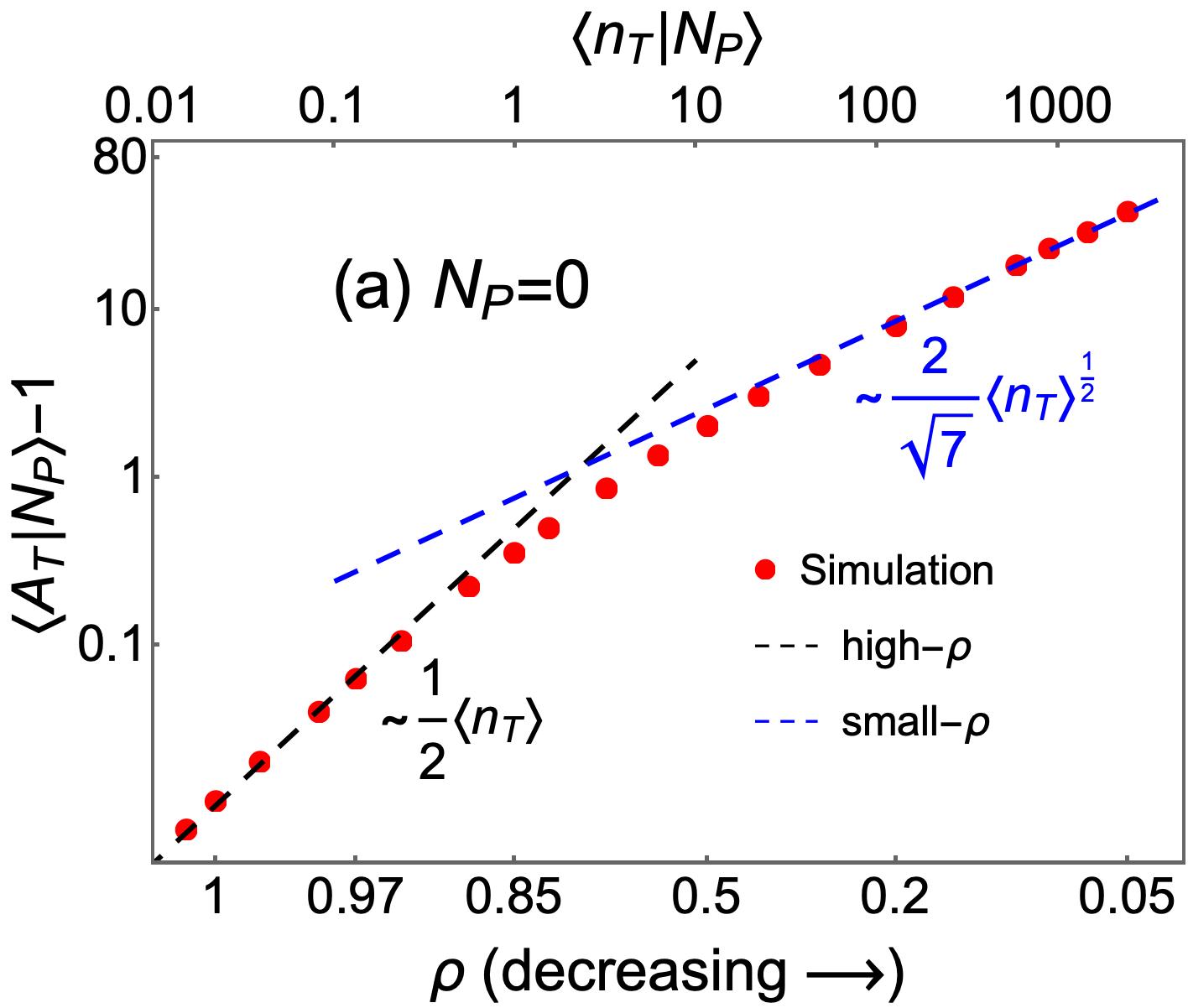}
	\includegraphics[scale=0.3]{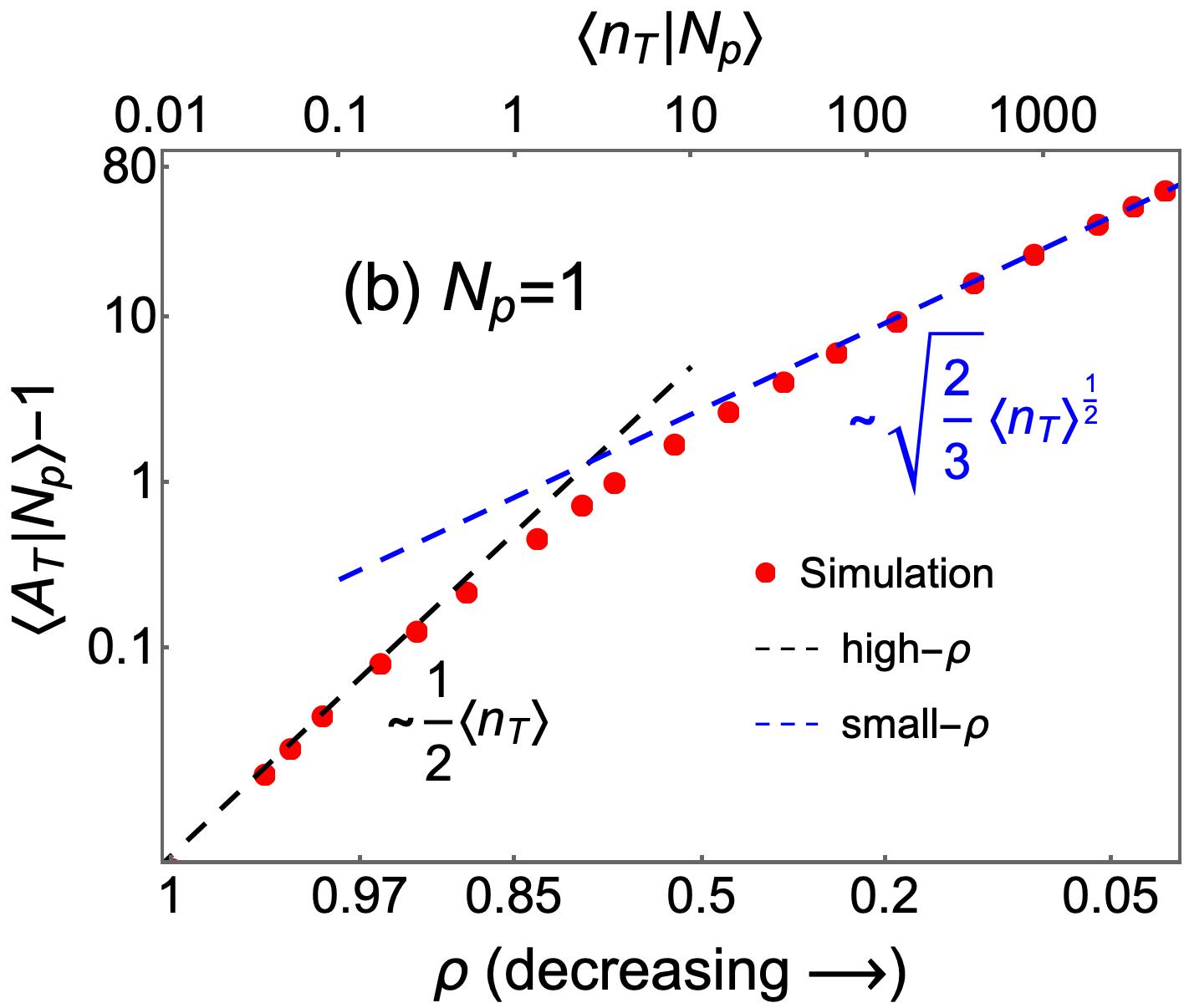}
	\caption{Parametric-plot of $\langle A _{\rm T}|N_{\rm P} \rangle $ and $\langle n _{\rm T}| N_{\rm P} \rangle $ parametrised by $\rho$, illustrating their scaling-change behavior for the $N_{\rm P}$-Sokoban walker with (a) $N_{\rm P}=0$ and (b) $N_{\rm P}=1$. In both panels, we plot density in the bottom axis while the averages $\langle A _{\rm T}|N_{\rm P} \rangle $ and $\langle n _{\rm T} |N_{\rm P} \rangle $ are plotted in the left and top axes respectively. The dashed lines are our theoretical results for small and high values of $\rho$ given in Eq.~\eqref{bhuqm}.}
	\label{fig-1d-crossover}
\end{figure*} 
 
\subsection{Averaging over disorder}
Let us now derive the probabilities of $L_1$ and $L_2$, which according to Eq.~\eqref{hagqo} are related to the positions of the $(N_{\rm P}+1)$-th obstacles on the positive and negative $x$-axes. For simplicity, we focus on the positive $x$-axis, where the position of this obstacle recall is denoted by $Y_{O_{N_{\rm P}+1}}^{+}$. Observe that starting from the lattice site $x=1$ to $x=Y_{O_{N_{\rm P}+1}}^{+}$, there are a total of $(N_{\rm P}+1)$ sites with obstacles while the remaining $\left( Y_{O_{N_{\rm P}+1}}^{+} -N_{\rm P}-1 \right)$ sites are free of obstacles (see Fig.~\ref{fig-new-1d-scheme}). Combining these two contributions, the total probability of observing $Y_{O_{N_{\rm P}+1}}^+ = y$ is given by
\begin{align}
q \left(  y \right) = \rho ^{N_{\rm P}+1}  (1-\rho)^{y-N_{\rm P}-1}  \frac{(y-1)!}{N_{\rm P}! \left(y-1-N_{\rm P} \right)!}, \label{eq-1d-sok-neiat-eq-1}
\end{align}
where $y$ can take possible integer values satisfying $y \ge (N_{\rm P}+1)$. In the expression, the factor $(y-1)!/ N_{\rm P}! \left(y-1-N_{\rm P} \right)! $ counts the number of distinct ways in which the first $N_{\rm P}$ obstacles can be placed anywhere from $x=1$ to $x=(y-1)$; all these arrangements lead to the same value $Y_{O_{N_{\rm P}+1}}^{+}=y$.

Now using Eq.~\eqref{hagqo}, we can write the probability of $L_2$ as
\begin{align}
q(L_2|N_{\rm P}) = \rho ^{N_{\rm P}+1}\left( 1-\rho \right)^{L_{2}}\frac{\left(  L_{2}+ N_{\rm P} \right)!}{N_{\rm P} !~ L_{2}!}, \label{gen-LDF-eq-1}
\end{align}
where $L_2 = 0,1,2, \cdots$. Similarly for $L_1$, we obtain
\begin{align}
q(L_1|N_{\rm P}) = \rho ^{N_{\rm P}+1}\left( 1-\rho \right)^{L_{1}}\frac{\left(  L_{1}+ N_{\rm P} \right)!}{N_{\rm P} !~ L_{1}!}. \label{gen-LDF-eq-1-SRS}
\end{align}
As is clear from Eq.~\eqref{hagqo}, the statistics of $L_1$ and $L_2$ depend on the number of obstacles the Sokoban walker is able to push. We therefore write the probabilities $q(L_1|N_{\rm P})$ and $q(L_2|N_{\rm P})$ as being conditioned on $N_{\rm P}$.

With these probabilities in hand, we now compute the disorder-averaged trap size and trapping time. Averaging Eq.~\eqref{eq-trapLL} with Eqs.~(\ref{gen-LDF-eq-1}-\ref{gen-LDF-eq-1-SRS}), we obtain
\begin{align}
\langle A _{\rm T} |  N_{\rm P} \rangle = 1+\frac{2 \left( N_{\rm P} +1\right) (1-\rho)}{\rho}. \label{new-pssR-eq-1}
\end{align}
As expected, the average trap size increases with $N_{\rm P}$, since a larger pushing capacity allows the walker to create larger gaps before becoming caged. Also, when $\rho=1$, all lattice sites are occupied by obstacles except the origin, which is occupied by the walker. Hence, $\langle A_{\rm T}| N_{\rm P}\rangle=1$ for $\rho=1$, independent of $N_{\rm P}$. As $\rho$ is decreased, the typical gaps between obstacles become larger, allowing the walker to explore a larger region before caging; accordingly, Eq.~\eqref{new-pssR-eq-1} implies that $\langle A_{\rm T}| N_{\rm P}\rangle$ grows as obstacle density decreases and the mean diverges as $\rho\to 0$.

To find the disorder-averaged trapping time, we first notice that $\langle L_1 \rangle \sim \langle L_2 \rangle \sim (1-\rho) /\rho$ following Eqs.~(\ref{gen-LDF-eq-1}-\ref{gen-LDF-eq-1-SRS}). In the small-density limit, $\rho \to 0$, both $L_1$ and $L_2$ are typically large enabling us to use the large-$L_{1,2}$ asymptotics in Eq.~\eqref{new-pssR-eq-4}. By contrast, in the high-density limit $\rho\to 1$, both $L_1$ and $L_2$ are typically zero. Such a configuration, as explained in Eq.~\eqref{abxiqbz}, gives $n_T(L_1=0,L_2=0)=0$. The leading order behavior will then be determined by $L_1=1,~L_2=0$ or  $L_1=0,~L_2=1$. This reasoning allows us to use the expression of the average trapping time in Eq.~\eqref{new-pssR-eq-4-high}. Employing Eqs.~(\ref{gen-LDF-eq-1}-\ref{gen-LDF-eq-1-SRS}) to perform the disorder averaging of Eqs.~(\ref{new-pssR-eq-4}-\ref{new-pssR-eq-4-high}), we obtain
\begin{align}
& \langle n_{\rm T}  | N_{\rm P} \rangle \simeq 
\begin{cases}
4(N_{\rm P}+1)(1 - \rho), & \text{for } \rho \to 1, \\
(7+5 N_{\rm P})(N_{\rm P}+1)/\rho^2, & \text{for } \rho \to 0.
\end{cases} \label{new-pssR-eq-1time}
\end{align}
The average trapping time vanishes as $\rho\to 1$ while it diverges as $\rho\to 0$. For $\rho=1$, as explained above, the walker is trapped immediately at the origin, whereas for $\rho\to 0$ the increasingly large gaps delay caging and lead to a divergence in the trapping time.

Combining Eqs.~(\ref{new-pssR-eq-1}-\ref{new-pssR-eq-1time}), we yield the parametric relation
\begin{align}
\langle A _{\rm T} | N_{\rm P}\rangle \simeq  \left\{ \begin{matrix}
& 1+\langle n _{\rm T} | N_{\rm P} \rangle \big/2,~~~~~~~~~~~\text{for }\rho \to 1,  \\ 
& 2\sqrt{\frac{(N_{\rm P}+1) }{(7+5 N_{\rm P})} }~ \langle n _{\rm T} | N_{\rm P} \rangle^{\frac{1}{2}}, ~~\text{for }\rho \to 0.
\end{matrix} \right. \label{bhuqm}
\end{align}
This equation reveals that, for Sokoban of any pushing capacity, the disorder-averaged trap size and trapping time exhibit a crossover in scaling -- from a linear relation for $\rho$ close to unity to a square-root dependence at small $\rho$. We later show that the linear high-density linearity persists in two dimensions, whereas the low-density regime changes drastically.

To validate our calculation, we perform numerical simulations following the method outlined in Appendix~\ref{appen-simulation}. For different values of $\rho$ , we obtain $\langle A_{\rm T}|N_{\rm P} \rangle$ and $\langle n_{\rm T}|N_{\rm P} \rangle $ from the simulations. These quantities are then presented in a parametric plot in Fig.~\ref{fig-1d-crossover} for (a) $N_{\rm P}=0$ and (b) $N_{\rm P}=1$. In each panel, $\left( \langle A_{\rm T}|N_{\rm P} \rangle - 1 \right)$ is shown on the left axis, $\langle n_{\rm T}|N_{\rm P} \rangle$ on the top axis, and the corresponding density $\rho$ along the bottom axis. We see that the change from the  high-density linear behavior to the low-density power-law behavior is completely consistent with the numerical simulations for both values of $N_{\rm P}$. Further notice that both $\langle A_{\rm T} |N_{\rm P} \rangle$ and $\langle n_{\rm T}|N_{\rm P}\rangle $ in Fig.~\ref{fig-1d-crossover} have a monotonic dependence on $\rho$. Upon decreasing density, both averages increase and diverge as $\rho \to 0$, which is an expected feature, as explained before. Changing the value of $N_{\rm P}$ does not qualitatively alter the monotonic behavior. This again is true only in $1d$ and one obtains a nonmonotonic behavior in the $2d$ setting due to the emergence of a self-trapping mechanism.

\section{Disorder-averaged survival probability}
\label{sec-1d-2}
Having looked at the mean trapping time and trap size, let us examine the disorder-averaged survival probability
\begin{align}
S(n|N_{\rm P}) = \langle Q \left( n|L_1, L_2 \right) \rangle _{L_1, L_2},  \label{ahbsu718}
\end{align}
with $Q(n|L_1, L_2)$ given in Eqs.~(\ref{sok-surv-eq-18}-\ref{sok-surv-eq-18kj08h1}). Since the statistics of $L_1$ and $L_2$ depend on $N_{\rm P}$ in Eq.~\eqref{hagqo}, we again denote the survival probability $S(n|N_{\rm P})$ as being conditioned on $N_{\rm P}$.

\subsection{Survival probability for $N_{\rm P}=0$}
We first look at the weak Sokoban limit, $N_{\rm P}=0$, where the walker does not push the obstacles. Therefore the values of $L_1$ and $L_2$ for a given initial obstacle configuration, as shown in Eq.~\eqref{hagqo}, are determined by the positions of the first obstacles in the $+x$ and $-x$ directions. From the long-time expression in Eq.~\eqref{sok-surv-eq-18}, it is clear that we require the joint probability $q\left(  L,L_i |N_{\rm P}=0\right)$ of $L=(L_1+L_2)$ and $L_i$ for $i \in \{1,2 \}$. This joint probability can be calculated using Eqs.~(\ref{gen-LDF-eq-1}-\ref{gen-LDF-eq-1-SRS}) for $N_{\rm P}=0$ as follows
\begin{align}
q\left(  L,L_i |N_{\rm P}=0\right) & = \sum _{L_{3-i}=0}^{\infty} ~q(L_i|N_{\rm P}=0)~q(L_{3-i} |N_{\rm P}=0 ) \nonumber \\
& ~~~~~~~~~~~~~~\times \delta _{L, L_{i}+L_{3-i}}, \nonumber \\ 
&= \rho ^2(1-\rho)^L~\Theta(L-L_i). \label{ahbaioi7153}
\end{align} 
Here $\Theta (z)$ stands for the Heaviside theta function, which takes the value $1$ if $z>0$ and $0$ otherwise. Using Eq.~\eqref{ahbaioi7153} to average Eq.~\eqref{sok-surv-eq-18}, we get
\begin{align}
S(n| N_{\rm P}=0) &  \simeq \frac{8 \rho ^2}{\pi} \sum _{L=1}^{\infty}~\left( 1-\rho \right)^{L} ~\left( \frac{1}{2}+\frac{2L}{\pi}\right)~ e^{  -\frac{ \pi ^2 }{8L^2}n }.
\end{align}
Performing this summation analytically is challenging. However, for large $n$, we can introduce a quasi-continuous variable $u=2 \sqrt{2} L / \pi \sqrt{n}$ and transform the summation above into an integral
\begin{align}
& S(n| N_{\rm P}=0) \simeq  \rho ^2 \sqrt{8n} \int _{0}^{\infty}du~\left( \frac{1}{2} + \frac{4 \alpha u}{\pi \lambda}  \right)~e^{-\frac{1}{u^2} - 2 \alpha u}, \label{1d-AIL-surv-L1l2-eq-1} \\
&~~~~~~~~~~\text{where } \alpha = \frac{\lambda \pi \sqrt{n}}{4 \sqrt{2}},~~\lambda = \mid \ln(1-\rho)  \mid. \label{sok-surv-eq-26}
\end{align}
The integration over $u$ can be carried out in terms of a Meijer-G function \cite{Andrews1985}
\begin{align}
\int _0^{\infty} du~u^{m}~\exp \left( -\frac{1}{u^2}-2 \alpha u  \right) = \frac{\MeijerG{3}{0}{0}{3}{-}{0,\frac{1+m}{2},\frac{2+m}{2}}{\alpha ^2}}{2 \sqrt{\pi} \alpha ^{1+m}} .
\label{sok-surv-eq-28}
\end{align} 
Eq.~\eqref{1d-AIL-surv-L1l2-eq-1} then becomes
\begin{align}
S(n| N_{\rm P}=0) \simeq & \sqrt{\frac{n \rho^4}{2 \pi \alpha ^2}} ~\left[   \MeijerG{3}{0}{0}{3}{-}{0,\frac{1}{2},1}{\alpha ^2}   \right. \nonumber \\
 & ~~~~\left.+ \frac{8}{ \pi \lambda} \MeijerG{3}{0}{0}{3}{-}{0,1,\frac{3}{2}}{\alpha ^2} \right]. \label{1d-AIL-surv-L1l2-eq-2} 
\end{align}
For $n \lambda^2 \gg 1$ (or equivalently $\alpha \gg 1$ from  Eq.~\eqref{sok-surv-eq-26}), a further simplified expression for the survival probability can be obtained by using the asymptotic form of the Meijer-G function. This form is given in Eq.~\eqref{sok-surv-eq-30} and using it we obtain

\vspace{0.5 cm}
\fbox{%
  \parbox{0.95\columnwidth}{
\begin{align}
 & S(n|N_{\rm P}=0) \simeq \mathcal{K}_{0} (n)\exp \left( -\frac{3 \pi ^{2/3}}{2^{5/3}} \left( \lambda ^{2} n \right)^{\frac{1}{3}} \right), \label{1d-AIL-surv-L1l2-eq-3} \\ 
& \text{with } \mathcal{K}_{0}(n) = \frac{4 \rho ^2}{ \sqrt{3 \pi ^3} \lambda ^2}~\left[ 16 \alpha + \left( \frac{68}{9}+2 \pi \lambda \right)  \alpha ^{1/3} \right], \label{1d-AIL-surv-L1l2-eq-4} 
\end{align}
} }

\vspace{0.5 cm}
\noindent
with $\alpha$ and $\lambda$ given in Eq.~\eqref{sok-surv-eq-26}. One can also get the same result by performing a saddle-point approximation directly in Eq.~\eqref{1d-AIL-surv-L1l2-eq-1}. This stretched-exponential result was announced before in Eq.~\eqref{main-result-eq-2}.

Notice that the prefactor scales as $\mathcal{K}_0(n) \sim \sqrt{n}$ at large times. Comparing our result with the BVDV formula in Eq.~\eqref{original-eqn}, the two survival probabilities, $S(n|N_{\rm P}=0)$ and $\phi(n)$, are characterized by the same stretched exponent proportional to $\lambda ^{2/3} n^{1/3}$, and share the same $\sqrt{n}$-scaling of the prefactor \cite{Anlauf1984}. The difference, however, arises in the proportionality constant inside the exponential term. It takes the value $ 3 \pi ^{2/3} / 2^{5/3} ~\left( \approx 2.0269 \right) $ for $S(n|N_{\rm P}=0)$ and $  3 \pi ^{2/3}/2 ~\left( \approx 3.2175 \right) $ for $\phi(n)$. A larger value for $\phi(n)$ means that it decays faster than $S(n|N_{\rm P}=0)$. The faster decay arises because $\phi(n)$ describes survival against a single obstacle, whereas $S(n\mid N_{\rm P}=0)$ corresponds to survival against caging, a mechanism that necessarily involves more than one obstacle.

%Even the prefactor $\mathcal{K}_{\rm AIL} $ which scales for large $n$ as $\mathcal{K}_{\rm AIL}  \sim \sqrt{n}$ in the leading order has the same temporal scaling for $\phi(n)$ \cite{Anlauf1984}.
%Therefore, the survival probability has a stretched-exponential decay with an exponent $1/3$. Furthermore, the prefactor $\mathcal{K}_{\rm AIL} $ also depends on time and it scales as $\mathcal{K}_{\rm AIL}  \sim \sqrt{n}$ in the leading order.

\begin{figure*}[t]
	\centering
    \includegraphics[scale=0.27]{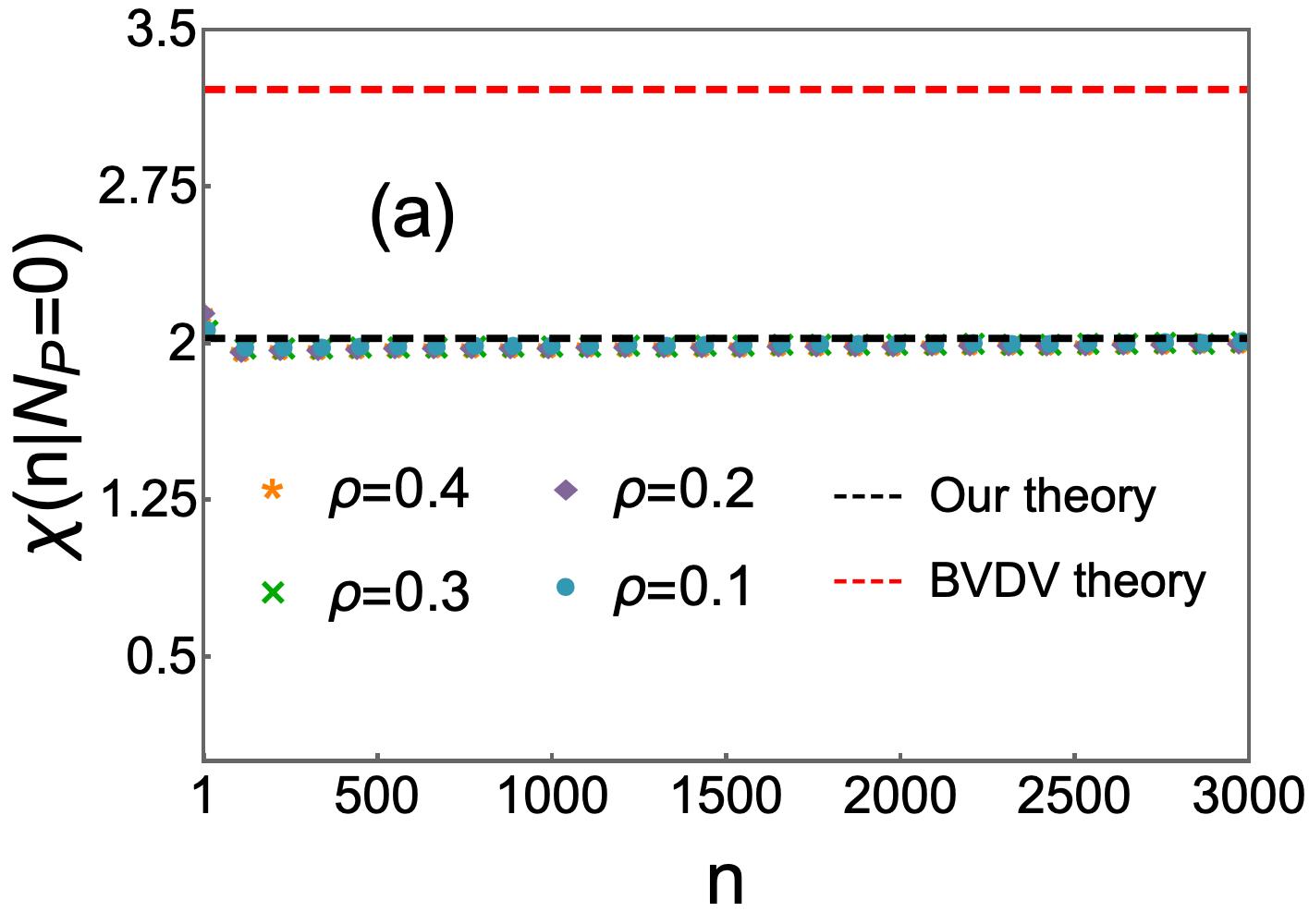}	\hspace{2 em}
	\includegraphics[scale=0.27]{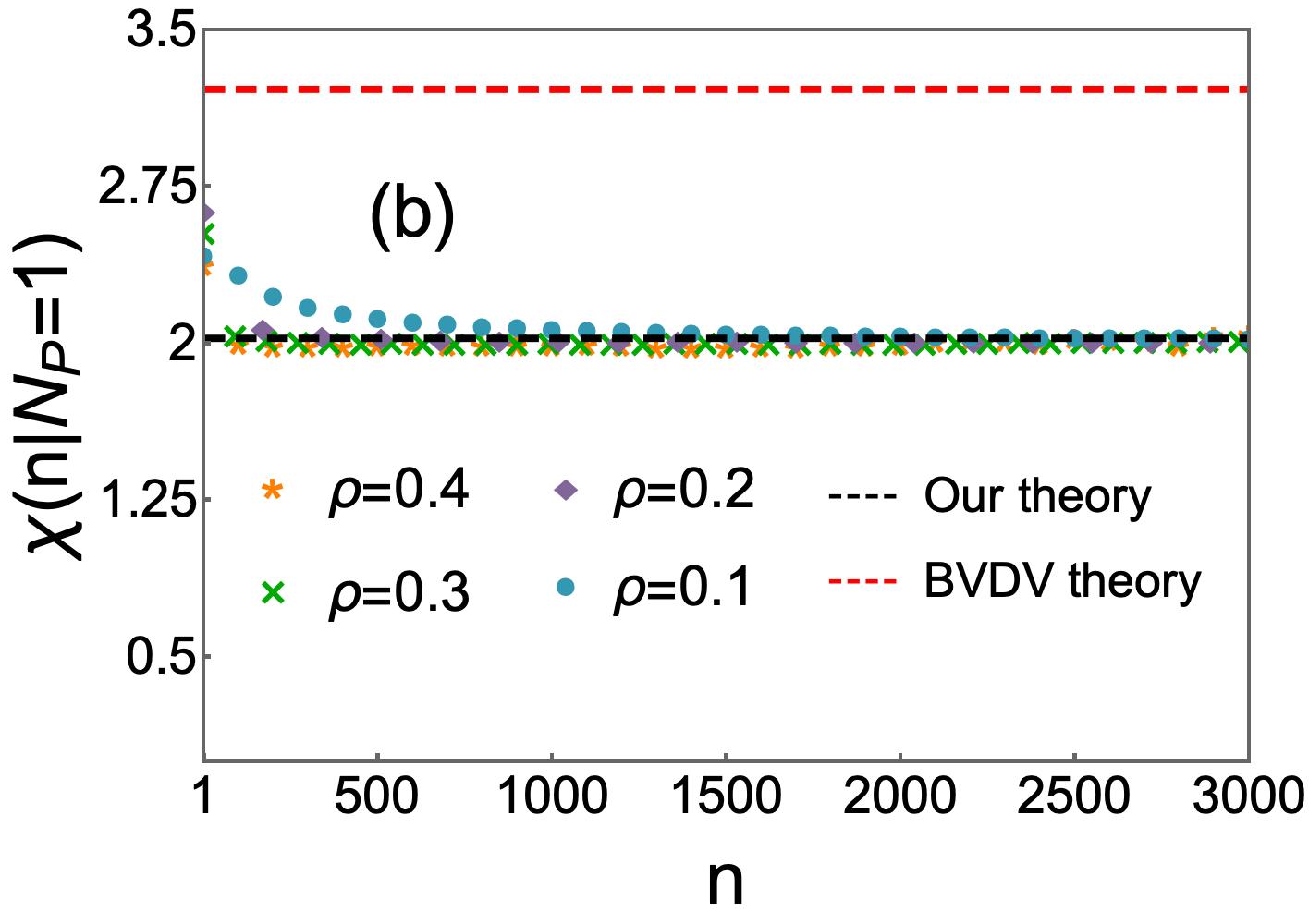}
	\caption{Illustration of the long-time stretched-exponential decay of the survival probability $S(n|N_{\rm P})$ for the one-dimensional $N_{\rm P}$-Sokoban model with (a) $N_{\rm P}=0$ and (b) $N_{\rm P}=1$. The function $\chi(n|N_{\rm P})$, defined in Eqs.~\eqref{1d-AIL-surv-L1l2-eq-5} and \eqref{sok-surv-eq-367}, is obtained from numerical simulation for different density values (represented by different symbols). For all values, the data, at large $n$, converge to a constant given by our theoretical calculation and shown by the black dashed line in both panels. For comparison, we have also presented in red dashed line the BVDV result in Eq.~\eqref{original-eqn}: $ \lim_{n \to \infty}\left[ -\ln \phi(n) / \lambda ^{2/3} n^{1/3} \right] = 3 \pi ^{2/3}/2  \approx 3.2175 $. We do not use any fitting parameter in this figure.}
	\label{fig-surv-1d}
\end{figure*}
To compare Eq.~\eqref{1d-AIL-surv-L1l2-eq-3} with numerical simulations, we plot
\begin{align}
\chi (n|N_{\rm P}=0) \equiv \frac{-\log  \left[ S(n |N_{\rm P}=0 ) \right]+\log \mathcal{K}_{0}(n)} {\lambda ^{2/3} n^{1/3}}  \label{1d-AIL-surv-L1l2-eq-5}
\end{align}
as a function of $n$ in Fig.~\ref{fig-surv-1d}(a). According to our theoretical result, $\chi (n |N_{\rm P}=0 )$ should asymptotically converge to a constant value $\chi (n \to \infty |N_{\rm P}=0) = 3 \pi ^{2/3} / 2^{5/3} \approx 2.0269$ independent of the density $\rho$. Indeed, in Fig.~\ref{fig-surv-1d}(a), we have shown the simulation results of $\chi(n |N_{\rm P}=0)$ for different density values, and for all cases, the simulation data converge to the predicted constant value, $ 2.0269$, indicated by the dashed black line. In this figure, we have also presented a comparison of our result with the BVDV formula in Eq.~\eqref{original-eqn}. According to this formula, $ \lim_{n \to \infty}\left[ -\ln \phi(n) / \lambda ^{2/3} n^{1/3} \right] = 3 \pi ^{2/3}/2 \approx 3.2175$, which is shown by the dashed red line. Clearly, the red line is situated above the black line.
%As observed, this value lies significantly above $\chi_{\rm AIL} (n \to \infty)$, indicating that $\phi(n)$ decays faster than $S _{\rm AIL}(n)$.

%The figure clearly shows that the stretched exponential behaviour is followed only at large $n$, with deviation appearing at small and moderate $n$. 

While the stretched-exponential decay is satisfied at long times, one can see deviations appearing at small and moderate values of $n$, see Fig.~\ref{fig-surv-1d}. To understand the behavior at moderate $n$, we will consider $Q \left( n|L_1, L_2 \right) $ in Eq.~\eqref{sok-surv-eq-18kj08h1}, and take its expectation with respect to Eqs.~(\ref{gen-LDF-eq-1}-\ref{gen-LDF-eq-1-SRS}). In Appendix~\ref{appen-surv-longn}, we derive a simplified expression for the survival probability when $n \rho ^2 \ll 1$

\vspace{0.3 cm}
\noindent\fbox{%
\parbox{0.95\columnwidth}{%
\begin{align}
S(n|N_{\rm P}=0) \simeq 1- n \rho ^2,~~~\text{for }n \rho ^2 \ll 1. \label{new-eqq-sal-eqew-1}
\end{align}
}%
}

\begin{figure*}[t]
	\centering
    \includegraphics[scale=0.27]{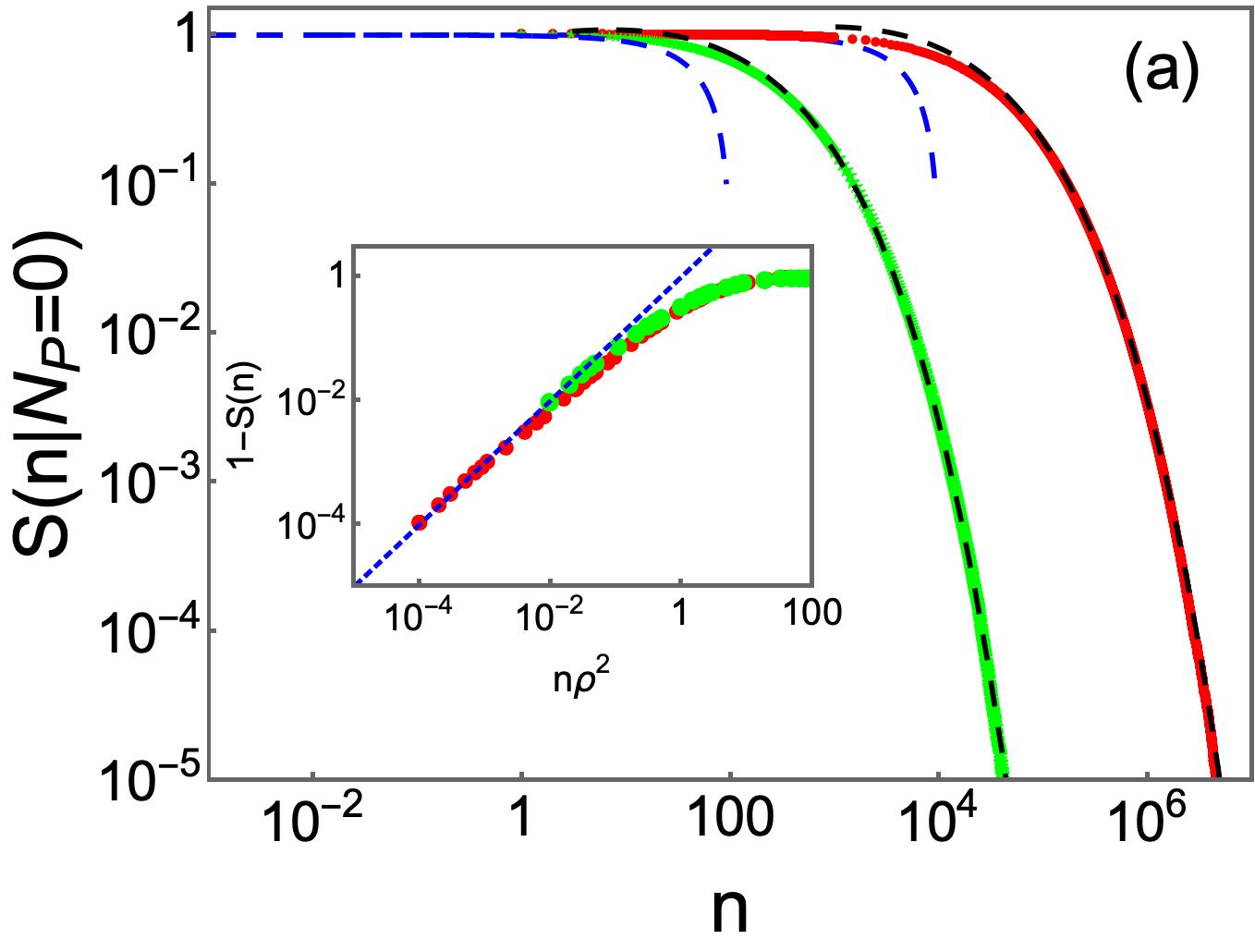} \hspace{2 em}	
	\includegraphics[scale=0.28]{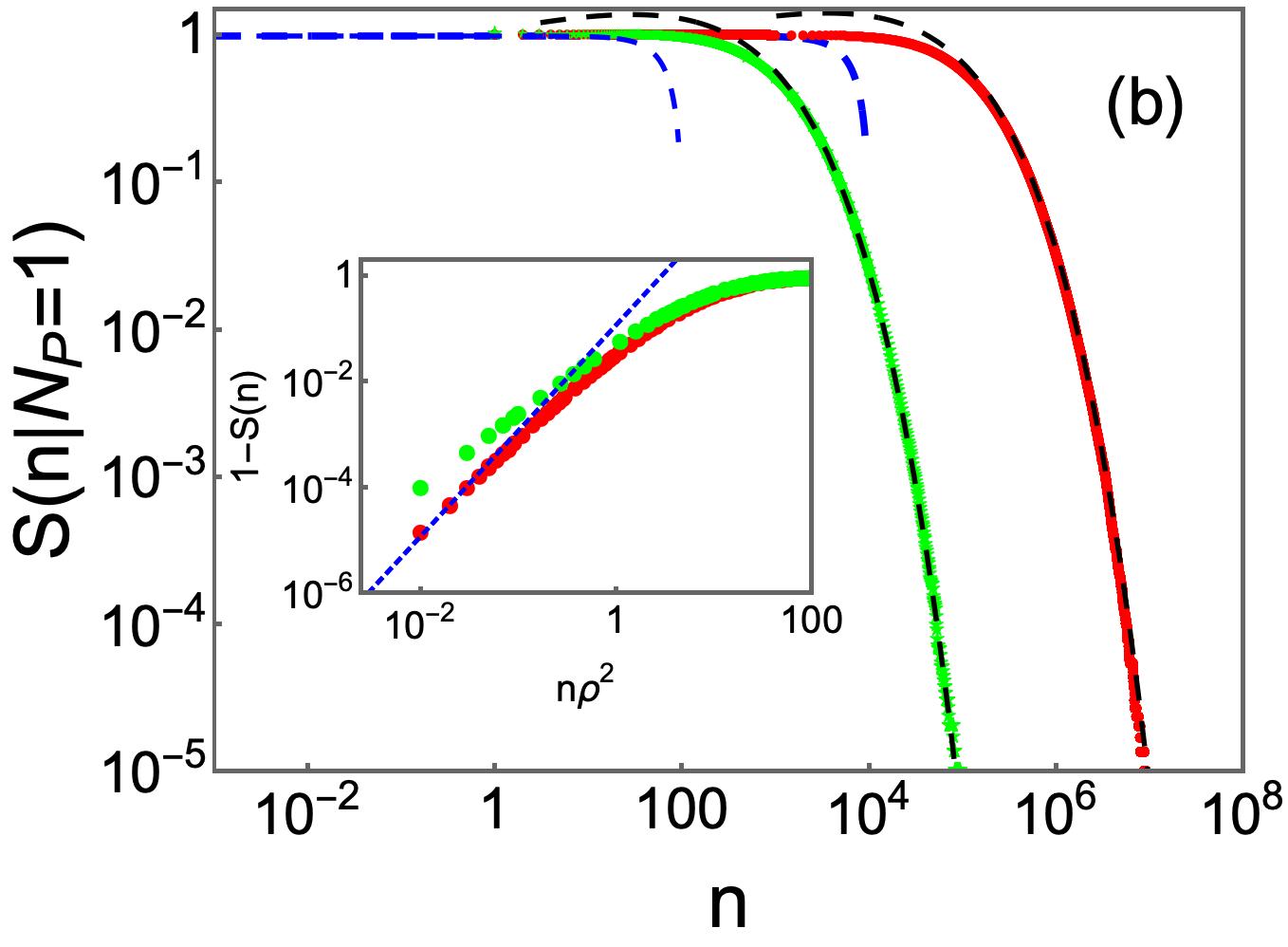}
	\caption{Survival probability $S(n|N_{\rm P})$ for the one-dimensional $N_{\rm P}$-Sokoban model with (a) $N_{\rm P}=0$ and (b) $N_{\rm P}=1$. Numerical simulations corresponding to $\rho = 0.01$ (red) and $\rho = 0.1$ (green) are compared with the theoretical results at moderate and long times. For $N_{\rm P}=0$, black dashed line corresponds to the long-time asymptotic expression in Eq.~\eqref{1d-AIL-surv-L1l2-eq-3} while the blue line represents the intermediate-time expression in Eq.~\eqref{new-eqq-sal-eqew-1}. Similarly, for $N_{\rm P}=1$, these are given in Eqs.~\eqref{sok-surv-eq-351993} and \eqref{sok-surv-eq-35} respectively. Insets in both panels present a magnified view of the intermediate-time comparison. Here, we have plotted $ \left( 1-S(n|N_{\rm P}) \right)$ vs $n \rho ^2$ for the same two densities. According to our theoretical analysis, $ \left( 1-S(n|N_{\rm P}) \right)$ grows linearly and quadratically with $n \rho ^2$ (with $n \rho ^2 \ll 1$) for $N_{\rm P}=0$ and $N_{\rm P}=1$, respectively. Indeed, for both cases, the simulation data (shown by symbols) agree well with the theoretical predictions. Deviation in the inset in panel~(b) for $\rho = 0.1$ arises because our derivation assumes the limit of small $n \rho^2$ while keeping $n$ moderately large. While this condition is easily satisfied at very low densities, it becomes increasingly difficult to achieve as $\rho$ increases.}
	\label{fig-short-n-1d}
\end{figure*}

\vspace{0.5 cm}
\noindent
In Fig.~\ref{fig-short-n-1d}(a), we show the survival probability in both the intermediate-time and long-time regimes. We find that the stretched-exponential form shows a good agreement with the numerical simulations when $S(n|N_{\rm P}=0)  \leq  0.1$. The agreement emerges even before entering into the regime of extremely rare events where $S(n|N_{\rm P}=0) $ is extremely small. This is largely due to the fact that the prefactor $\mathcal{K}_{0}(n)$ is known exactly, as given in Eq.~\eqref{1d-AIL-surv-L1l2-eq-4}, allowing for a good agreement even when $n$ is not too large.

For $S(n|N_{\rm P}=0) $ close to unity, however, the stretched-exponential expression is not accurate and we observe deviations in Fig.~\ref{fig-short-n-1d}(a). In this regime, the moderate-time expression in Eq.~\eqref{new-eqq-sal-eqew-1} governs the survival dynamics, especially at small densities. This is demonstrated in the inset of the figure where we have plotted $\left(1-S(n|N_{\rm P}=0) \right)$ vs $n \rho ^2$ for two different densities. Our analytical result predicts a linear relationship that is independent of the density when $n \rho ^2 \ll 1$. Indeed, the simulation data is consistent with this prediction.

Before closing this discussion, we compare our intermediate time result with Rosenstock's approximation for $\phi(n)$ in Eq.~\eqref{original-eqn-2}. While $\left( 1-\phi(n) \right) \sim \sqrt{n \rho^2}$, our result is clearly different. Thus, although, both $S(n|N_{\rm P}=0) $ and $\phi(n)$ show similar stretched-exponential decay at large $n$ as far as the time dependence is concerned, their intermediate time behaviors are quite different.

\subsection{Survival probability for $N_{\rm P}=1$}
\label{sec-1d-3}
We next look at the Sokoban with pushing parameter $N_{\rm P}=1$, the one-dimensional version of the model introduced by Reuveni et al. \cite{Shlomi-1, Shlomi-2} (since $N_{\rm P}=0$ is not really a Sokoban). As explained in Eq.~\eqref{hagqo}, the values of $L_1$ and $L_2$ are now determined by the positions of the second obstacles in the $+x$ and $-x$ directions. For this case also, the survival probability can be computed in the same manner. As shown in Appendix~\ref{appen-intermediate-time}, the moderate-time behavior of the disorder-averaged survival probability is

\vspace{0.2 cm}
\noindent\fbox{%
\parbox{0.95\columnwidth}{%
\begin{align}
S(n |N_{\rm P}=1) \simeq 1-\frac{n^2 \rho ^4}{8},~~~\text{for }n \rho ^2 \ll 1, \label{sok-surv-eq-351993}
\end{align}
}%
}

\vspace{0.5 cm}
\noindent
It is important to emphasize that this expression is derived using $Q(n|L_1, L_2)$ in Eq.~\eqref{sok-surv-eq-18kj08h1}, which itself holds only at intermediate times. Consequently, Eq.~\eqref{sok-surv-eq-351993} is expected to be valid only for moderate values of $n$ but with small $\rho$ so that $n \rho ^2 \ll 1$. Under these conditions, $S(n|N_{\rm P}=1)$ decreases quadratically which is very different from the linear decrease for the weak Sokoban limit $(N_{\rm P}=0)$ in Eq.~\eqref{new-eqq-sal-eqew-1}. This implies that the survival probability remains closer to unity for the $N_{\rm P}=1$ case than for the $N_{\rm P}=0$ case. The enhanced survival in the former arises from the enhanced pushing capability, which allows the walker to avoid getting caged in a situation that would otherwise cage a weak Sokoban walker.

The late-time survival probability for $N_{\rm P}=1$ can also be obtained by following the same steps as for $N_{\rm P}=0$ in Eq.~\eqref{1d-AIL-surv-L1l2-eq-3}. We therefore relegate the details to Appendix~\ref{appen-surv-1d-sok}, focusing on the regime $n\lambda^2\gg 1$, where we find

\vspace{0.5 cm}
\noindent\fbox{%
\parbox{0.95\columnwidth}{%
\begin{align}
S(n|N_{\rm P}=1) \simeq \mathcal{K}_{1}(n) \exp \left( -\frac{3 \pi ^{2/3}}{2^{5/3}} \left(\lambda ^{2} n \right)^{\frac{1}{3}} \right). \label{sok-surv-eq-35}
\end{align}
}%
}

\vspace{0.5 cm}
\noindent
The prefactor $\mathcal{K}_{1}(n)$ also depends on $n$ and has a rather lengthy expression which is given in Eq.~\eqref{sok-surv-eq-36}. Its leading order expression in $n$ is
\begin{align}
\mathcal{K}_{1}(n) \simeq  \frac{2^{\frac{19}{6}} (4-\pi) \rho ^4}{\sqrt{3} \pi ^{\frac{7}{6}} \lambda ^{\frac{5}{3}}}~n^{\frac{7}{6}}, \label{sok-surv-eqajo-35}
\end{align}
with $\lambda = |\ln (1-\rho)|$. We again see the survival probability $S(n|N_{\rm P}=1)$ decays in a stretched-exponential manner at late times, with the exponent $1/3$, same as the BVDV formula for $\phi(n)$ in Eq.~\eqref{original-eqn}. Increasing the pushing strength from $N_{\rm P}=0$ to $N_{\rm P}=1$ does not change the long-time stretched-exponential relaxation. The effect of $N_{\rm P}$, however, can be found in the prefactor outside of this stretched-exponential decay. While it scales as $\mathcal{K}_{0}(n) \sim n^{1/2}$ for $N_{\rm P}=0$, the prefactor has a different scaling of the form $\mathcal{K}_{1}(n)  \sim n^{7/6}$ for the Sokoban model with $N_{\rm P}=1$. 
%Thus, although $S(n|N_{\rm P})$ for both models belongs to the BVDV trapping universality class as far as the stretch exponent is concerned, the time-scaling of the prefactor in front of the exponential term is very different. 

We next present the comparison of Eq.~\eqref{sok-surv-eq-35} for $S(n|N_{\rm P}=1)$ with numerical simulations. For this, we again define the quantity
\begin{align}
\chi(n|N_{\rm P}=1) \equiv  \frac{ -\log \left[ S(n|N_{\rm P}=1) \right]+\log \mathcal{K}_{1}(n) }{\lambda ^{2/3} n^{1/3}}.  \label{sok-surv-eq-367}
\end{align}
and plot it as a function of $n$. According to our calculation, this quantity should converge to $\chi(n \to \infty |N_{\rm P}=1) = 3 \pi ^{2/3} / 2^{5/3} \approx 2.0269$ independent of the density. Fig.~\ref{fig-surv-1d}(b) presents the simulation results for $\chi(n|N_{\rm P}=1)$ for different values of density $\rho$. For all densities considered, the simulation data converge to the same constant value in the large-$n$ limit. This observation is in excellent agreement with the theoretical result. Also, a comparison with the BVDV result in Eq.~\eqref{original-eqn} is shown in this figure by the red dashed line. This line is significantly above $\chi (n \to \infty |N_{\rm P}=1) $, indicating that $\phi(n)$ has a faster decay than $S(n|N_{\rm P}=1)$.

We have also compared the moderate-time and long-time behaviors with numerical simulations in Fig.~\ref{fig-short-n-1d}(b). This comparison, similar to that performed for the $N_{\rm P}=0$ case, shows that the long-time expression in Eq.~\eqref{sok-surv-eq-35} works quite well for the survival dynamics except when $S(n|N_{\rm P}=1)$ is close to unity. There, our intermediate-time result in Eq.~\eqref{sok-surv-eq-351993} gives a better approximation specially for small densities, see the inset in the figure. Notice that in the inset, the simulation data for $\rho = 0.1$ do not match with Eq.~\eqref{sok-surv-eq-351993} while for $\rho = 0.01$, they match. This occurs because our derivation assumes the limit of $n \rho^2 \ll 1$ while keeping $n$ moderately large. While this condition is easily satisfied at very low densities, it becomes increasingly difficult to achieve as $\rho$ increases.

\begin{figure*}[t]
	\centering
   \includegraphics[scale=0.2]{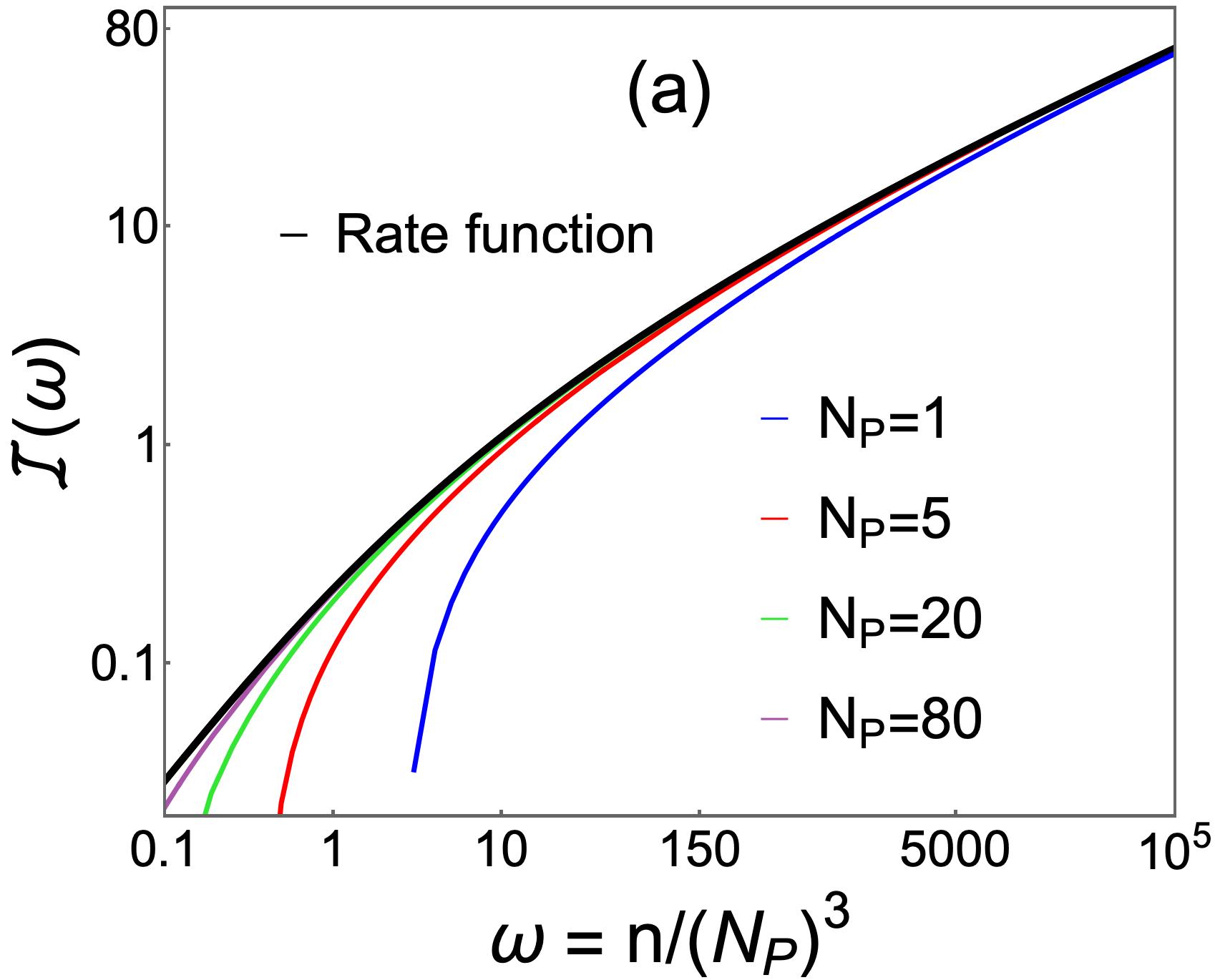}
   \includegraphics[scale=0.2]{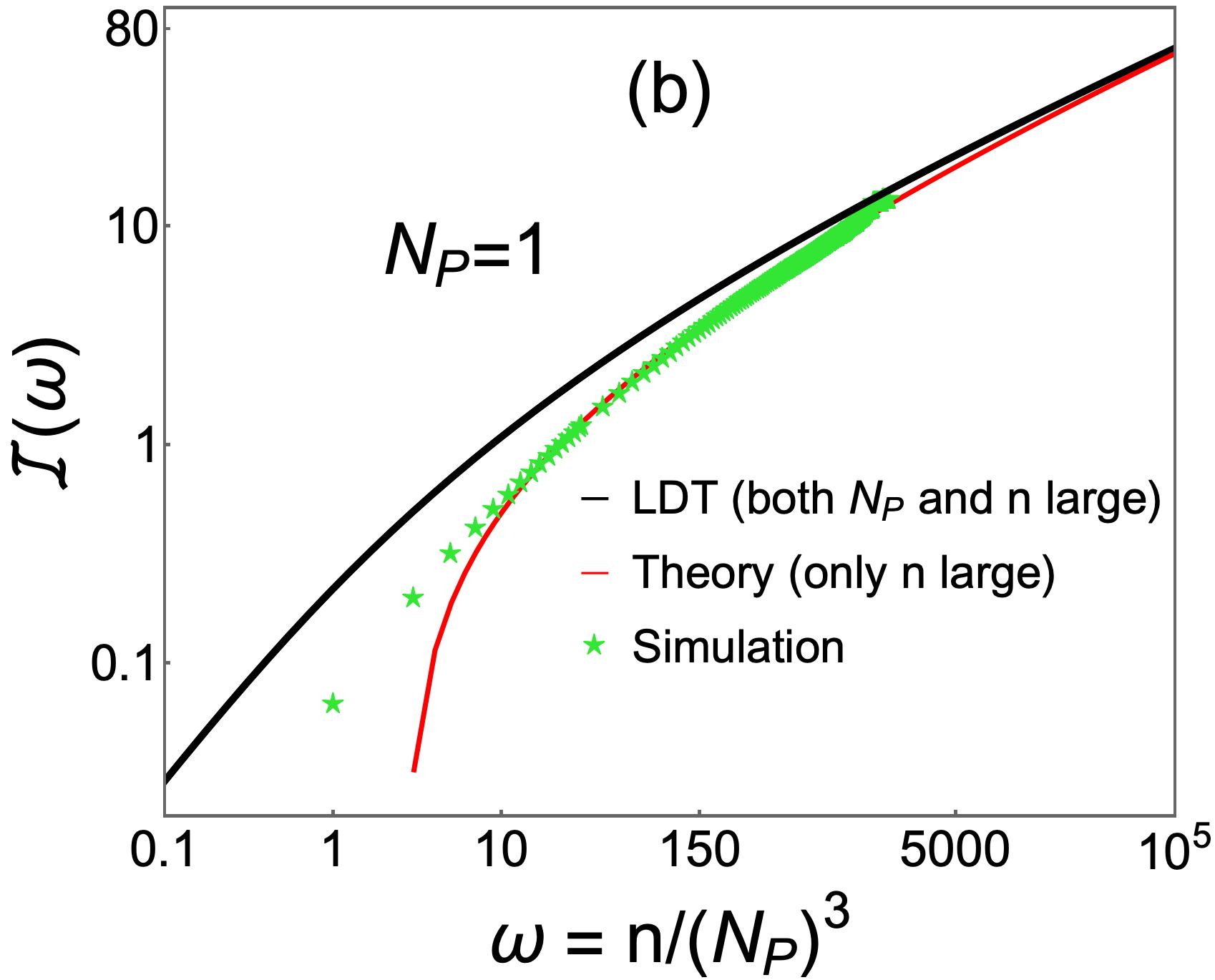}
   \\[1.5em]
   \includegraphics[scale=0.18]{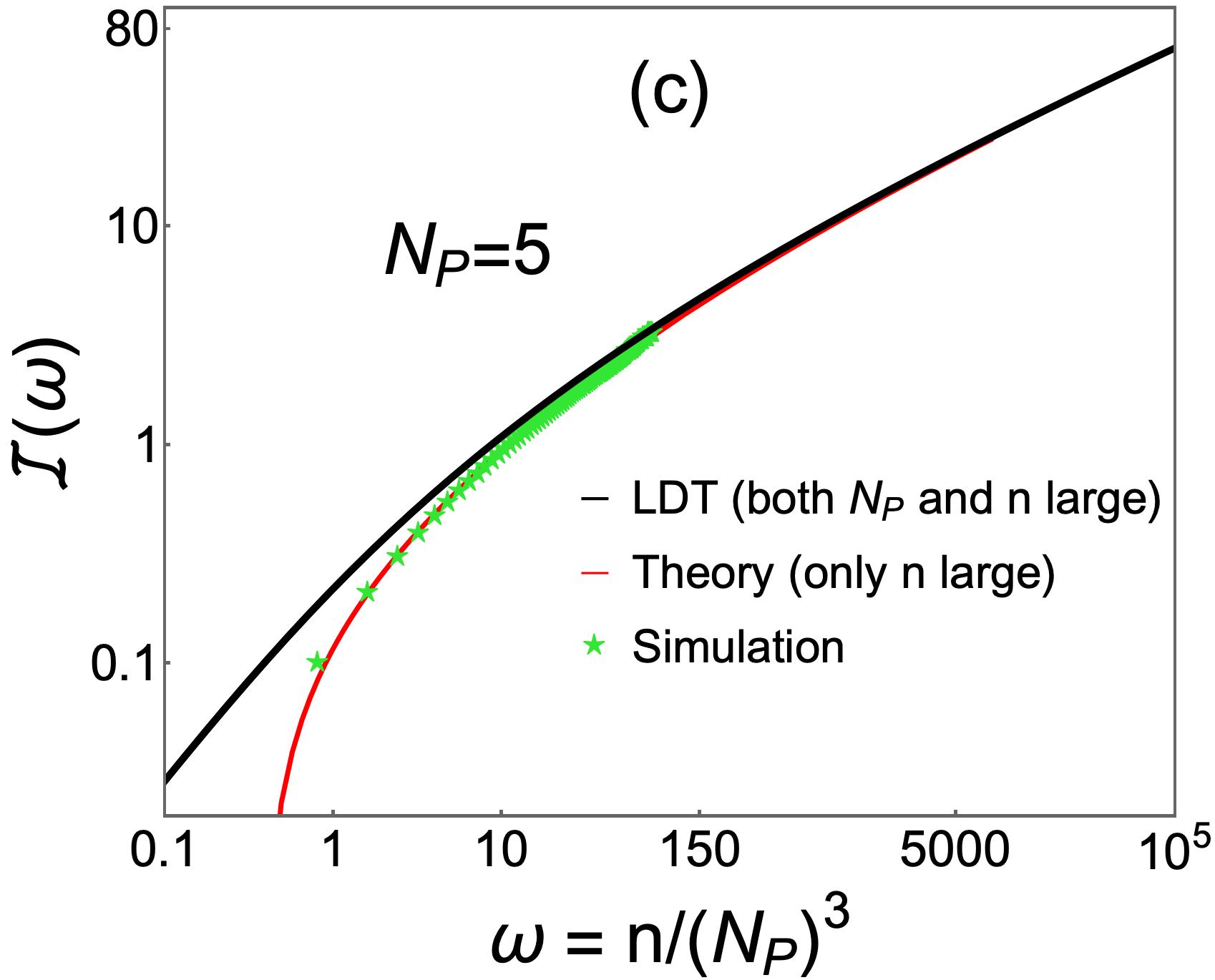}
   \includegraphics[scale=0.18]{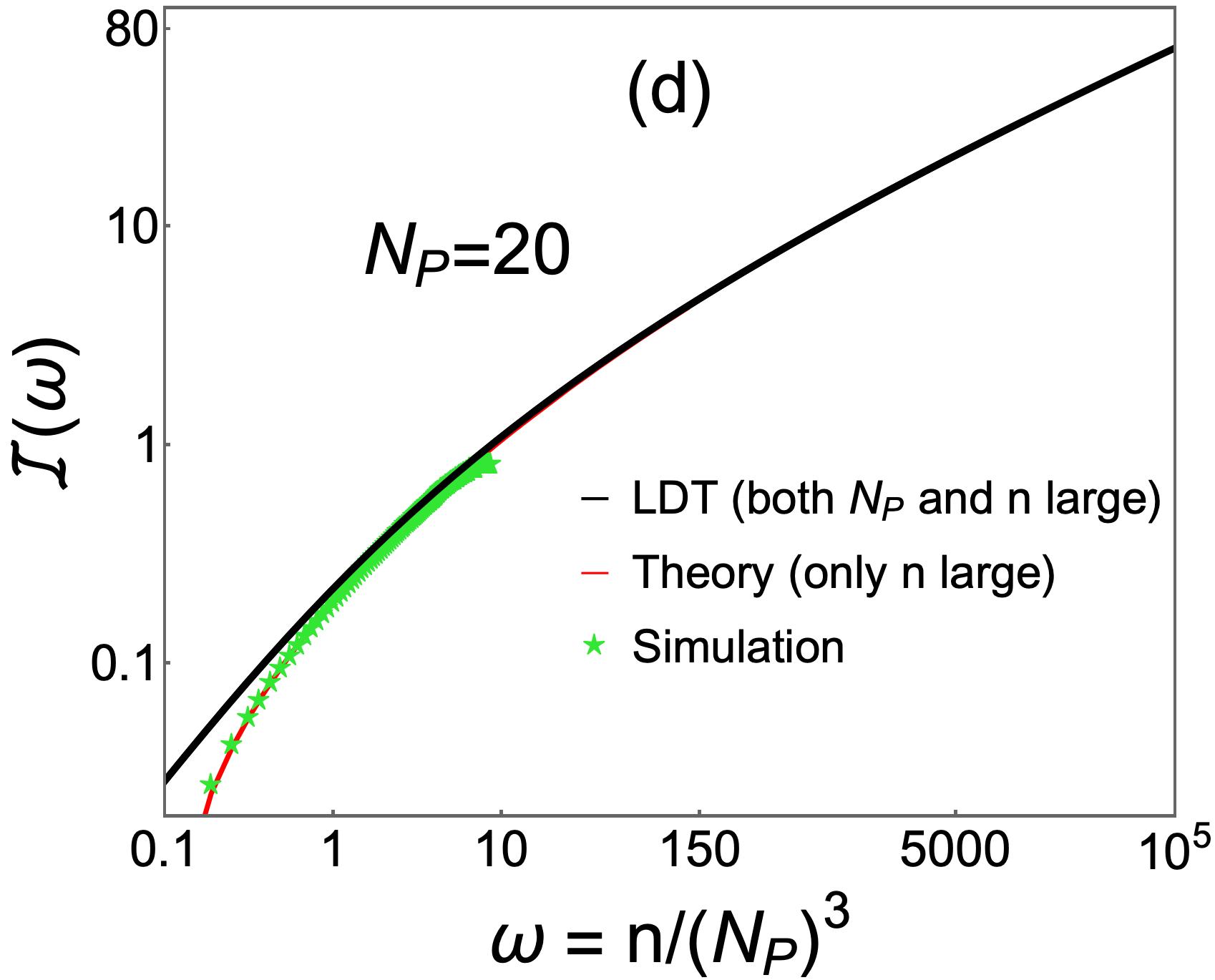}
   \includegraphics[scale=0.18]{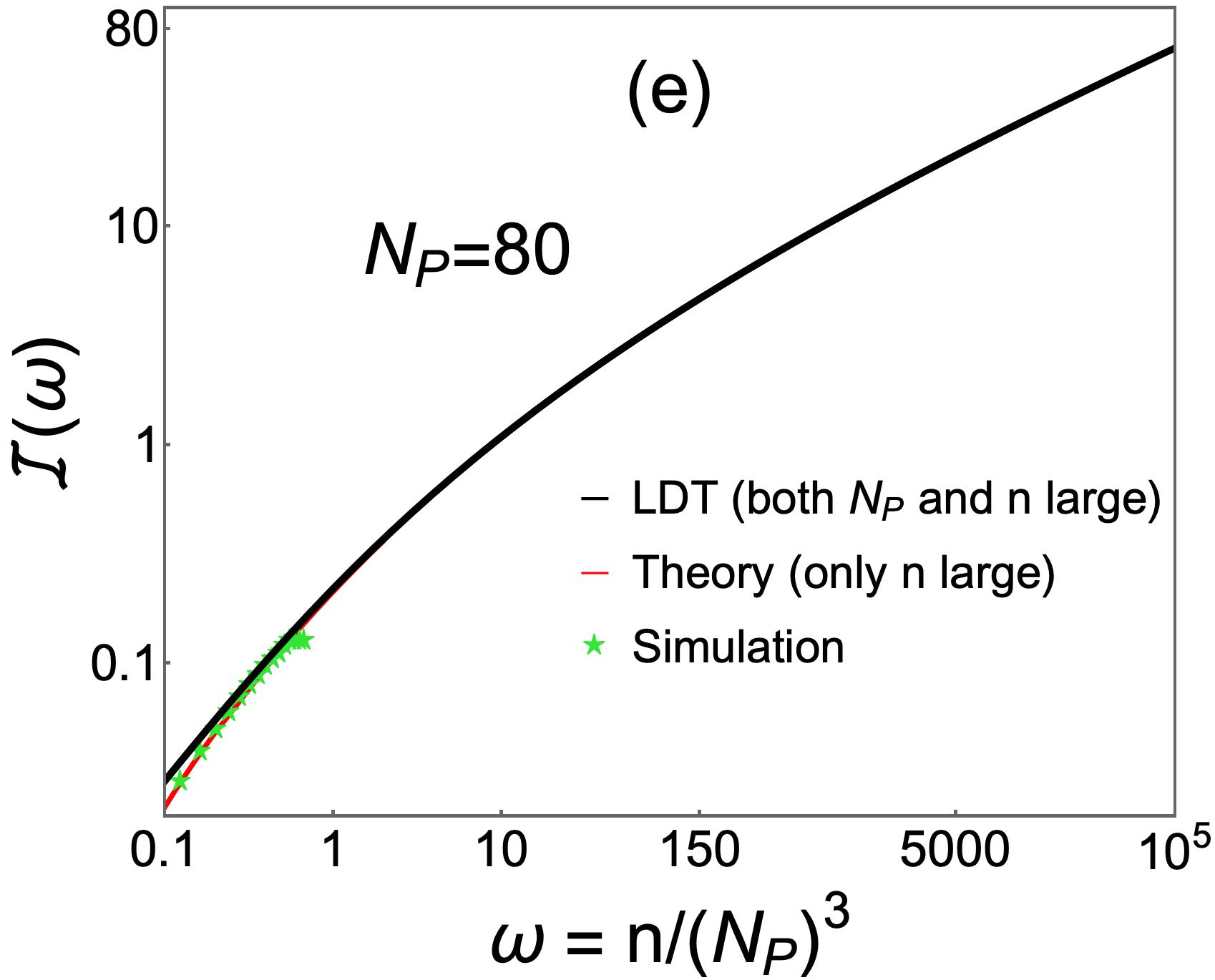}	
	\caption{Rate function $\mathcal{I}(\omega)$ is plotted as a function of $\omega = n /\left( N_{\rm P} \right)^3$ for $\rho = 0.5$. In panel (a), the solid black line represents the rate function in Eq.~\eqref{gen-LDF-eq-9} based on the large-deviation theory (LDT), which is valid for $n \gg 1$, $N_{\rm P} \gg 1$ with $\omega$ fixed. It is compared with $ \left[ -\ln S \left( n| N_{\rm P} \right) \big/ N_{\rm P} \right]$ for different $N_{\rm P}$ (shown by coloured lines) and $S \left( n|N_{\rm P} \right)$ determined in Eq.~\eqref{gen-LDF-eq-44}. Recall that this expression is valid for large $n$ but arbitrary $N_{\rm P}$. Therefore, we see a departure between them at small and moderate $N_{\rm P}$. However, as $N_{\rm P}$ is increased, the two expressions show a better agreement. In panels (b)-(e), the two theoretical results in Eqs.~\eqref{gen-LDF-eq-44} and \eqref{gen-LDF-eq-9} are compared with the numerical simulations (shown by green symbols).}
	\label{fig-rate-function}
\end{figure*}

\subsection{General survival probability: a large-deviation analysis}
\label{sec-1d-4}
For both chosen values of $N_{\rm P}$, we saw that the long-time exponent characterizing the survival probability is same as the BVDV formula. This naturally raises the question of what happens for general $N_{\rm P}$. As mentioned before in Eq.~\eqref{main-result-eq-3}, the survival probability for $N_{\rm P}\gg 1$ admits a large-deviation form. We now derive this result.

For general $N_{\rm P}$, the probability of $L_i$ with $i \in \{1,2 \}$ is given in Eqs.~(\ref{gen-LDF-eq-1}-\ref{gen-LDF-eq-1-SRS}). Substituting $ L_{i} = N_{\rm P} y_{i} $ and then performing the Stirling approximation
\begin{align}
\ln N_{\rm P}  ! = N_{\rm P} \ln N_{\rm P}-N_{\rm P},~~~~\text{for } N_{\rm P} \gg 1,
\end{align}
we can write a large-deviation form \cite{TOUCHETTE20091, LDT-2, LDT-3}
\begin{align}
q(L_{i} = N_{\rm P} y_{i} | N_{\rm P} \gg 1) \sim \exp \left[ -N_{\rm P} \Lambda(y_i)\right], \label{gen-LDF-eq-2}
\end{align}
with function $\Lambda(y)$ defined as
\begin{align}
\Lambda(y) = y \ln y-(1+y)\ln (1+y)-y \ln(1-\rho)-\ln \rho.  \label{gen-LDF-eq-3}
\end{align}
Next, we average the conditional survival probability $Q(n|L_1, L_2)$ in Eq.~\eqref{sok-surv-eq-18} over $L_1$ and $L_2$
\begin{align}
S \left( n| N_{\rm P} \right) = \sum _{L_1=0}^{\infty} \sum _{L_2=0}^{\infty} Q(n|L_1, L_2)~ q(L_1|N_{\rm P}) ~q(L_2|N_{\rm P}).  \label{gen-LDF-eq-44}
\end{align}
This expression, with $Q(n|L_1, L_2)$ from Eq.~\eqref{sok-surv-eq-18}, is valid for large $n$ but any arbitrary value of $N_{\rm P}$. For $N_{\rm P} \gg 1$, it can be expressed as
\begin{align}
S \left( n| N_{\rm P} \gg 1 \right) \sim & \int_{0}^{\infty} dy_1~\int_{0}^{\infty} dy_2~\sin \left( \frac{\pi y_1}{2(y_1+y_2)} \right) \nonumber \\
& \times \exp \left[ - N_{\rm P} \Psi \left( \frac{n}{ N_{\rm P} ^3},y_1, y_2 \right) \right], \label{gen-LDF-eq-4}
\end{align}
where
\begin{align}
\Psi(\omega, y_1, y_2)= \frac{\pi ^2 \omega}{8(y_1+y_2)^2}+ \Lambda(y_1) +\Lambda(y_2).\label{gen-LDF-eq-5}
\end{align}
From Eq.~\eqref{gen-LDF-eq-4}, it now follows that the survival probability admits a large-deviation form in the joint limit $N_{\rm P} \gg 1 , n \gg 1 $ keeping the ratio $n/\left( N_{\rm P} \right)^3 $ fixed

\vspace{0.5 cm}
\noindent\fbox{%
\parbox{0.95\columnwidth}{%
\begin{align}
\lim_{\substack{N_{\rm P} \gg 1 , n \gg 1 \\    \omega=n/\left( N_{\rm P} \right)^3  \text{ fixed}         }} \frac{-\ln S\left( n| N_{\rm P}\right)}{N_{\rm P}} = \mathcal{I}\left( \omega \right), \label{gen-LDF-eq-6}
\end{align}
}%
}

\vspace{0.5 cm}
\noindent
which was also announced before in Eq.~\eqref{main-result-eq-3}. The rate function $\mathcal{I}(\omega)$ is given by the minimization of $\Psi \left( \omega ,y_1, y_2 \right)$  with respect to both $y_1$ and $y_2$
\begin{align}
\mathcal{I}(\omega) = \min _{y_1, y_2} \Psi \left( \omega ,y_1, y_2 \right). \label{gen-LDF-eq-7}
\end{align}
By taking the derivative of Eq.~\eqref{gen-LDF-eq-5} and setting it to zero, it turns out that the values of $y_1$ and $y_2$ (say $\tilde{y}_1$ and $\tilde{y}_2$) corresponding to the minimization are exactly the same $(\tilde{y}_1 = \tilde{y}_2 = \tilde{y})$ and it is governed by the equation
%\begin{align}
%-\frac{\pi ^2 \omega}{32 \tilde{y}^3} + \ln \tilde{y} -\ln(1+\tilde{y})+\lambda = 0. \label{gen-LDF-eq-8}
%\end{align}
\begin{align}
\frac{\pi ^2 \omega}{32  \lambda \tilde{y}^3} + \frac{1}{\lambda}\ln \left( \frac{1+\tilde{y}}{\tilde{y}} \right)  = 1. \label{gen-LDF-eq-8}
\end{align}
This equation can be numerically solved to generate $\tilde{y}(\omega)$ as a function of $\omega$ and inserting this solution in Eq.~\eqref{gen-LDF-eq-7} gives the rate function
\begin{align}
\mathcal{I}(\omega) = \frac{\pi ^2 \omega}{32 \tilde{y}(\omega)^2}+2 \Lambda \left(\tilde{y}(\omega) \right),\label{gen-LDF-eq-9}
\end{align}
\noindent
where $\Lambda\left( \tilde{y}(\omega) \right)$ is given in Eq.~\eqref{gen-LDF-eq-3}. Eqs.~\eqref{gen-LDF-eq-8} and \eqref{gen-LDF-eq-9} completely characterize the exact large-deviation rate function describing the survival probability $S(n|N_{\rm P})$. 

\bluepw{One can find the asymptotic expressions of the rate function analytically in different limits of $\omega$, which were also quoted before in Eq.~\eqref{main-result-eq-4}. To derive them, we look at the two terms on the left-hand-side of Eq.~\eqref{gen-LDF-eq-8}. Depending on whether $\omega$ is large or small, either of these terms contribute. For example, when $\omega \ll \frac{32 \tilde{y}^3}{ \pi ^2} \ln \left( \frac{1+\tilde{y}}{\tilde{y}} \right) $, then the first term on the left-hand-side of Eq.~\eqref{gen-LDF-eq-8} is negligible in comparison to the second term. We then obtain
\begin{align}
\tilde{y}(\omega) & \simeq \frac{(1-\rho)}{\rho}. \label{gen-LDF-eq-10s}
\end{align}
By contrast, in the opposite limit, $\omega \gg \frac{32 \tilde{y}^3}{ \pi ^2} \ln \left( \frac{1+\tilde{y}}{\tilde{y}} \right) $, the second term can be dropped relative to the first term, yielding
\begin{align}
\tilde{y}(\omega)  \simeq \frac{\pi ^{2/3} \omega^{1/3}}{2^{5/3} \lambda^{1/3}}. \label{gen-LDF-eq-10}
\end{align}}
%\begin{align}
%\tilde{y}(\omega) & \simeq \frac{(1-\rho)}{\rho},~~~~~~~~~\text{for }\omega \to 0 , \nonumber \\
%& \simeq \frac{\pi ^{2/3} \omega^{1/3}}{2^{5/3} \lambda^{1/3}}, ~~~~~~~\text{for }\omega \to \infty. \label{gen-LDF-eq-10}
%\end{align}
Substituting them in Eq.~\eqref{gen-LDF-eq-9}, we obtain the asymptotic expressions of $\mathcal{I}(\omega)$ in Eq.~\eqref{main-result-eq-4}. In terms of $n$, this translates to

\vspace{0.5 cm}
\noindent\fbox{%
\parbox{1.02\columnwidth}{%
\begin{align}
S( n| N_{\rm P} \gg 1 ) \sim
\begin{cases}
\exp \left( -\frac{\pi ^2 \rho ^2 n}{32(1-\rho)^2 \left( N_{\rm P}\right) ^2} \right), & \text{fo }  n \ll \left( N_{\rm P} \right)^3 \\[6pt]
\exp\left( -\frac{3 \pi ^{2/3} \lambda ^{2/3}}{2^{5/3}} n^{1/3} \right),
 & \text{for }  n \gg \left( N_{\rm P} \right)^3
\end{cases}
\label{gen-LDF-eq-11}
\end{align}
}%
}

\vspace{0.5 cm}
\noindent
indicating a change from an exponential decay of the survival probability to the stretched-exponential decay at very large times. Interestingly, the long-time stretched-exponential form is independent of $N_{\rm P}$ and is exactly the same as for the Sokoban models with $N_{\rm P}=0$ and $N_{\rm P}=1$; see Eqs.~\eqref{1d-AIL-surv-L1l2-eq-3} and \eqref{sok-surv-eq-35} respectively. Therefore, in $1d$, the long-time stretched-exponential decay of the survival probability is completely universal for any $N_{\rm P}$ and belongs to the BVDV universality class of the classical trapping problem in Eq.~\eqref{original-eqn}.

Let us heuristically try to understand this universality. Recall that for any value of $N_{\rm P}$ much smaller than the overall system size (due to the finite
pushing ability), the motion is confined to a finite interval of length $L$ for a given initial obstacle configuration. The statistics of $L$, in turn, depends on how many obstacles the walker is allowed to push. For a given $N_{\rm P}$, the probability of $L$ can be written as $h(\rho, L, N_{\rm P})~e^{-\lambda L}$ where $\lambda = |\ln(1-\rho)|$ and $h(\rho, L, N_{\rm P})$ is a polynomial function in $L$ that depends on the value of $N_{\rm P}$. One can verify this form using the exact probability in Eq.~\eqref{gen-LDF-eq-1} also. The factor $e^{-\lambda L}$, on the other hand, does not depend on $N_{\rm P}$. For small gaps, $L \ll 1/ \lambda$, this factor is almost unity and only $h(\rho, L,N_{\rm P})$ controls the behavior of the survival probability. However, for large $L$, which dictates the long-time behavior, the exponential factor is dominant and the function $h(\rho, L,N_{\rm P})$ gives only the sub-leading effect. For a given value of $L$, we saw in Eq.~\eqref{sok-surv-eq-18} that the survival probability at long time decays as $\sim \exp \left(- \pi ^2 n /8 L^2 \right)$. \bluepw{When we average this exponential decay over $e^{-\lambda L}$, the resulting disorder-averaged survival probability takes the form in terms of the Meijer-G function as indicated in Eq.~\eqref{sok-surv-eq-28}. Using the asymptotic form of this function in Eq.~\eqref{sok-surv-eq-30}, as also done before in deriving Eq.~\eqref{1d-AIL-surv-L1l2-eq-3}, we obtain the stretched-exponential relaxation for the disorder-averaged survival probability as quoted in the second line of Eq.~\eqref{gen-LDF-eq-11}.}

%dictates the long-time behaviour of the survival probability.
%For the walker to survive being caged at long times, the interval $L$ over which it moves also has to be significantly large.
%This indicates that in one dimension the long time trapping properties are primarily governed by the initial configuration of obstacles, rather than the specific dynamical rules such as the walker's pushing ability. In particular, survival at large times is dominated by rare regions that contain unusually large obstacle-free segments at the initial time.

%In one dimension, our results indicate that the long-time behavior of the survival probability in caging scenarios is primarily governed by the initial spatial configuration of obstacles, rather than the specific dynamical rules such as the walker's pushing ability. In particular, survival at large times is dominated by rare regions that contain unusually large obstacle-free segments at the initial time. These regions allow the walker to remain mobile for extended periods before eventually becoming caged. As a result, even models with different local dynamics—such as the AIL model and the Sokoban model—exhibit similar asymptotic behavior, highlighting the geometric origin of the trapping mechanism in one dimension.
%It is quite natural to ask - what leads to this robustness? 

Fig.~\ref{fig-rate-function}(a) represents a comparison of our rate function $\mathcal{I}(\omega)$ in Eq.~\eqref{gen-LDF-eq-9}
with $ \left[ -\ln S\left( n| N_{\rm P} \right) \big/ N_{\rm P}\right]$ where $S \left( n | N_{\rm P} \right)$ is determined using Eq.~\eqref{gen-LDF-eq-44}. Recall that the latter expression is valid for large $n$ but arbitrary $N_{\rm P}$. On the other hand, Eq.~\eqref{gen-LDF-eq-9} is based on the large-deviation calculation which is valid under the joint limit $n \gg 1,~N_{\rm P} \gg 1$ with the ratio $\omega = n /\left( N_{\rm P} \right)^3$ held fixed. Therefore, the two expressions, as expected, show deviations at small values of $N_{\rm P}$. However, as $N_{\rm P}$ is increased, they show better agreement, establishing the validity of Eq.~\eqref{gen-LDF-eq-9}. Next in panels (b)-(e) of the same figure, we have also compared them with the numerical simulations. The simulation data, shown in green, is completely consistent with Eq.~\eqref{gen-LDF-eq-44}, which is shown in red. Deviations appear only for small $n$, since our theoretical expression is not valid there. From these figures, it is also clear that sampling larger $\omega$ values for large $N_{\rm P}$ in simulation is computationally challenging [see panel (e)]. This requires performing Monte Carlo simulations at time scales much greater than $ \left( N_{\rm P} \right)^3$, where the survival probability is extremely small. As a result, a very large sample size is needed to accurately capture these rare events. The range of $\omega$ over which we can compare the simulation with our theoretical expressions quickly decreases as we increase $N_{\rm P}$. However, within this range, we see a good agreement between simulation and Eq.~\eqref{gen-LDF-eq-44} and both of them converge to the rate function $\mathcal{I}(\omega)$ in Eq.~\eqref{gen-LDF-eq-9} on increasing $N_{\rm P}$ (shown in black).

\section{Sokoban model in two dimension}
\label{sec-2d-1}
In one dimension, we have a universal long-time stretched-exponential relaxation of the survival probability for any $N_{\rm P}$. The stretch exponent $1/3$ matches with the BVDV theory in Eq.~\eqref{original-eqn}.
%the the survival probability $S(n)$ for the Sokoban model exhibits a universal stretched-exponential relaxation at late times for any value of $N_{\rm P}$. %Moreover, this behavior is completely universal against variations in the pushing rules of the walker. 
In this section, we analyze the trapping aspects of the two-dimensional Sokoban model and its variant, the G-Sokoban model, introduced in Sec.~\ref{sec-model} and in Fig.~\ref{fig-update}. 
%For comparison, we also present corresponding results for the de Gennes' AIL model.
%extend the analysis to two dimensions and investigate whether similar universal behavior persists. We first begin with the study of the trapping time and trap size. The goal is to use these observables to identify the precise trapping mechanism governing the Sokoban model. For comparison, we also present corresponding results for the de Gennes' AIL model.
\begin{figure}[t]
	\centering
	\includegraphics[scale=0.25]{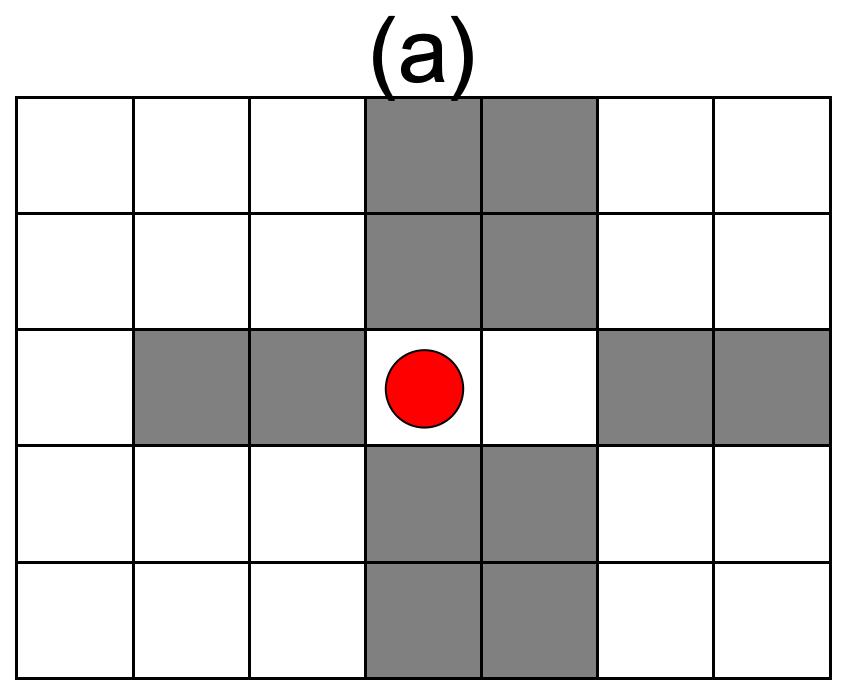} \hspace{2 em}
	\includegraphics[scale=0.25]{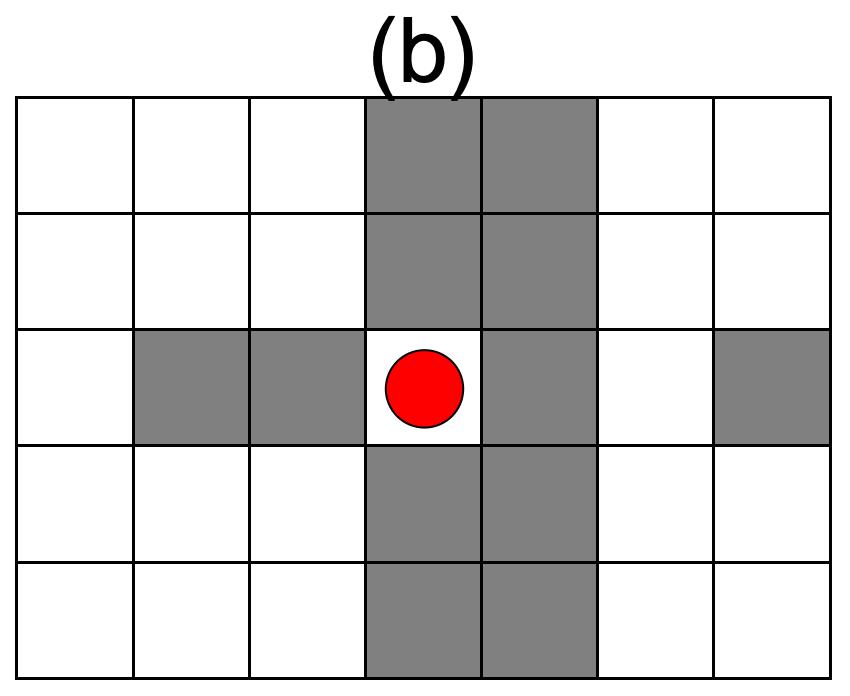}
	\caption{Panels (a) and (b)  schematically show the initial configurations of obstacles as gray blocks and walker in red. A single vacancy located either adjacent to the walker or at a next-nearest-neighbor site determine the trapping time of the Sokoban model in Eq.~\eqref{2d-eq-5} for $\rho \to 1$.}
	\label{fig-vacncy}
\end{figure}
\subsection{Mean trapping time}
Let us first look at the average of \( n_{\rm T} \). Recall that, by our definition, the walker, for a given realization of the initial obstacle configuration, is considered trapped when the number of distinct visited lattice sites saturates indefinitely to a finite value. The observable \( n_{\rm T} \) denotes the time at which this saturation is first reached. Analytical calculations in two dimensions turn out to be considerably more challenging than in the one-dimensional case. However, for densities close to unity (\( \rho \to 1 \)), we can calculate the leading order behavior of $\langle n_{\rm T} \rangle$ in $(1-\rho)$. 

\begin{figure*}[t]
	\centering
	\includegraphics[scale=0.35]{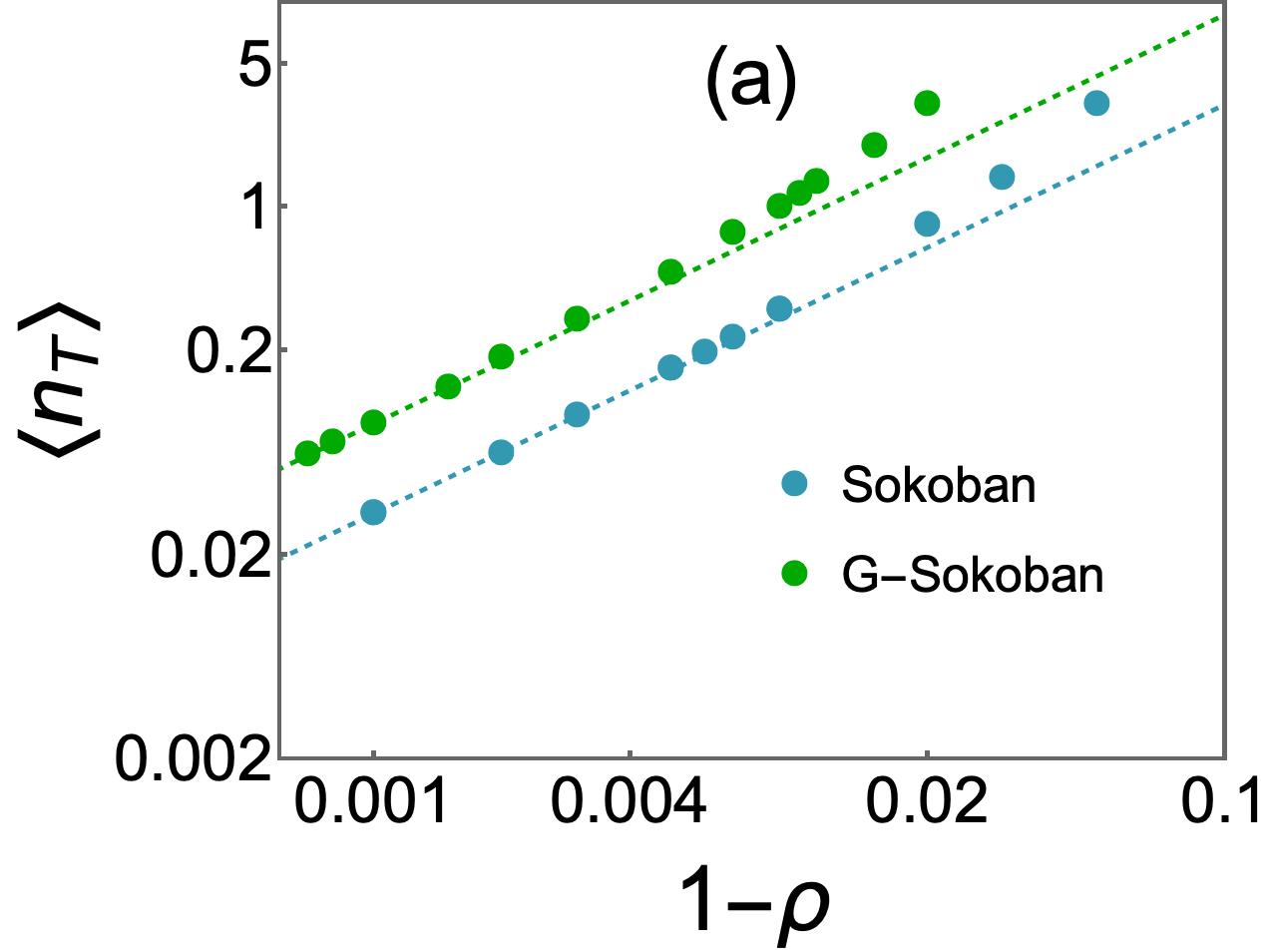} \hspace{2 em}
	\includegraphics[scale=0.34]{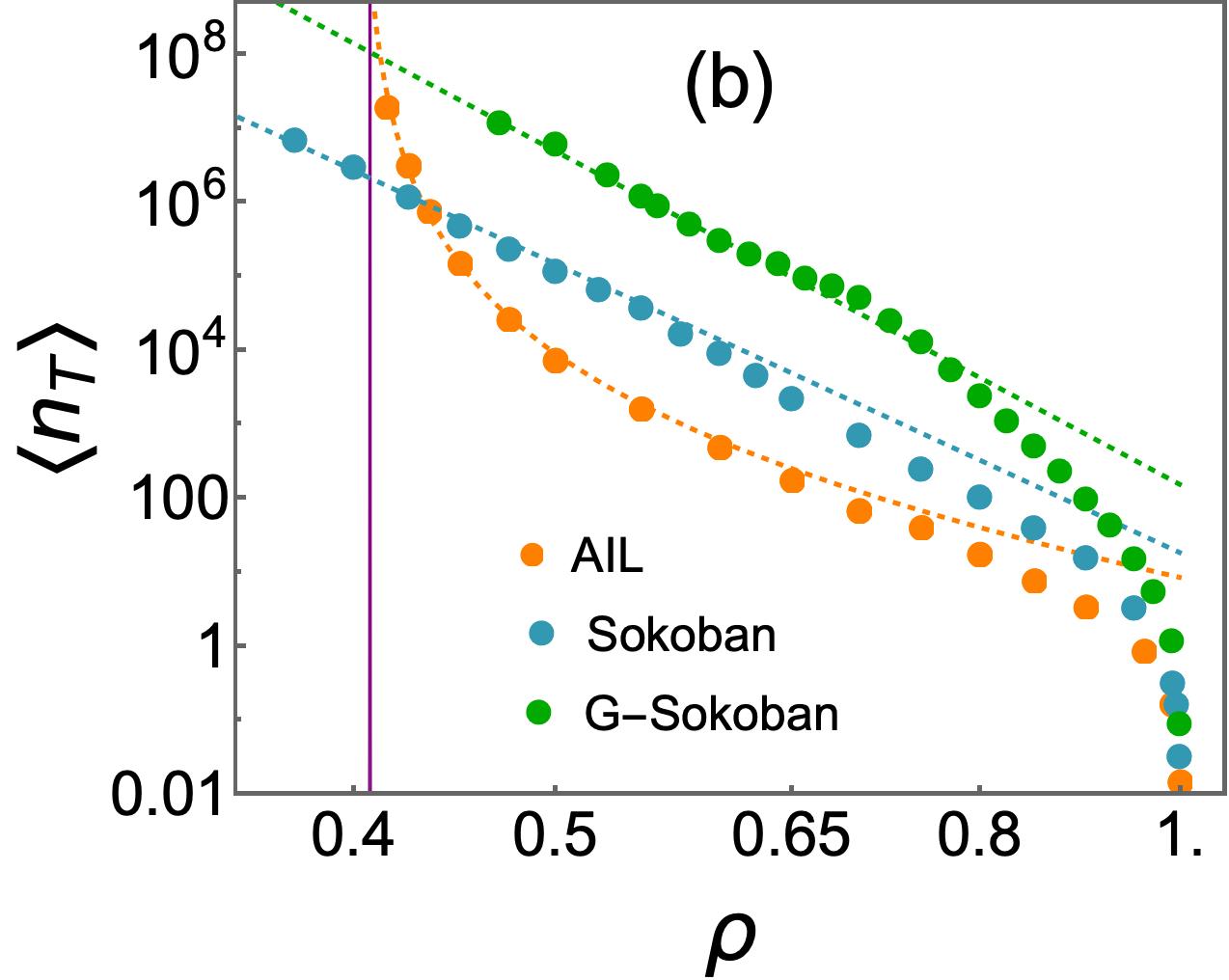}
	\caption{(a) Mean trapping time $\langle n _{\rm T} \rangle$ as a function of $(1-\rho)$ for the two-dimensional Sokoban and G-Sokoban models when $\rho$ close to unity. The dots are obtained from the simulations while dashed lines are the high-density expressions in Eq.~\eqref{2d-eq-5} for Sokoban and in Eq.~\eqref{2d-eq-6} for G-Sokoban. (b) Here, we show $\langle n _{\rm T} \rangle$ across all the accessible density range. Once again, the dots are from the numerical simulation. The blue and the green lines are fits to the simulation data, as explained in Eqs.~\eqref{2d-eq-8} and \eqref{2d-eq-9} respectively. For comparison, we have also shown in orange the simulation results for the AIL model, which diverges around the percolation threshold, $\rho \to \rho_c$. \bluepw{In Sec.~\ref{panid-neq}, we present a heuristic scaling argument suggesting that, near $\rho_c$, the leading divergence of $\langle n_{\rm T} \rangle_{\rm AIL}$ is described by $\langle n_{\rm T} \rangle_{\rm AIL} \sim (\rho-\rho_c)^{-3.8373}$, shown by the orange line in panel~(b).}}
	\label{fig-2d-nT}
\end{figure*}

To see this, let us consider $\rho = 1$ so that all lattice sites -- except the origin where the walker is initially placed -- are occupied with obstacles. So the Sokoban is trapped right at the beginning, \emph{i.e.} $\langle n_{\rm T} \rangle =0$. On the other hand, when \( \rho \) is close to, but not exactly equal to unity, the leading-order behavior is determined by the presence of a single vacancy. Two type of configurations can arise, as shown schematically in Fig.~\ref{fig-vacncy}(a) and (b). In both cases, one needs at least twelve lattice sites with obstacles and one site that is vacant. The vacant site may be adjacent to the walker [panel (a)] or located at a next-nearest-neighbor position [panel (b)], such that the walker can jump by pushing the obstacle into this vacant site. Moreover, this vacant site can be located in any of the four nearest neighbors of the walker, which gives a degeneracy of $4$. Combining everything together, the probability of observing such initial configurations of obstacles is
\begin{align}
\mathbb{P} =8 \rho  ^{12} (1-\rho). \label{2d-eq-4}
\end{align} 
Given this configuration, the trapping time can, in principle, take all possible positive integer values. For example, $n_{\rm T}=1$, when the Sokoban successfully jumps on the very first attempt. The probability of such a successful jump is \(1/4\). Similarly $n_{\rm T}=2$ when the first jump attempt is unsuccessful but the second attempt succeeds. The probability of this is $3/4 \times 1/4$. Proceeding in this way, we can write the mean of $n_{\rm T}$ as
\begin{align}
\langle n_{\rm T} \rangle _{\rm sok}  & \simeq \mathbb{P} \left [ \frac{1}{4} \times 1 + \frac{3}{4}  \times \frac{1}{4} \times 2   + \cdots  \right. \nonumber \\
&~~~~~~~~ \left. +\cdots+ \left( \frac{3}{4} \right)^{m-1} \times \frac{1}{4} \times m + \cdots  \right], \label{2d-eq-2}
\end{align}	
where we have used the subscript `sok' to indicate the model (this notation will be adopted throughout the remainder of the paper). It is possible to perform the above summation analytically and obtain
\begin{align}
\langle n_{\rm T} \rangle _{\rm sok} \simeq 32 (1-\rho),~~~~\text{as }\rho \to 1. \label{2d-eq-5}
\end{align}
The same treatment for the G-Sokoban model yields
\begin{align}
& \langle n_{\rm T} \rangle _{\rm Gsok} \simeq 88 (1-\rho), ~~~~\text{as }\rho \to 1. \label{2d-eq-6}
\end{align}
%For $\rho = 1$, the average correctly goes to zero and it approaches this value linearly in $(1-\rho)$. We next compare this with the Sokoban model, where obstacles can be pushed if the site next to it is vacant; see Fig~\ref{fig-update}. Under this scenario, the vacant site can either be situated directly adjacent to the walker, as shown in Fig.~\ref{fig-vacncy}(b), or located at a next-nearest neighbor position, as shown in Fig.~\ref{fig-vacncy}(c). In the latter case, the walker can push the obstacle, move into its position, and become trapped.Now the probability of observing each of these configurations initially is $4 \rho ^{12}(1-\rho)$ and their sum gives

%As in the AIL model, the walker in this case also becomes trapped upon making a successful jump in the direction of the vacant site. Therefore, for the trapping time to be $n_{\rm T} = m$, the walker must remain stationary till the $(m-1)$--th time step and jump only in the $m$-th time step. The probability of this happening is $(3/4)^{m-1} \times 1/4$. Summing over all possible values of $m$, the mean trapping time can be found to be
%\begin{align}
%\langle n_{\rm T} \rangle _{\rm sok} \simeq 32 (1-\rho),~~~~\text{as }\rho \to 1. \label{2d-eq-5}
%\end{align}
%Beyond the original Sokoban model introduced in \cite{Shlomi-1}, we also study its variant, referred to as the G-Sokoban model, where the obstacles can be pushed in three possible directions, as explained in Sec.~\ref{sec-model}. Employing the same treatment as before yields

\noindent
Our analysis thus reveals that the mean trapping time for both models vanishes linearly in $(1-\rho)$, albeit with different prefactors.	For a given value of $\rho$ close to unity, we find that the G-Sokoban model has the highest trapping time, followed by the Sokoban walker. Due to the highest mobility of the obstacles in the G-Sokoban model, the traps can be relatively large, and the walker thus requires a longer time to get trapped in these larger traps. In Fig.~\ref{fig-2d-nT}(a), we present a comparison between our theoretical result and the numerical simulations, and find an excellent agreement between them for both models.

%We have compared these expressions with the numerical simulations in Fig.~\ref{fig-2d-nT}. For a given value of $\rho$ close to $1$, we find that G-Sokoban model has the highest trapping time, followed by Sokoban walker and then AIL model.

%\begin{figure}[t]
%	\centering
%	\includegraphics[scale=0.35]{fig-avg-nT-rho1p0.jpeg}
%	\includegraphics[scale=0.22]{fig-avg-nT-all-rho.jpeg}
%	\includegraphics[scale=0.34]{fig-avg-nT-all-rho-both.jpeg}
%	\caption{\textit{Left panel:} Mean trapping time $\langle n _{\rm T} \rangle$ as a function of $(1-\rho)$ for two dimensional models with $\rho$ close to unity. The dots are obtained from the simulations while dashed lines are the high-density expressions in Eq.~\eqref{2d-eq-3} for AIL, Eq.~\eqref{2d-eq-5} for Sokoban and in Eq.~\eqref{2d-eq-6} for generalized-Sokoban. \textit{Middle panel:} Here, we show $\langle n _{\rm T} \rangle$ across all the accessible density range. Once again the dots are from simulation. The blue and the green lines are fits to the data in the moderate density regime, as explained in Eqs.~\eqref{2d-eq-8} and \eqref{2d-eq-9} respectively. The orange line, on the other hand, represents our analysis in Eq.~\eqref{2d-eq-7}. \textit{Right Panel:} The high-density expression is combined with the moderate-density expression, inferred from the fitting analysis, to obtain a description in the entire density range accessible through numerical simulations. The combined expressions are given in Eqs.~\eqref{new-pBe-eq-1}, \eqref{new-pBe-eq-2} and \eqref{new-pBe-eq-3} for the three models.
%	}
%	\label{fig-2d-nT}
%\end{figure}

Beyond $\rho \to 1$, we have performed numerical simulations to study the average trapping time for other values of $\rho$. We refer to Appendix~\ref{appen-simulation} for details on the numerical simulations. Fig.~\ref{fig-2d-nT}(b) shows a plot of $\langle n_{\rm T} \rangle $ as a function of $\rho$ for both models. For comparison, we have also shown the simulation results for the AIL model. One can clearly observe that $\langle n_{\rm T} \rangle $ diverges for the AIL model (shown in orange) as the density approaches the percolation threshold, $\rho \to \rho _c ^+~(\approx 0.407)$. However it remains finite for the Sokoban model even past the percolation threshold, at least within the range of density values for which we can perform our simulations. The minimum density for which we could reliably run our simulation is  $\rho = 0.375$ for the Sokoban model and $\rho = 0.47$ for the G-Sokoban model. At these densities, the average trapping time is on the order of $\sim 5 \times 10^6$, and to accurately probe such an average time, we need a total simulation time of the order of $\tau_{\rm sim} \sim 10^8 $. Decreasing $\rho$ further increases $\tau_{\rm sim}$, and to avoid the finite system size effects at such large time values, we also need to increase the total lattice size (see Appendix~\ref{appen-simulation}). Both of these factors contribute to making the simulations significantly more expensive.

Within the accessible densities, we do not see any divergence of $\langle n_{\rm T} \rangle$ for the Sokoban and G-Sokoban models, while it diverges for the AIL model. Let us first explain why it diverges for the AIL model.

{\color{black}
\subsubsection{Divergence of $\langle n_{\rm T} \rangle $ for the AIL model}
\label{panid-neq}
%For $\rho \leq \rho _c$, a spanning cluster of vacant sites of infinite size emerges, and it takes an infinite time for the AIL walker to sample all sites within this cluster \cite{Havlin-2}. This leads to the divergence of  $\langle n_{\rm T} \rangle $ for $\rho \leq \rho _c$ (see Appendix~\ref{sec-appen-diverge-AIL} for more details).

We explained before that when $\rho=1$, all sites except the origin, which is initially occupied by the walker, have obstacles. When $\rho$ is close but not equal to unity, the lattice is mostly occupied by the obstacles, with only small, sparsely distributed clusters of vacant sites. As $\rho$ is decreased further, these clusters start to merge and their typical size increases. This growth continues till the percolation threshold, and for $ \rho \leq \rho _c$, a giant cluster of infinite average size emerges.

%Below critical density, 

Let us next understand the trapping time in this context.
Remember that the trapping time $n_{\rm T}$ is defined as the time at which the number $\Omega(n)$ of distinct visited sites attains its saturation value for the first time; see Fig.~\ref{fig-trap-size}(d). For a given realization of the AIL model at $\rho >\rho_c$, $\Omega(n)$ will saturate once the walker has visited all the sites of the finite-sized cluster containing the walker. Therefore, in the AIL model, $n_{\rm T}$ is similar to the cover time. For a general stochastic process moving on a lattice, cover time is defined as the shortest time needed to visit every lattice site of a given domain at least once \cite{Aldous1983, Chupeau2015}. In standard studies of cover time, the domain is usually fixed for all realizations, and the randomness in the cover time comes only from stochastic nature of the process. In the AIL model, however, the domain is determined by the initial configuration of the obstacles, which itself is random. Thus, one has two sources of randomness determining $n_{\rm T}$: the stochastic motion of the AIL walker  on a fixed cluster, and the randomness of the initial obstacle configuration, which generates clusters of different sizes.

%since it represents the time required to explore all sites in the cluster at least once \cite{Aldous1983, Chupeau2015}. However, unlike standard studies of cover time in a domain, where the domain size is fixed for all realizations, the cluster size in the AIL model varies from realization to realization.

For $\rho \leq \rho_c$, the cluster size can be infinite in some realizations, and naturally the corresponding time to visit all sites in the cluster also diverges. This leads to the divergence of the average trapping time observed in Fig.~\ref{fig-2d-nT}(b) for $\rho \leq \rho_c$. 

The interpretation of the trapping time as the cover-time-like quantity is also useful to heuristically understand how $\langle n_{\rm T} \rangle _{\rm AIL}$ diverges around percolation threshold. To this end, let us first recall the simpler case of a disorder-free $(\rho =0)$ two-dimensional random walk moving in a square $ \mathbb{L} \times \mathbb{L}$ domain where $\mathbb{L}~\gg 1$. We further assume that the domain is fixed for all realizations. In this case, the average cover time scales with the system size as $\sim \mathbb{L}^2\ln^2 \mathbb{L}$ \cite{Aldous1983,Chupeau2015}. Notice that the scaling is different from the characteristic equilibration time of the walk, for example the saturation time of the mean-square displacement (MSD), which scales as $\sim \mathbb{L}^2$. Thus, the scalings of the cover time and the MSD saturation time differ by a $\ln^2\mathbb{L}$ factor. Since this factor varies slowly with $\mathbb{L}$, the MSD-based scaling can still capture the leading algebraic dependence of the cover time on the system size.

Motivated by this observation, we turn to the AIL model with $\rho \gtrsim \rho _c$ and check whether the MSD saturation captures the divergence of $\langle n_{\rm T}\rangle_{\rm AIL}$ near the percolation threshold. At short times, the MSD scales sub-diffusively as $\langle r_n^2 \rangle \sim n^{2/d_W} $ with the exponent $d_W \approx 2.878$, while at late times, it saturates to a time-independent value $\langle r_n^2 \rangle \sim 1/(\rho - \rho _c)^{8/3} $ \cite{Havlin-2}. From these two limits, the MSD saturation timescale is $n_{\rm MSD} \sim  1  \big/ \left( \rho - \rho _c \right)^{\gamma _{\rm AIL}} $ with  $\gamma _{\rm AIL} = 4d_W/3 \approx 3.8373 $. In Fig.~\ref{fig-2d-nT}(b), we have plotted this MSD-based scaling (shown by orange dashed line) along side the simulation data for $\langle n_{\rm T}\rangle_{\rm AIL}$ (shown by orange dots). We see that the agreement between them is quite good, especially close to the percolation threshold. This provides suggestive indication that the leading order behavior of the mean trapping time is given by  $\langle n_{\rm T}\rangle_{\rm AIL}\sim(\rho-\rho_c)^{-\gamma_{\rm AIL}}$ for $\rho \gtrsim \rho _c$. Nevertheless, a rigorous theory for the divergence of $\langle n_{\rm T}\rangle_{\rm AIL}$ remains an open problem.}

\begin{figure*}[]
	\centering
	\includegraphics[scale=0.22]{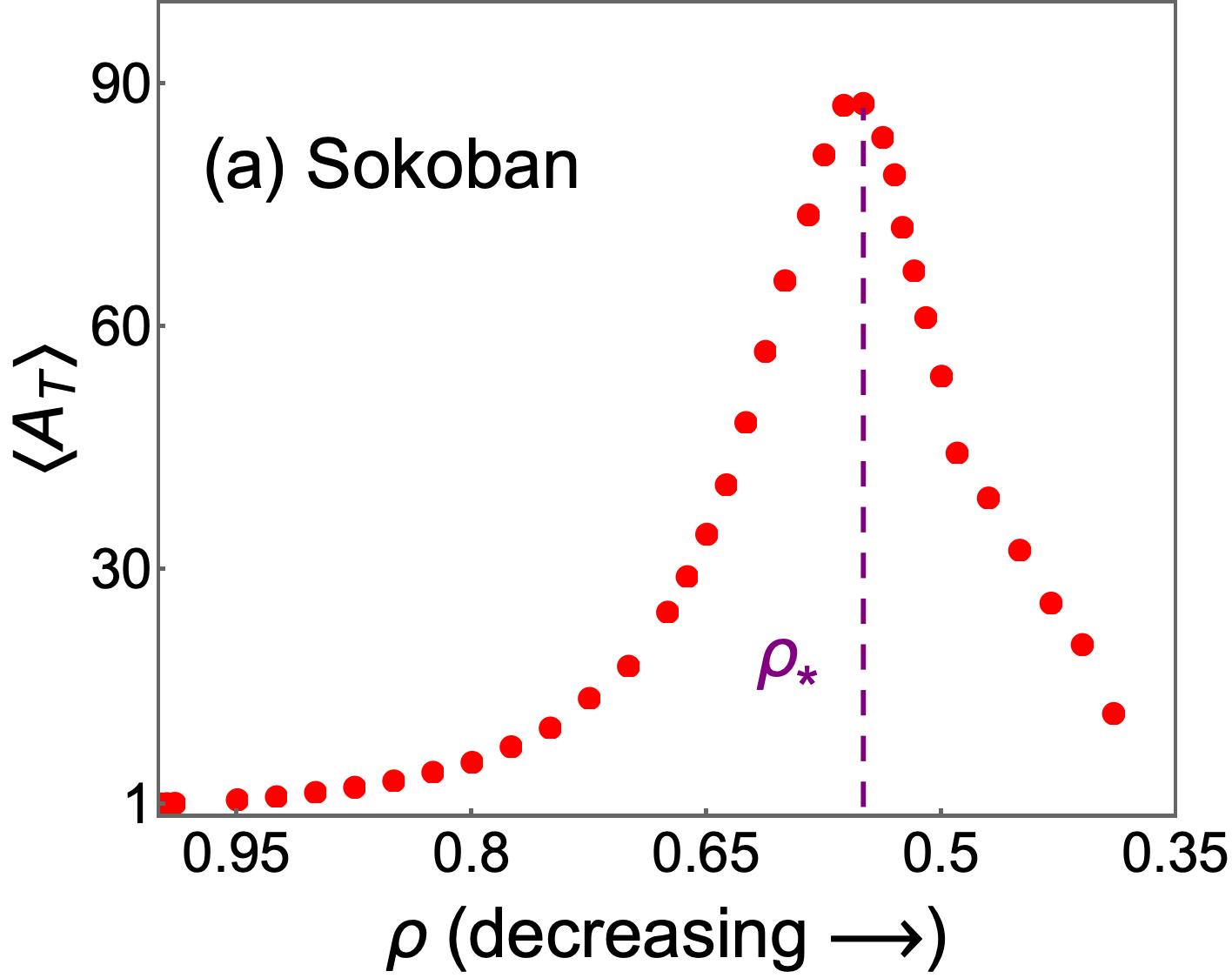} \hspace{2 em}
	\includegraphics[scale=0.21]{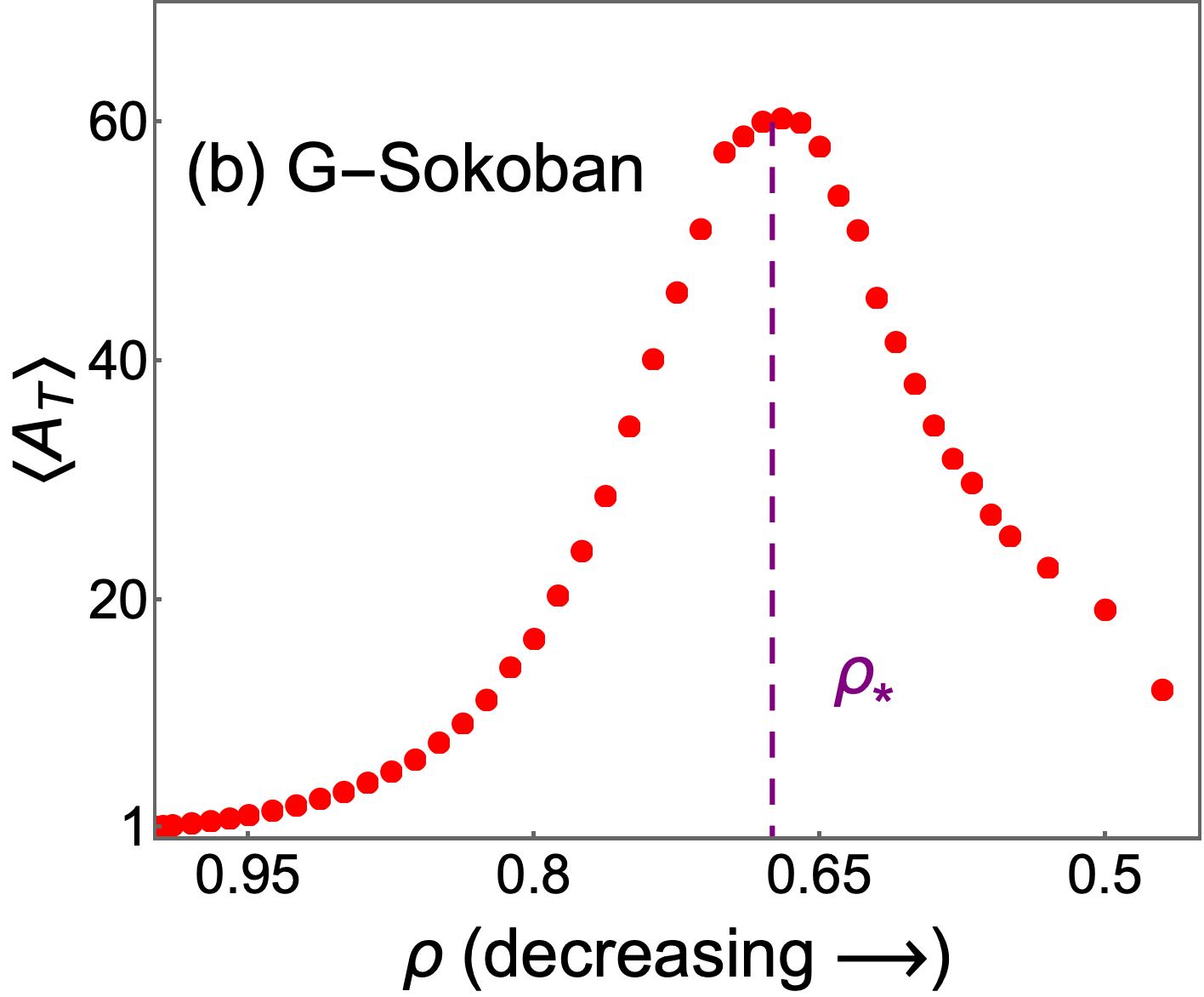}
	\includegraphics[scale=0.21]{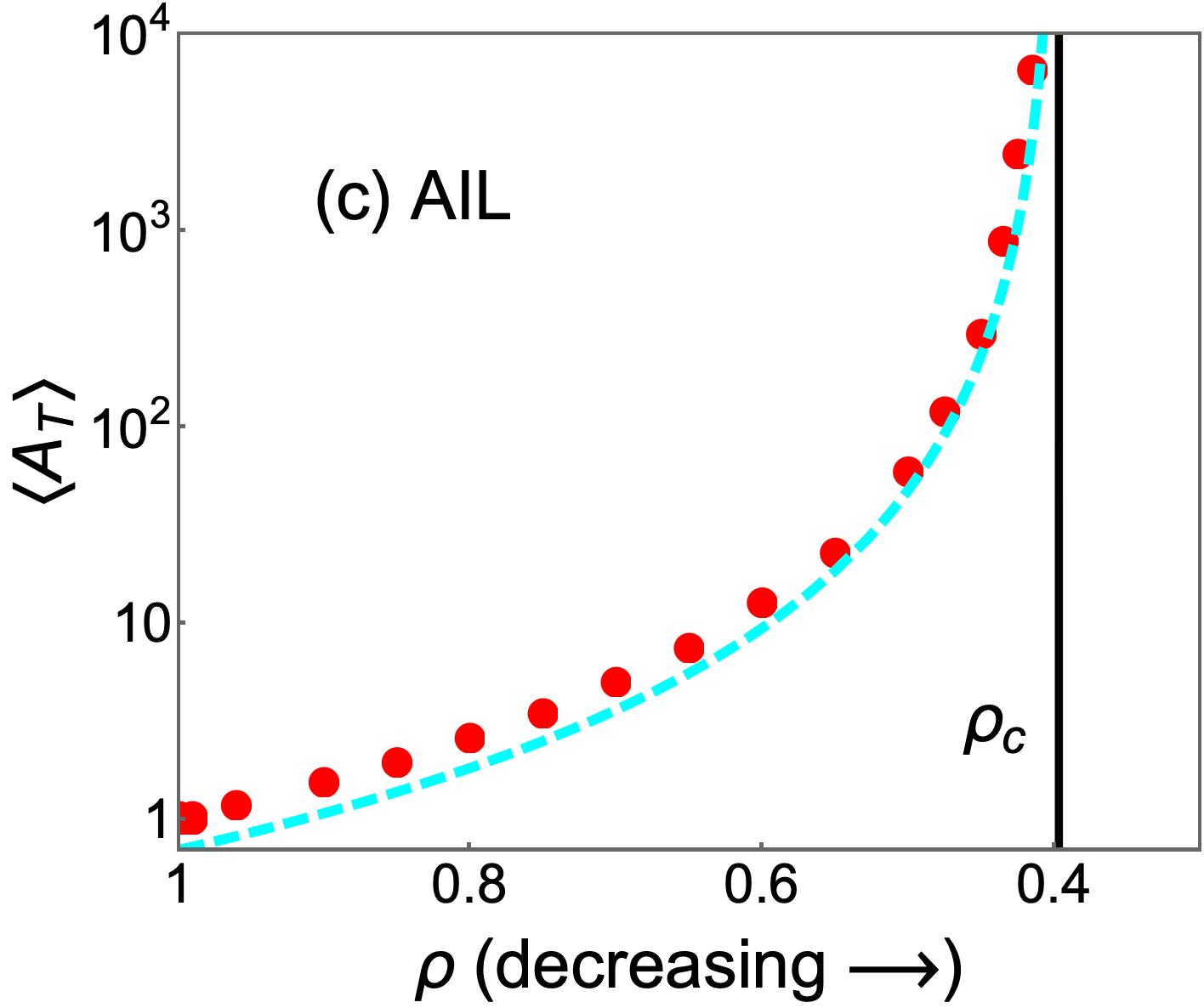} \hspace{2 em}
	\caption{Average trap size $ \langle A_{\rm T} \rangle $ is plotted as a function of $\rho$ for the two-dimensional (a) Sokoban, (b) G-Sokoban and (c) AIL models. Across all panels, the red dots are the simulation data. For the Sokoban and G-Sokoban models, the plot is nonmonotonic with the maximum value occurring at $\rho _*$ (shown by purple dashed line). Based on simulations, we estimate $\rho _* \approx 0.55$ for the Sokoban model and $\rho _* \approx 0.675$ for the G-Sokoban model, where the corresponding maximal mean trap sizes are approximately $87$ and $60$, respectively. In contrast, the plot for the AIL model is monotonic with $ \langle A_{\rm T} \rangle $ diverging at $\rho _c \approx 0.407$. As explained in Sec~\ref{ajbus}, the divergence around $\rho \gtrsim \rho _c$ is described by $\langle A_{\rm T} \rangle \sim (\rho-\rho_c)^{-43/18}$  and shown by the cyan line in panel~(c).}
	\label{fig-2d-area-neww}
\end{figure*}

\subsubsection{Finite $\langle n_{\rm T} \rangle$ for the Sokoban and G-Sokoban models}
Next, we turn to the Sokoban random walk. As remarked before, the average trapping time is finite for all the accessible density values and it does not diverge even past the percolation threshold; see Fig.~\ref{fig-2d-nT}(b). This supports the hypothesis that the percolation transition is lost for the two-dimensional Sokoban random walk. \cite{Shlomi-1}.

\bluepw{Since the Sokoban can push obstacles, the cluster of vacant sites evolves with time. Due to its dynamical restructuring, we find that our heuristic analysis based on the saturation of the MSD does not yield the correct scaling of the trapping time for the Sokoban case. This can be verified using the result from \cite{Shlomi-1} for intermediate densities. In this regime, it was demonstrated that the MSD is diffusive at short time scales, but eventually saturates to a value proportional to $\left[\left(1 - \rho\right)/\rho\right]^{10.989}$ at long times. Then, the saturation time scale should be of the order of $\sim\left[ \left(1-\rho \right) /\rho\right] ^{10.989}$.}

\bluepw{However, a fitting analysis of the simulation data within the accessible density regime reveals that the mean trapping time for the Sokoban model satisfies a scaling
\begin{align}
\langle n_{\rm T} \rangle _{\rm sok} \sim  \frac{(1-\rho)}{\rho^{\gamma _{\rm sok}}}, \label{2d-eq-8}
\end{align}
with the exponent estimated to be $\gamma _{\rm sok} \approx 13.17 \pm 0.1 $. The error is estimated through least squares fitting of the simulated data. This fitting result is shown by the blue dashed line in Fig.~\ref{fig-2d-nT}(b). In this figure, we clearly see that $\langle n_{\rm T} \rangle _{\rm sok}$ remains finite even past the percolation threshold. Moreover, the scaling in Eq.~\eqref{2d-eq-8} is different from the saturation timescale of the MSD. Together these observations reflect the dynamical restructuring of the environment in the Sokoban model, and suggest that trapping in the Sokoban model does not occur merely through the exploration of a cluster of vacant sites. Later, we will show that, at low densities, a self-trapping mechanism emerges for the Sokoban walker, in which it gets dynamically localized by a creating a cage for itself.}

%In two dimensions, the random walk being recurrent can visit a site multiple times given sufficient time. Every time it visits a site, it can also bring an obstacle along. The combined effect of the recurrence and the pushing ability can enable the walker to dynamically create its own trap by accumulating the obstacles along its path. As a result, although there might exist a giant vacancy cluster for density values close to the percolation threshold, the walker can still reshape its local environment to generate confining traps. This self-trapping mechanism causes the walker to get trapped even past the percolation threshold.

We have repeated the same fitting treatment for the G-Sokoban model and found
\begin{align}
\langle n_{\rm T} \rangle _{\rm Gsok} \sim \frac{(1-\rho)}{\rho^{\gamma _{\rm Gsok}}},  \label{2d-eq-9}
\end{align}
with a different exponent $\gamma _{\rm Gsok} \approx 14.93 \pm 0.4$. This is again shown in Fig.~\ref{fig-2d-nT}(b). Once again, we do not find any divergence of the average trapping time within the considered density regime.

\subsection{Trap size}
\label{sec-2d-2}
We now turn our focus to the size of the trap. Remember that the trap size is defined as the number of vacant sites enclosed by the trap. As before, the case of $\rho \to 1$ turns out to be perturbatively solvable for $\langle A_{\rm T} \rangle$. When $\rho = 1$, all lattice sites except the origin are occupied by obstacles, resulting in a trap size of $\langle A_{\rm T} \rangle = 1$. When $\rho \to 1$, the leading order correction to this is determined by one single vacancy, as indicated in Fig.~\ref{fig-vacncy}. These configurations are characterized by $A_{\rm T} = 2$ and their probability is given in Eq.~\eqref{2d-eq-4}. Following this expression, we obtain the mean trap size 
 \begin{align}
& \langle A_{\rm T} \rangle_{\rm sok} \simeq 1 + 8(\rho -1),~~\text{as }\rho \to 1. \label{2d-eq-10}
\end{align} 
Similarly, for the G-Sokoban model, we find
\begin{align}
& \langle A_{\rm T} \rangle_{\rm Gsok} \simeq 1 + 12(\rho -1), ~~\text{as }\rho \to 1.
\end{align}
Combining these results with our previous analysis of $\langle n_{\rm T} \rangle$, we find a linear relation for both models
\begin{align}
& \langle A_{\rm T} \rangle_{\rm sok} \simeq 1+ \frac{ 1}{4} \langle n_{\rm T} \rangle _{\rm sok},  \label{2d-eq-11} \\
& \langle A_{\rm T} \rangle_{\rm Gsok} \simeq 1+ \frac{ 3}{22} \langle n_{\rm T} \rangle _{\rm Gsok}
\end{align}	
These relations are valid only for $\rho$ close to unity, where both averages are small. It is also reminiscent of what was rigorously proven in one dimension (see Eq.~\eqref{bhuqm} and Fig.~\ref{fig-1d-crossover}). 
%To verify it in simulations, we have again generated $\langle A_{\rm T} \rangle$ and $\langle n_{\rm T} \rangle $ for different values of $\rho$ using numerical simulations and then plotted them in Fig.~\ref{fig-2d-area-neww}, see inset in the left panel. We notice that for all three models, the simulation data approaches the predictions of Eq.~\eqref{2d-eq-11} in the appropriate regime.
The universality of the high-density linearity across different models and dimensions suggests that the underlying trapping mechanism in the high density regime is the same for all of them. In particular, the Sokoban walker explores the small void surrounding its initial position. It can slightly enlarge this void by displacing nearby obstacles, but the extent of this expansion is limited due to the high obstacle density. Ultimately, the walker becomes trapped once all vacant sites within this locally reshaped region have been visited. In contrast, a qualitatively different self-trapping mechanism emerges for both models at low densities, as we now illustrate.
%dynamical trapping mechanism emerges, which especially becomes relevant at moderate to low densities. This mechanism is also responsible for the finiteness of $\langle n_{\rm T} \rangle $ past the percolation threshold for these models.

\subsubsection{Self-trapping and nonmonotonic $\langle A_{\rm T} \rangle$}
\label{ajbus}
We plot \( \langle A_{\rm T} \rangle \) as a function of $\rho$ in Fig.~\ref{fig-2d-area-neww} using numerical simulations. For both models, we see in panels~(a-b) that the \( \langle A_{\rm T} \rangle \) vs $\rho$ plot is nonmonotonic. Starting from densities close to unity, the mean trap size starts to increase on decreasing the density. The growth of the mean trap size continues only till a characteristic density $\rho _*$, at which point \( \langle A_{\rm T} \rangle \) achieves its (finite) maximum value. Below this threshold, $\rho < \rho _*$, we find a turnover behavior in which the mean trap size decreases with decreasing $\rho$. Based on numerical simulations, we find the turnover density to be $\rho_* \approx 0.55$ for the Sokoban model and $\rho_* \approx 0.675$ for the G-Sokoban model. The corresponding maximum values of the average trap size are approximately $87$ and $60$, respectively.

\bluepw{By contrast, for the AIL model, the dependence of \( \langle A_{\rm T} \rangle \) on $\rho$ is monotonic with divergence at the percolation threshold $\rho _c$; see Fig.~\ref{fig-2d-area-neww}(c). As explained before, trapping in the AIL model simply corresponds to the exploration of a random cluster, so the trap size is just the size of the cluster itself. Since the average cluster size diverges at the percolation threshold as $\sim 1/(\rho-\rho_c)^{43/18}$ \cite{Havlin-2}, the average trap size therefore diverges as $\langle A_{\rm T} \rangle_{\rm AIL} \sim 1/(\rho-\rho_c)^{43/18}$; see the cyan line in Fig.~\ref{fig-2d-area-neww}(c).}

%Looking at Fig.~\ref{fig-2d-area} for the Sokoban and the G-Sokoban models, we see that \( \langle A_{\rm T} \rangle \) and \( \langle n_{\rm T} \rangle \) possess a non-monotonic dependence, parametrized by $\rho$. While, the initial dependence between them is still described by the high-density linear relation in Eq.~\eqref{2d-eq-11}, the power-law behavior in Eq.~\eqref{2d-eq-12} appears only in the intermediate density regime. It is shown by dashed lines in the middle and right panels of Fig.~\ref{fig-2d-area}. Unlike in the previously studied models, the power-law ceases to hold beyond a certain point, and the mean trap size instead starts to decrease with density while the mean trapping time still increases. This change happens at density $\rho_* \approx 0.55$ for the Sokoban model and at $\rho_* \approx 0.675$ for the G-Sokoban model, also see Fig.~\ref{turnover-fig}. This is fundamentally different from the other models. 

%Although the percolation transition is lost for the $2d$ Sokoban and generalized Sokoban models, they exhibit a turnover regime which, as we now argue, is a consequence of the pushing ability. 

\begin{figure}
\begin{center}
\includegraphics[width=0.3\textwidth]{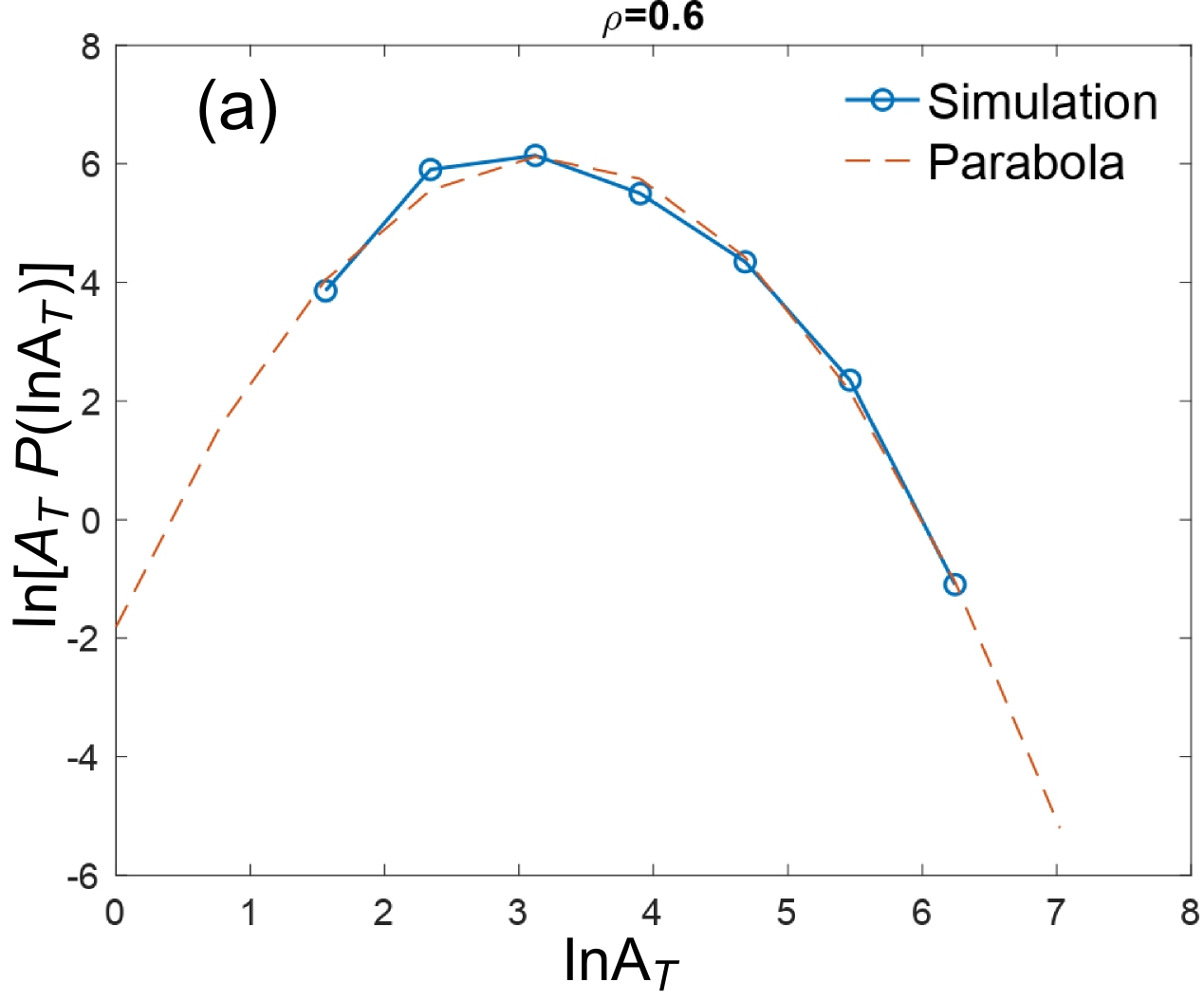} 

\vspace{1.5 em}
\includegraphics[width=0.3\textwidth]{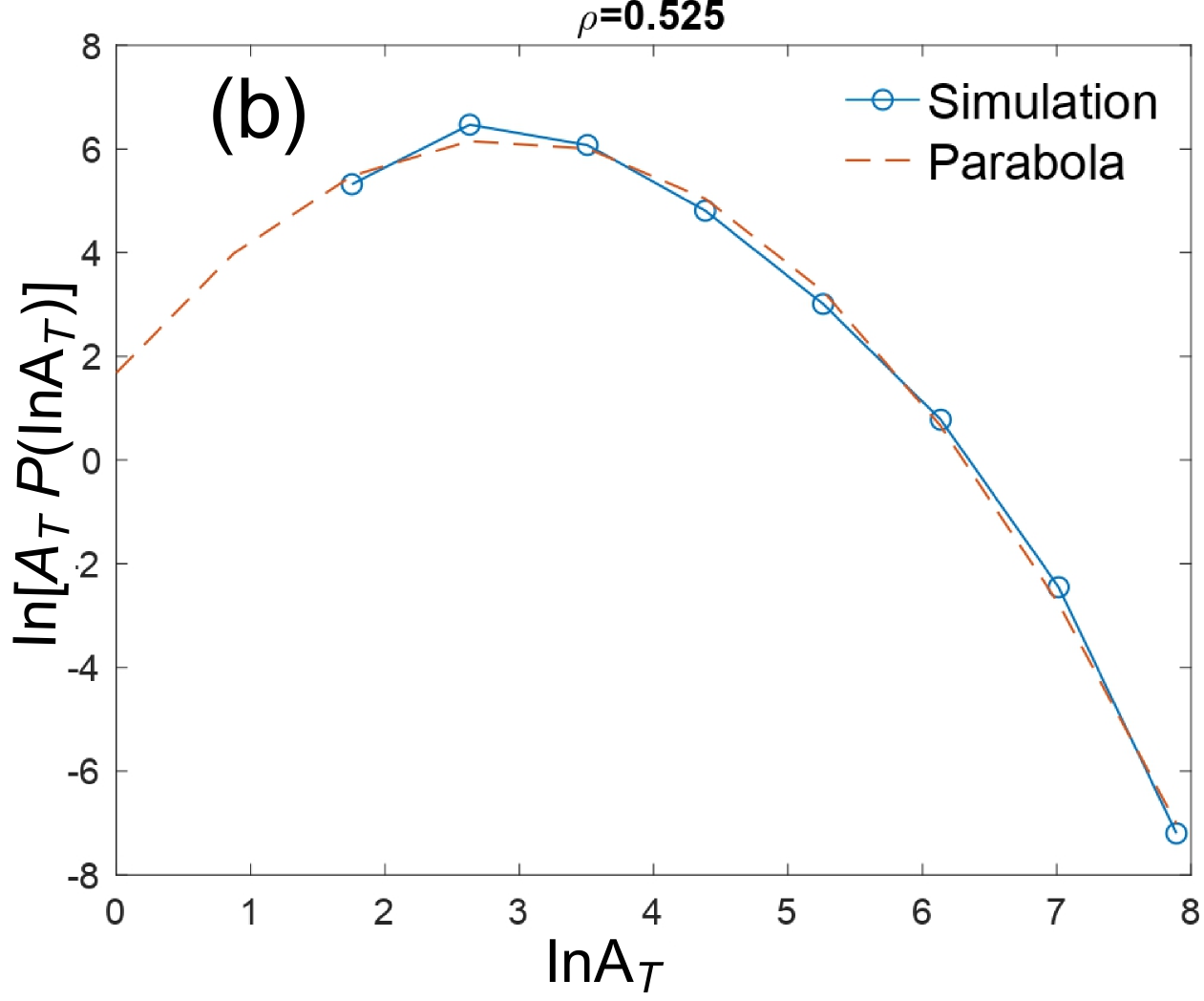}
\end{center}
\caption{\bluepw{The trap size distribution: Shown is the measured $\ln [A_T P(\ln A_T)]$ vs. $\ln A_T$
for (a) $\rho=0.6$ and (b) $\rho=0.525$ for two-dimensional Sokoban model. Also shown is the best-fit parabola in the two cases, showing excellent agreement.}}
\label{fig-trap-size-dist}
\end{figure}

\begin{figure}[]
	\centering
	\includegraphics[scale=0.25]{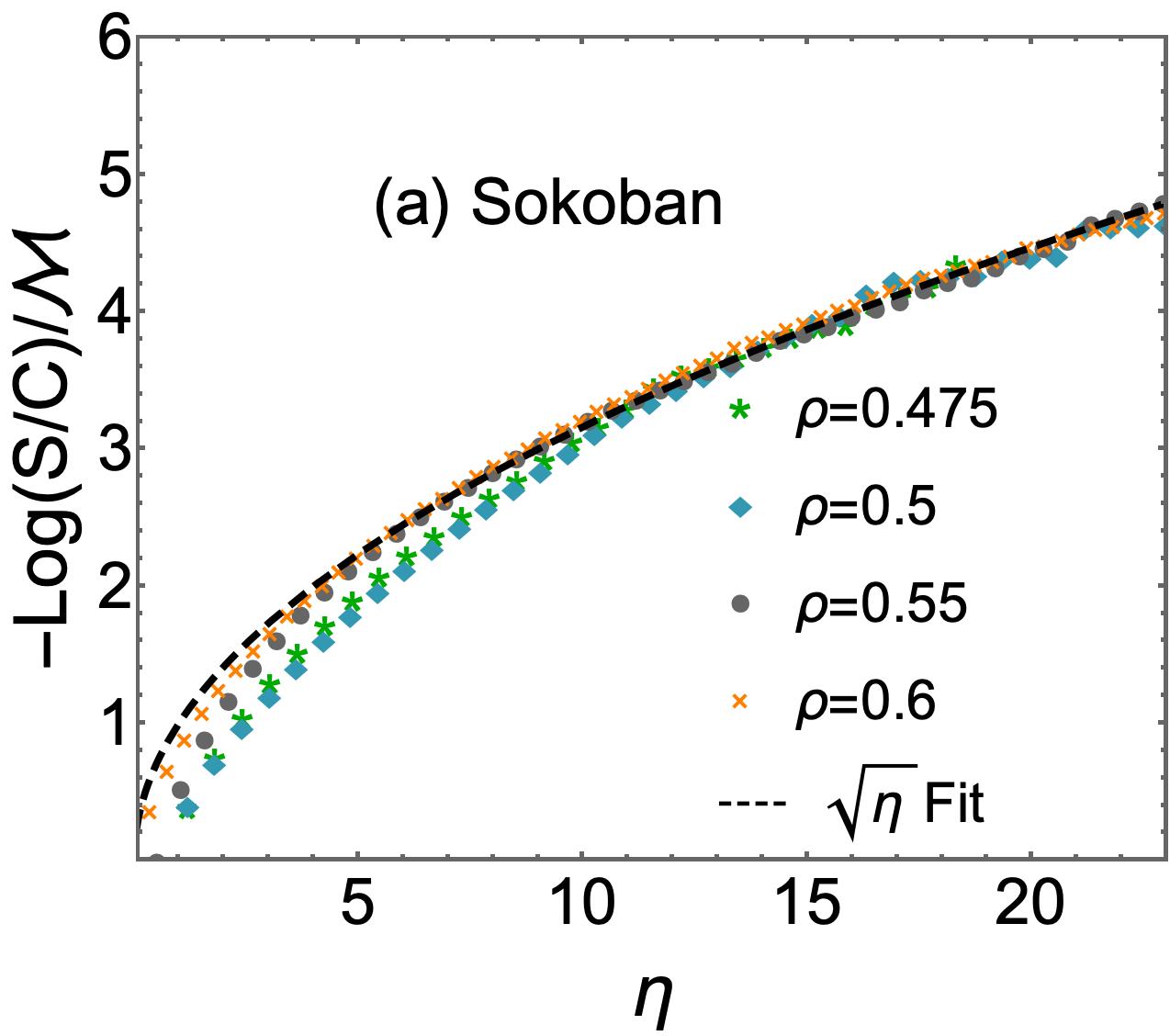} 
	\includegraphics[scale=0.25]{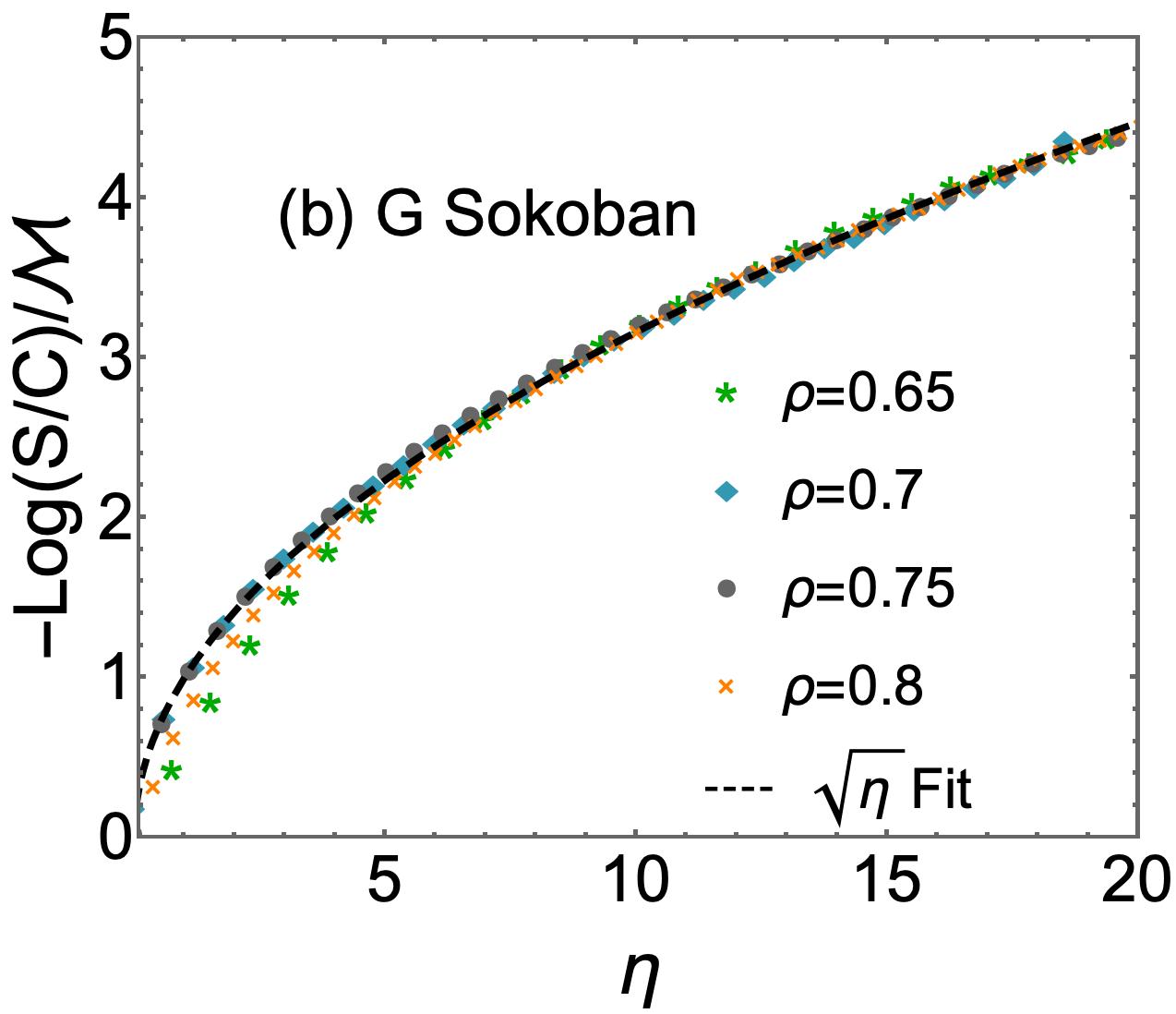}
	\caption{Demonstration of the late-time stretched-exponential decay of the survival probability for two-dimensional Sokoban and G-Sokoban models. We have plotted $-\text{log}\left[ S(\eta) \big/  C(\eta ) \right] \big/ \mathcal{M}(\rho)$ as a function of $\eta = n /\langle n_{\rm T} \rangle $ using numerical simulations for different densities (shown by symbols). For both models at all densities, the numerical data converge to $\sqrt{\eta}$ form, as indicated by the dashed line when $\eta$ is sufficiently large. To generate these plots, we have determined the average $\langle n_{\rm T} \rangle$ from simulations, while the functions $C(\eta)$ and $\mathcal{M}(\rho)$
are estimated using a fitting procedure as explained in Appendix~\ref{appen-fit-2d}.}
\label{fig-2d-survival}
\end{figure}

\begin{figure}[]
	\centering
	\includegraphics[scale=0.3]{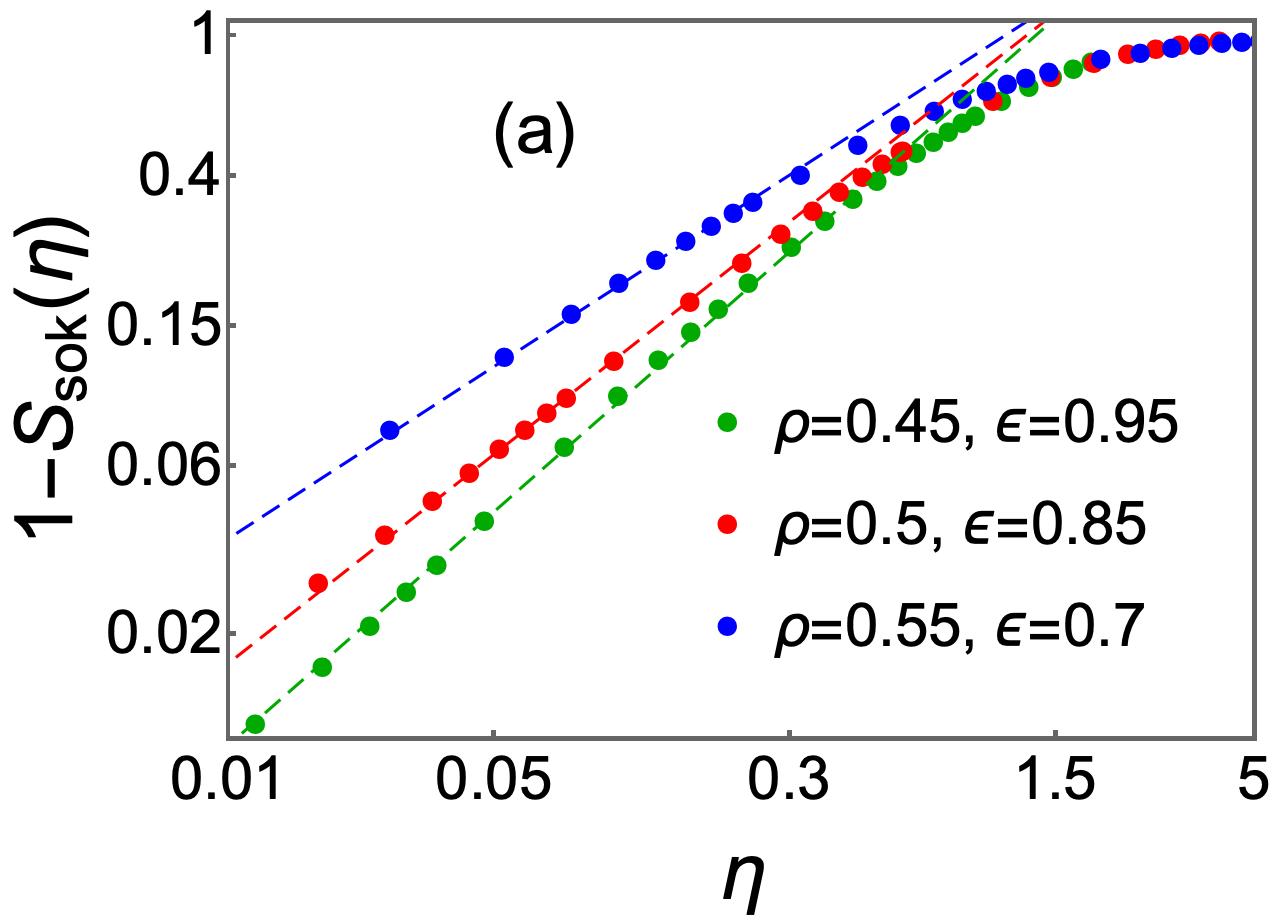} \hspace{1 em}
	\includegraphics[scale=0.3]{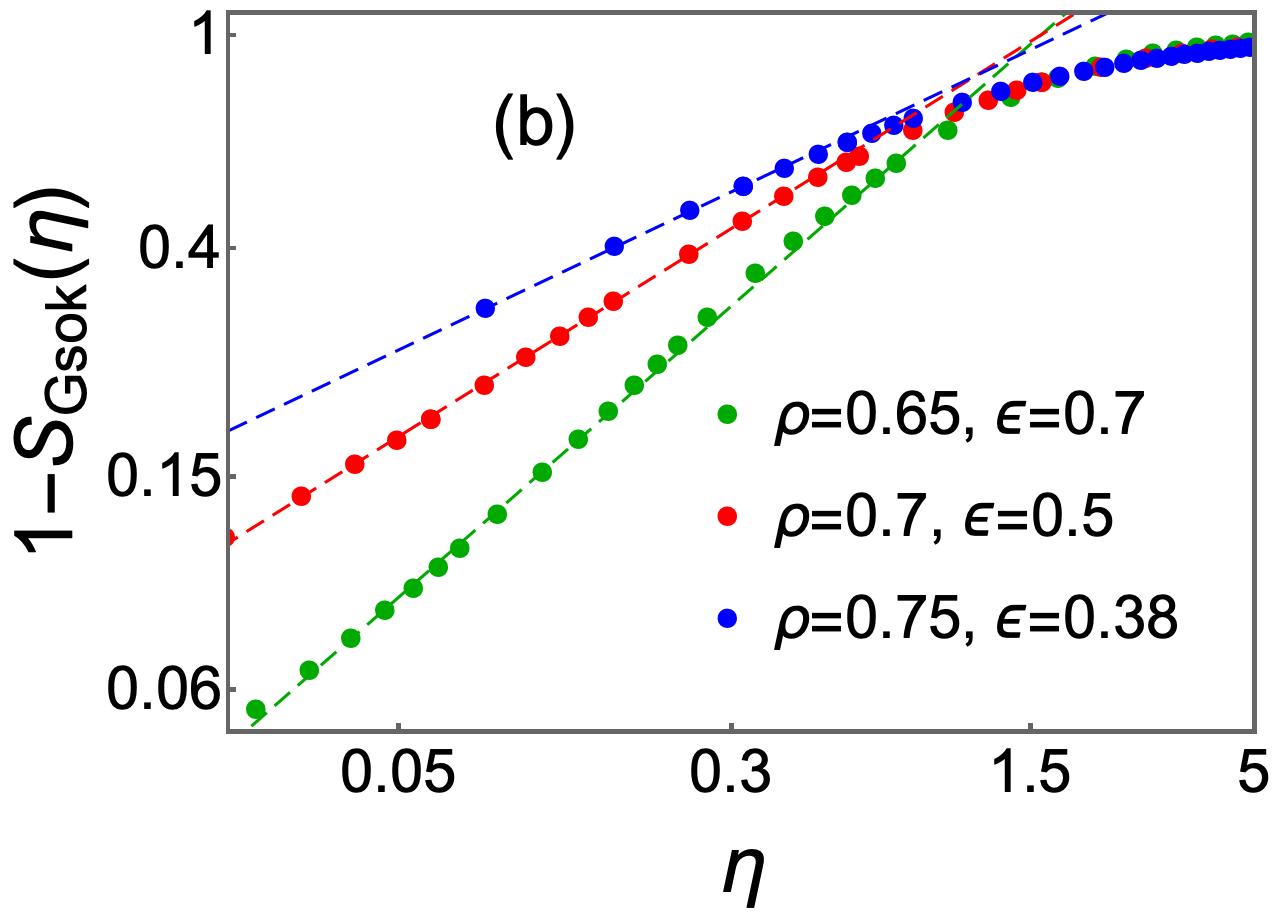}
	\caption{Moderate-time behavior of the survival probability for two-dimensional (a) Sokoban  and (b) G-Sokoban models. We have plotted $ \left( 1-S(\eta) \right)$ vs $\eta$ using numerical simulations where $\eta = n / \langle n_{\rm T} \rangle $. The dashed lines are the fits to the simulation data, and from these fits, we infer $\left( 1-S(\eta) \right) \sim \eta^{\epsilon (\rho)}$ when $\eta \ll 1$. The exponent $\epsilon$, as quoted in the legends, depends on the density as well as on the models. Contrast this with $1d$ where the exponent is independent of the density for a given model. 
	}
\label{2s-short-time}
\end{figure}

So what gives rise to nonmonotonicity for models with pushing ability? Consider again the high-density limit $\rho \to 1$. In this regime, all sites, except for the origin occupied by the walker, are filled with obstacles. Consequently, the average trap size $\langle A_{\rm T} \rangle =1$. On slightly decreasing the density, a void of vacancies is created around the walker which increases the mean trap size $\langle A_{\rm T} \rangle$. As $\rho$ decreases further, the surrounding void grows larger, resulting in a continued increase in $\langle A_{\rm T} \rangle$. For the AIL model, this trend will continue all the way till $\rho \to \rho _c^{+}$ at which point $\langle A _{\rm T} \rangle $ diverges.

However, the situation changes dramatically for the Sokoban and G-Sokoban models, where the walker can minimally push the obstacles. The growth trend continues only till the threshold $\rho _*$. Below this threshold, $\rho < \rho _*$, a qualitatively different trend emerges. Even when large voids are available for motion, the walker may encounter a rare region containing a manipulable arrangement of obstacles. This region, through successive pushing by the walker, can be gradually reorganized into a confining structure; see Fig.~\ref{fig-trap-size}. This leads to the formation of a localized trap that was not pre-existing but dynamically constructed by the walker itself. As $\rho$ continues to decrease, there are fewer manipulable obstacles, so that traps are harder to create, and smaller on average. This results in a turnover to decreasing trap size $\langle A_{\rm T} \rangle$, making the overall relation nonmonotonic.

%Due to the recurrent nature of the walker, it can visit a lattice site multiple times. During each visit, it may also carry an obstacle along its path, gradually modifying the local environment. Given sufficient time, it can become trapped by gathering the obstacles around itself. This mechanism, however, is dominant only when $\rho < \rho _*$. As the obstacle density continues to decrease below this threshold, there are fewer available obstacles to gather giving rise to smaller trap size. This results in a turnover to decreasing trap size $\langle A _{\rm T} \rangle $, making the overall relation nonmonotonic. 

Note that at high densities, self-trapping is suppressed, as the Sokoban cannot substantially rearrange the obstacles due to the limited available space. However, at low densities, the available space allows the walker to move and push obstacles to create configurations that result in self-trapping. As a result, the Sokoban walker cannot percolate, as observed in \cite{Shlomi-1}, or equivalently, in the finiteness of $\langle 	A_{\rm T} \rangle$ and $\langle n_{\rm T} \rangle$ in our case. Nonetheless, it exhibits a trapping crossover, marked by a characteristic density $\rho_*$ that separates two distinct regimes: a self-trapping regime at low density, where the walker becomes dynamically localized within a self-formed trap, and a pre-existing trapping regime at high density, where confinement arises from the initial arrangement of obstacles. This crossover is also observed in the G-Sokoban model highlighting its robustness and suggesting that self-trapping is a generic dynamical mechanism across a broad class of Sokoban-type walkers.

%Finally, we turn to the question of whether the power-law relation between $\langle A_{\rm T} \rangle$ and $\langle n_{\rm T} \rangle$ given in Eq.~\eqref{2d-eq-12} persists even in the Sokoban and G-Sokoban models. For this, we obtain a parametric plot between these two observables in Fig.~\ref{fig-2d-area-neww}, parametrised by $\rho$. Looking at the insets in the middle and right panels, we see that the high density regime, as expected, is described by the linear relation between $\langle A_{\rm T} \rangle$ and $\langle n_{\rm T} \rangle$ in Eq.~\eqref{2d-eq-11}. This is then followed by the power-law relation $\langle A_{\rm T} \rangle \sim \langle n_{\rm T} \rangle ^{0.622}$ in Eq.~\eqref{2d-eq-12}. Unlike in the AIL model, this power-law behavior, however, appears only in the intermediate density regime for the Sokoban and G-Sokoban models. This is shown by the green dashed lines in the respective panels.  Beyond this regime, the system, as explained above, enters a turnover region in which $\langle A_{\rm T} \rangle$ decreases with decreasing $\rho$ while $\langle n_{\rm T} \rangle$ continues to grow. This makes the relation between $\langle A_{\rm T} \rangle$ and $\langle n_{\rm T} \rangle$ nonmonotonic.

\subsection{Trap Size Distribution}
Above we considered the behavior of the mean trap size as a function of $\rho$.  It is also interesting to consider the distribution of trap sizes. Running many realizations of the two-dimensional Sokoban model till trapping for a given $\rho$, we constructed the histogram of trap sizes and discovered that the distribution is well described by a log-normal distribution
\begin{equation}
P(A_{\rm T}) = \frac{1}{A_{\rm T} \sigma \sqrt{2\pi}} \exp\!\left[ -\frac{1}{2\sigma^2} \left( \ln \left(\frac{A_{\rm T}}{\langle A_{\rm T} \rangle } \right)  + \frac{\sigma^2}{2} \right)^2 \right],
\end{equation}
where $\sigma$ is the variance of $\ln A_{\rm T}$. To see the log-normal distribution, we plot $\ln [A_T P(\ln A_T)]$ against $\ln A_T$. For a log-normal distribution, this would yield a parabola.  We show in Fig. \ref{fig-trap-size-dist} the results of implementing this for $\rho=0.6$ and $\rho=0.525$, together with the best fit parabola.  We see that for both values of $\rho$
the parabola gives an excellent fit. We have done this for densities in the range $0.475<\rho<0.6$ and found excellent fits to a log-normal distribution in all cases.  It is an interesting challenge to understand the origin of this log-normal distribution.

\subsection{Survival probability}
\label{sec-2d-3}
We now turn our attention to the survival probability that the Sokoban random walker has not been trapped in dimension two. 
For the $1d$ case, we saw that the survival probability has the long-time stretched-exponential relaxation with exponent proportional to $\lambda ^{2/3} n^{1/3}$ and belonged to the BVDV trapping universality class. For two-dimensional models, we first rescale time by the average trapping time and define
%Adopting the simulation method described in \ref{appen-simulation}, we now proceed to study the survival probability for the two-dimensional models. To this end, we first rescale time by the average trapping time and define
 \begin{align}
  \eta = n  \big/ \langle n_{\rm T} \rangle. \label{2d-st-surv-eq-1}
 \end{align}
Based on extensive numerical simulations, we show in Fig.~\ref{fig-2d-survival} that the late-time decay of the survival probability for both Sokoban and G-Sokoban models turns out to be

\vspace{0.5 cm}
\noindent\fbox{%
\parbox{\columnwidth}{%
\begin{align}
S_{\#}\left(\eta \right) \simeq~C_{\#}(\eta)\exp \left( - \mathcal{M}_{\#}(\rho) ~\eta^{\frac{1}{2}} \right),~~~\text{for }  \eta \gg 1. \label{2d-st-surv-eq-2}
\end{align}
}%
}

\vspace{0.5 cm}
\noindent 
Here, $\#$ stands for the sok or Gsok depending on the model. Comparing this form with the $\phi(n)$ in Eq.~\eqref{original-eqn}, we observe that the stretch exponent $1/2$ is same as the BVDV formula for both models. However unlike this formula, we do not have any analytical prediction of the functions $C_{\#}(\eta)$ and $\mathcal{M}_{\#}(\rho)$. Instead, both quantities are extracted numerically through a fitting procedure. For this, we plot $\log S(\eta)$ versus $\eta$ for various densities and find that the resulting curves can be fitted with $\text{log}S_{\#} = \text{log}C_{\#} - \mathcal{M}_{\#}(\rho) ~\eta ^{1/2}$. The coefficient $\mathcal{M}_{\#}(\rho)$ is then determined from the asymptotic slope in the large-$\eta$ regime, while the prefactor $C_{\#}(\eta)$ is obtained by adjusting the vertical intercept to ensure a good fit across the appropriate range. More details of this fitting procedure are given in Appendix~\ref{appen-fit-2d}. In particular, for $\mathcal{M}_{\#}(\rho)$, following this fitting procedure, we estimate
 \begin{align}
\mathcal{M}_{\#}(\rho) \approx 
\begin{cases}
%~~1.255, & ~~~\text{for AIL}, \\[4pt]
~~0.0785\, \langle n_{\rm T} \rangle^{1/3}, &~~~ \text{for Sokoban}, \\[4pt]
~~0.55\, \langle n_{\rm T} \rangle^{1/10}, & ~~~\text{for G-Sokoban}.
\end{cases}
\label{2d-st-surv-eq-3}
\end{align}  
With these estimates, we now proceed to plot $-\text{log}\left[ S(\eta) \big/  C(\eta ) \right] \big/ \mathcal{M}(\rho)$ as a function of $\eta $ in Fig.~\ref{fig-2d-survival} using numerical simulations for different values of density. Regardless of the value, the rescaled curves for both models converge to the same asymptotic $\sqrt{\eta}$ form when $\eta$ is large

Plugging the density dependence of the average trapping time via Eqs.~\eqref{2d-eq-8} and \eqref{2d-eq-9} and converting from $\eta$ back to the actual time $n$, we get
\begin{align}
S (n) \sim \exp \left( -f_2(\rho) n^{1/2} \right),~~~~\text{for } n \gg  \langle n_{\rm T} \rangle,
\end{align}
with
\begin{align}
f_2(\rho) \sim
\begin{cases}
%~~\left( \rho -\rho _c\right)^{ \frac{ \gamma _{\rm AIL}  }{   2}} \big/ \left( 1-\rho \right)^{\frac{1}{2}}, & ~~~\text{for AIL},~\rho > \rho _c \\[4pt]
~~ \rho ^{ \frac{ \gamma _{\rm sok}  }{   6}} \big/ \left( 1-\rho \right)^{\frac{1}{6}}, &~~~ \text{for Sokoban}, \\[4pt]
~~\rho ^{ \frac{ 2\gamma _{\rm Gsok}  }{   5}} \big/ \left( 1-\rho \right)^{\frac{2}{5}}, & ~~~\text{for G-Sokoban}.
\end{cases} \label{anuqon16}
\end{align} 
Recall that $\gamma _{\rm sok} \approx 13.17 \pm 0.1$ and $\gamma _{\rm Gsok} \approx 14.93 \pm 0.4$.
Therefore, we see that even in two dimensions, for both Sokoban and G-Sokoban models, $S(n)$ bears a resemblance to $\phi(n)$ in Eq.~\eqref{original-eqn}, as far as the exponent $1/2$ is concerned. This suggests the BVDV trapping universality class even in two dimensions.

Finally, we have analyzed the intermediate time behavior of the survival probability and plotted $\left( 1-S(\eta) \right)$ vs $\eta$ in Fig.~\ref{2s-short-time} for the two models. In $1d$, we analytically proved that $\left( 1-S (\eta) \right)$ scales as $\sim n \rho^2 $ for $N_{\rm P}=0$ and as $\sim n^2 \rho ^4$ for $N_{\rm P}=1$. In $2d$, in contrast, our simulation reveals
\begin{align}
1-S(\eta) \sim \eta ^{\epsilon(\rho)}
\end{align}
where the exponent $\epsilon$ depends on $\rho$. For densities considered in Fig.~\ref{2s-short-time}, the value of the exponent is quoted in the legend. Moreover, for a given density, $\epsilon$ also depends on the model. 
%For example, in the left and middle panels of this figure, when $\rho = 0.5$, we observe $\epsilon \approx 0.3$ for AIL and $\epsilon \approx 0.85$ for Sokoban. 
Thus, the short-time behavior is sensitive to both density as well as the details of the underlying model.

\section{Conclusion and Outlook}
\label{sec-conclusion}
In conclusion, we analyzed the trapping behavior of the Sokoban-type random walk in a disordered medium with obstacle density $\rho$, where the walker can locally modify its environment by pushing a few obstacles that block its path. Our goal was to understand how the limited pushing dynamics affects the nature of trapping in one and two dimensions. For this, we studied the following three quantities: (i) the disorder-averaged survival probability $S(n)$ that the walker has not yet been trapped until time $n$, (ii) the average trap size $\langle A_{\rm T} \rangle $ and, (iii) the average trapping time $\langle n_{\rm T} \rangle $.

We showed that the survival probability $S(n)$ for the Sokoban model has a stretched-exponential relaxation at late times, with stretch exponents $1/3$ and $1/2$ in one and two dimensions, respectively; see Figs.~\ref{fig-short-n-1d} and \ref{fig-2d-survival}. The exponents are similar to the Balagurov-Vaks-Donsker-Varadhan (BVDV) formula for the reactive-trapping-based survival probability $\phi(n)$ described in Eq.~\eqref{original-eqn}. In one dimension, we proved that this result is valid for the general $N_{\rm P}$-Sokoban model, in which the walker is capable of pushing up to an arbitrary $ N_{\rm P}$ number of obstacles. For $ N_{\rm P}  \gg 1$, the survival probability $S(n)$ is characterized by a large-deviation rate function and it exhibits an exponential decay for not too large $n$, before reverting to the same stretched-exponential decay at long times; see Eq.~\eqref{gen-LDF-eq-11}.

In two dimensions, we studied a variant of the Sokoban model, referred to as the G-Sokoban model, which features a modified pushing dynamics. Despite this modification, numerical simulations suggest that the survival probability $S(n)$ still exhibits a long-time stretched-exponential decay with a stretch exponent of $1/2$.

While the long-time behavior of both survival probabilities, $S(n)$ and $\phi(n)$, display similarities, placing them in the same universality class, there are some differences also. In one dimension, both of them decay with an exponent proportional to $\lambda^{2/3}n^{1/3}$ at long times, but with markedly different proportionality constants: for $\phi(n)$ in Eq.~\eqref{original-eqn}, the constant is $3\pi^{2/3}/2 \approx 3.2175$, whereas for $S(n)$ in Eq.~\eqref{gen-LDF-eq-11}, it is $3\pi^{2/3}/2^{5/3} \approx 2.0269$ independent of $N_{\rm P}$. Differences also appear in the algebraic prefactor multiplying the stretched-exponential term. For instance, in Eq.~\eqref{sok-surv-eqajo-35} for $S(n)$ with $N_{\rm P}=1$ this prefactor scales as $n^{7/6}$, while for $\phi(n)$ it scales as $\sqrt{n}$ \cite{Anlauf1984}.

 %$\chi (n |N_{\rm P}=0 )$ should asymptotically converge to a constant value $\chi (n \to \infty |N_{\rm P}=0) = 3 \pi ^{2/3} / 2^{5/3} \approx 2.0269$ independent of the density $\rho$. Indeed, in Fig.~\ref{fig-surv-1d} (left panel), we have shown the simulation results of $\chi(n |N_{\rm P}=0)$ for different density values, and for all cases, the simulation data converge to the predicted constant value, $ 2.0269$, indicated by the dashed black line. In this figure, we have also presented a comparison of our result with the BVDV formula in Eq.~\eqref{original-eqn}. According to this formula, $ \lim_{n \to \infty}\left[ -\ln \phi(n) / \lambda ^{2/3} n^{1/3} \right] = $

The moderate-time behaviors of $S(n)$ and $\phi(n)$ are also qualitatively different.  For $\phi(n)$, the decay at moderate times is well described by the Rosenstock approximation: $\left(1 - \phi(n)\right) \sim \sqrt{n \rho^2}$ for $n \rho^2 \ll 1$ in one dimension (with moderate $n$), see Eq.~\eqref{original-eqn-2}. Contrarily, we find that $S(n)$ for the Sokoban model exhibits a much slower decrease in this regime, with $\left(1 - S(n)\right) \sim n \rho^2$ for $N_{\rm P}=0$ and $\left(1 - S(n)\right) \sim n^2 \rho^4$ for $N_{\rm P}=1$. This highlights that despite similarities in the long-time exponents, the survival probabilities $S(n)$ for different values of $N_{\rm P}$ are very different.

We also reported a dynamical crossover in the underlying trapping mechanisms that replaces the classical percolation transition in the two-dimensional Sokoban and G-Sokoban models. Unlike in the AIL model, where trapping occurs due to geometric constraints imposed by the static disorder, the above crossover occurs due to the limited pushing ability of the walker. As a result, the average trap size exhibits a nonmonotonic dependence on the obstacle density $\rho$, as shown in Fig.~\ref{fig-2d-area-neww}.  Starting from high density, it increases as $\rho$ is decreased, achieving a maximum near a characteristic density $\rho_*$, and then decreases beyond this value. Based on numerical simulations, we estimated $\rho _* \approx 0.55$ for the Sokoban model and $\rho _* \approx 0.675$ for the G-Sokoban model. The crossover density $\rho _*$ separates two qualitatively distinct trapping mechanisms: at high densities , the walker is predominantly trapped by pre-existing obstacles in the environment, whereas at low densities, trapping arises primarily through a self-trapping mechanism, where the walker dynamically constructs its own confinement by reorganizing its local environment. This self-trapping mechanism prevents long-range transport in the Sokoban model.

Going forward, our work paves the way for several directions for future investigation. While analytical treatment is feasible in one dimension, extending such a treatment to two dimensions remains a significant open challenge.  Many of our results in two dimensions are based on the fitting procedure and verifying these results through a more rigorous method is an important future challenge. On the computational side, although numerical simulation offers valuable insights in two dimensions, it becomes increasingly expensive at small densities and at long times. 
Standard numerical techniques based on simple sampling become inefficient in this regime due to the rarity of trapping events. Therefore, it remains an important open problem to develop efficient computational methodologies to probe the Sokoban model in the small-density regime. Another interesting future direction would be to replace the unbiased diffusive motion in the Sokoban model with sub-diffusive dynamics (e.g., via a continuous-time random walk framework) or directed diffusive motion or pushy random motion, and investigate how this affects the trapping behavior and the associated survival probability \cite{newadd2,Bonomo2026}.

%\textit{Experiments:} Finally, we discuss potential experimental routes to verify the main results of our paper. In particular, recent experiments on bristle robots \cite{expt-1} or active colloids \cite{expt-2, expt-3} offer promising platforms for experimental verification of our results. Even though these systems differ from our model by evolving in continuous space, caging can nevertheless be defined operationally by discretizing the arena into grid cells and monitoring how many distinct cells are visited. In studies of glassy dynamics and biological motion, another standard indicator of caging is the time-averaged mean-squared displacement \cite{CT-2007,CT-2020, CT-2022}: when this quantity develops a plateau, the system is understood to be caged, with the plateau value providing an estimate of the cage size. A further approach involves analyzing the convex hull, which quantifies the overall spread of the particle trajectory \cite{CH-2017, CH-2009, CH-2010}. For the Sokoban walker, a saturation of the hull area can be interpreted as the onset of caging, and the largest hull diameter gives a measure of the cage scale.

\section*{Acknowledgment}
The support of Israel Science Foundation's grant 2311/25 is gratefully acknowledged.

\appendix
\section{Details of the numerical simulations}
\label{appen-simulation}
Here we will present details of the numerical simulation followed to obtain various results in the paper. We take a square lattice of size $\mathcal{N}^{d}$, $d$ being the dimension, and place a random walker at the center. The system size is typically chosen in the range $\mathcal{N} = 301\text{--}1001$ for $1d$ and $\mathcal{N} = 601\text{--}5001$ for $2d$ depending on the obstacle density $\rho$. Except at the center, which contains the walker, every other site $(i,j)$ is characterized by an indicator function
\begin{align}
\mathbb{I}(i,j) & =1,~~~\text{with probability }\rho, \nonumber \\
&=0,~~~\text{with probability }(1-\rho) \nonumber.
\end{align}
to indicate the presence or absence of an obstacle at that particular site. For a given initial realization of the obstacles, we then evolve the system (walker + obstacles) according to the Sokoban dynamics, as discussed in Sec.~\ref{sec-model}. To see whether the walker is trapped or not, we monitor the number of distinct sites $\Omega(n)$ that it has visited. For a given realization, it is a monotonically increasing function of time and saturates only if the Sokoban is trapped. Otherwise, it continues to grow indefinitely. Therefore, by looking at $\Omega _{\rm sat} = \Omega \left(n \to \infty \right)$, we can say whether the walker is trapped or not. When the walker is trapped, we define the trapping time, $n_{\rm T}$, as the time at which $\Omega(n)$ saturates to its $\Omega \left(n \to \infty \right)$ value for the first time (see Fig.~\ref{fig-trap-size}). In simulation, we cannot have an infinite run time. We typically run our simulation for a total time of $\tau_{\rm sim} \sim 100 \langle n_{\rm T} \rangle $ such that varying $\tau_{\rm sim}$ does not measurably change the statistics of $n_{\rm T}$. 

In two dimensions, we perform an additional check for caging by looking at the snapshot of the configuration at time $\tau_{\rm sim}$ and ensuring that the observed caging is irreversible, \emph{i.e.,} the sequence of allowed Sokoban moves can carry the walker outside the cage.

%\blueww{A more systematic way to identify caging would be to explicitly inspect the local obstacle configuration around the walker. However, implementing such a geometric check in our simulations is computationally expensive, and we do not pursue it here. Instead, we infer caging from the behavior of $\Omega(n)$.}

Let us denote the whole trajectory by $\{ x(1), x(2), x(3),\cdots, x \left(\tau_{\rm sim} \right)  \}$ and the number of distinct visited sites by $\{ \Omega(1), \Omega(2), \Omega(3),\cdots, \Omega \left(\tau_{\rm sim} \right)  \}$. Note that by definition $\Omega (n) = \Omega _{\rm sat}$ for $n \geq n_{\rm T}$. In our simulation, we look at the quantity $\left[ \Omega _{\rm sat} -\Omega (n)  \right]$ which is positive for $n<n_{\rm T}$ and zero for $n \geq n_{\rm T}$. The first time it becomes zero is the trapping time $n_{\rm T}$. Next we repeat this procedure for around $10^3\text{--}10^4$ realizations (depending on $\rho$) and then take the ensemble average to obtain mean $\langle n_{\rm T} \rangle$

We are also interested in the trap size $A_{\rm T}$ which refers to the number of vacant sites in the trap. For this, we take the part of trajectory from $n_{\rm T} \leq n \leq \tau_{\rm sim}$ and count the number of new sites visited using this part. The saturated value of this number represents the trap size $A_{\rm T}$ for this realization. We then repeat this for other realizations also and obtain the mean $\langle A_{\rm T} \rangle$.

Finally, in order to calculate the survival probability of not being trapped, we take 
\begin{align}
S(n)& = 1,~~~\text{if } n < n_{\rm T} , \nonumber \\
& =0,~~~\text{if } n \geq n_{\rm T} , \nonumber
\end{align}
and then repeat the same analysis for $10^5 \text{--}10^6$ realizations and yield the averaged survival probability.

\section{First-passage visit to $-L_1$ without hitting $L_2$ for one-dimensional random walk}
\label{appen-FPT-1d}
In this appendix, we consider a random walker moving inside a finite interval $[-L_1, L_2]$ and initially located at the origin. At each time step, it moves symmetrically with probability of $1/2$ to either of its neighboring sites. We are interested in calculating the first-passage probability $\mathcal{F}_1 \left( m|x_0 = 0, \{ x_{j} \} \big|_{j=1}^{m-1} < L_2 \right)$ to $-L_1$ given that the walker has not touched $L_2$ at intermediate times, \emph{i.e.,} $x_1, x_2, \ldots,x_{m-1}$ are all smaller than $L_2$. To derive this, we first consider the propagator $\mathcal{P}_{\rm aa}(x,n|0)$ of finding the walker at some position $x$ inside the interval $[-L_1, L_2]$ after a time step $n$, with absorbing boundary conditions at $x = -L_1$ and $x = L_2$. The propagator is governed by the evolution equation
\begin{align}
\mathcal{P}_{\rm aa}(x,n+1|0) = \frac{1}{2} \mathcal{P}_{\rm aa}(x+1,n|0) + \frac{1}{2} \mathcal{P}_{\rm aa}(x-1,n|0), \label{1d-probaa-eq-1}
\end{align} 
with the absorbing boundary conditions 
\begin{align}
\mathcal{P}_{\rm aa}(-L_1, n|0) = 0,~~~\mathcal{P}_{\rm aa}(L_2, n|0) = 0. \label{1d-probaa-eq-2}
\end{align}
Our first-passage probability is related to this propagator by \cite{Rednerbook}
\begin{align}
  \mathcal{F}_1 \left( m| \right. & \left.   x_0 = 0, \{ x_{j} \} \big|_{j=1}^{m-1} < L_2 \right) = \nonumber \\
  &  \left\{ \begin{matrix}
& \frac{1}{2}~\mathcal{P}_{\rm aa} \left(-L_1+1, m-1 \right),~~~\text{for }L_1 \neq 0, \\
& \delta _{m,0},~~~~~~~~~~~~~~~~~~~~~~~~~~~~\text{for }L_1 = 0,
\end{matrix} \right.
\label{1d-probaa-eq-2192}
\end{align}
Since the walker starts from the origin, the first-passage time to $L_1 = 0$ is simply zero. This corresponds to the second relation in Eq.~\eqref{1d-probaa-eq-2192}. On the other hand, for any non-zero value of $L_1$, the walker must reach $x = -L_1+1$ at the $(m-1)$-th time step and then make a successful jump to $x = -L_1$ with probability $1/2$ at the $m$-th step. This gives rise to the first relation in Eq.~\eqref{1d-probaa-eq-2192}. Our goal now is to solve Eq.~\eqref{1d-probaa-eq-1} with appropriate boundary conditions and then use Eq.~\eqref{1d-probaa-eq-2192} to calculate the first-passage probability.

Given the nature of the boundary conditions, we expand $\mathcal{P}_{aa}(x,n|0)$ as
\begin{align}
\mathcal{P}_{\rm aa}(x,n|0) = \sum _{\theta = 1}^{L}~\sin \left[ \frac{ \pi \theta  (x+L_1)}{L}  \right]~\bar{\mathcal{P}}_{\rm aa}(\theta,n|0),
\label{1d-probaa-eq-3}
\end{align}
with $L = (L_1+L_2)$. The inverse transformation can be written as
\begin{align}
\bar{\mathcal{P}}_{\rm aa}(\theta,n|0) = \frac{2}{L} \sum _{x=-L_1}^{L_2} \sin \left[ \frac{\pi  \theta  (x+L_1)}{L}  \right]~\mathcal{P}_{\rm aa}(x,n|0). \label{1d-probaa-eq-4}
\end{align}
To arrive at this inverse transformation, we have used the following property
\begin{align}
\sum _{x=-L_1}^{L_2} \sin \left[ \frac{ \pi \theta_1  (x+L_1)}{L}  \right]~\sin \left[ \frac{  \pi \theta_2  (x+L_1)}{L}  \right] = \frac{L}{2}~\delta _{\theta _1, \theta _2}. \label{1d-probaa-eq-5}
\end{align}
Substituting Eq.~\eqref{1d-probaa-eq-3} in Eq.~\eqref{1d-probaa-eq-1} yields the recurrence relation
\begin{align}
\bar{\mathcal{P}}_{\rm aa}(\theta,n+1|0) =  \cos \left( \frac{\pi \theta}{L} \right) ~\bar{\mathcal{P}}_{\rm aa}(\theta,n|0),
\end{align}
which can be solved as
\begin{align}
\bar{\mathcal{P}}_{\rm aa}(\theta,n|0)  =  \cos ^n \left( \frac{\pi \theta}{L} \right) ~\bar{\mathcal{P}}_{\rm aa}(\theta,0|0). \label{sjaiuq}
\end{align}
We now need the initial value $\bar{\mathcal{P}}_{\rm aa}(\theta,0|0)$. For this, we use the initial condition $\mathcal{P}_{\rm aa}(x,0|0) = \delta _{x,0}$ in Eq.~\eqref{1d-probaa-eq-4} and obtain
\begin{align}
\bar{\mathcal{P}}_{\rm aa}(\theta,0|0) = \frac{2}{L}~\sin \left( \frac{\pi \theta L_1}{L} \right).
\end{align}
Plugging this in Eq.~\eqref{sjaiuq} completely specifies $\bar{\mathcal{P}}_{\rm aa}(\theta,n|0)$. The final form of the propagator can now be obtained by using the inverse transformation in Eq.~\eqref{1d-probaa-eq-4}
\begin{align}
\mathcal{P}_{\rm aa}(x,n|0) = & \frac{2}{L}~\sum _{\theta = 1}^{L} \cos ^n \left( \frac{\pi \theta}{L} \right)~ \sin \left( \frac{\pi \theta L_1}{L} \right)  \nonumber \\ 
&~~~~~ \times  \sin \left[ \frac{ \pi \theta  (x+L_1)}{L}  \right]. \label{ahOO92Y}
\end{align}
Having obtained the propagator, we now turn to Eq.~\eqref{1d-probaa-eq-2192} and write the first-passage probability as
\begin{align}
& \mathcal{F}_1 \left( m|  x_0 = 0, \{ x_{j} \} \big|_{j=1}^{m-1} < L_2 \right) =  \frac{\left(1-\delta_{m,0} \right)~\left(1-\delta_{L_1,0} \right)}{L}~ \nonumber \\
& \times \sum _{\theta = 1}^{L-1} \cos ^{m-1} \left( \frac{\pi \theta}{L} \right)~ \sin \left( \frac{\pi \theta L_1}{L} \right)  ~\sin \left(  \frac{ \pi \theta }{L} \right) +  \delta_{L_1,0} \delta _{m,0} . \label{ajkw861}
\end{align} 
This expression has been quoted in the main text. In addition to this, the propagator $\mathcal{P}_{\rm aa}(x,n|0) $ can also be employed to evaluate the other first-passage probability $\mathcal{F}_2 \left( m|x_0 = 0, \{ x_{j} \} \big|_{j=1}^{m-1} >- L_1 \right)$ to $L_2$ given that the walker has not hit $-L_1$
\begin{align}
\mathcal{F}_2 \left( m| \right. & \left. x_0 = 0, \{ x_{j} \} \big|_{j=1}^{m-1} >- L_1 \right) =  \nonumber \\
& \left\{ \begin{matrix}
& \frac{1}{2}~\mathcal{P}_{\rm aa} \left(L_2-1, m-1 \right),~~~\text{for }L_2 \neq 0, \\
& \delta _{m,0},~~~~~~~~~~~~~~~~~~~~~~~~~~\text{for }L_2 = 0.
\end{matrix} \right.
\label{1d-probaa-eq-211x92}
\end{align}
Utilizing the expression of $\mathcal{P}_{\rm aa}(x,n|0) $ in Eq.~\eqref{ahOO92Y} gives us
\begin{align}
& \mathcal{F}_2 \left( m|  x_0 = 0, \{ x_{j} \} \big|_{j=1}^{m-1} >- L_1 \right)  =  \frac{\left(1-\delta_{m,0} \right)\left(1-\delta_{L_2,0} \right)}{L} \nonumber \\
& \times \sum _{\theta = 1}^{L-1} \cos ^{m-1} \left( \frac{\pi \theta}{L} \right)~ \sin \left( \frac{\pi \theta L_2}{L} \right)  ~\sin \left(  \frac{ \pi \theta }{L} \right)+\delta _{L_2,0} \delta _{m,0}, \label{ajkw862}
\end{align}
which again has been quoted in the main text.

\bluepw{We now make a few comments about our derivation. Our derivation relied on the propagator $\mathcal{P}_{\rm aa}(x,n+1|0)$ which satisfied the time-evolution operator in Eq.~\eqref{1d-probaa-eq-1}. Observe that this equation is a forward equation, since the transition operator acts on the final position $x$, while the initial position $x_0=0$ is held fixed. Equivalently, one may formulate the problem using a backward equation, in which the transition operator acts on the initial position $x_0$, while the final position is held fixed \cite{SM2005}. However, for symmetric nearest-neighbor random walk, the transition operator is self-adjoint and therefore both the forward and the backward equations have the same sine eigenfunctions, as observed in Eq.~\eqref{1d-probaa-eq-3}. Thus,  the same derivation can equivalently be carried out using the backward equation.}

\bluepw{Another recently developed approach for computing first-passage time probabilities in a finite interval with two absorbing boundaries is the filtration method \cite{Sakamoto2024} (see Appendix~\ref{app:filtration} for details). In this method, the two-boundary first-passage problem is constructed from the corresponding one-boundary first-passage problem on a semi-infinite domain. Consider our problem with two absorbing boundaries at $-L_1$ and $L_2$ and the computation of $\mathcal{F}_1 \left( m|  x_0 = 0, \{ x_{j} \} \big|_{j=1}^{m-1} < L_2 \right)$. The central idea of \cite{Sakamoto2024} is to first consider all trajectories that are absorbed at $-L_1$ while ignoring the other boundary at $L_2$. Let us denote this ensemble of trajectories by $\mathcal{R}$. However, among these trajectories, some may have already crossed or touched the $L_2$ boundary at an earlier time, and therefore should not contribute to $\mathcal{F}_1 \left( m|  x_0 = 0, \{ x_{j} \} \big|_{j=1}^{m-1} < L_2 \right)$. Therefore, one needs to systematically subtract or filter out these unwanted trajectories from the ensemble $\mathcal{R}$. The filtration method is a technique to carry out this subtraction recursively, using one-boundary first-passage probabilities. Eventually, the first-passage probability $\mathcal{F}_1 \left( m|  x_0 = 0, \{ x_{j} \} \big|_{j=1}^{m-1} < L_2 \right)$ with two absorbing boundaries is expressed completely in terms of the one-boundary first-passage probabilities; see Eq.~\eqref{appen-FM-eq-3} below . In Appendix~\ref{app:filtration}, we have provided further details on this method and shown that it essentially leads to the same expression of $\mathcal{F}_1 \left( m|  x_0 = 0, \{ x_{j} \} \big|_{j=1}^{m-1} < L_2 \right)$ as given in Eq.~\eqref{ajkw861} using the spectral expansion.}

%\bluepw{Another recently developed approach for computing first-passage time probabilities in a finite interval with two absorbing boundaries is the filtration method \cite{Sakamoto2024}. In this method, the two-boundary first-passage problem is constructed from the corresponding one-boundary first-passage problem on a semi-infinite domain. Consider our problem with two absorbing boundaries at $-L_1$ and $L_2$ and the computation of $\mathcal{F}_1 \left( m|  x_0 = 0, \{ x_{j} \} \big|_{j=1}^{m-1} < L_2 \right)$. The central idea of \cite{Sakamoto2024} is to first consider all trajectories that are absorbed at $-L_1$ while ignoring the other boundary at $L_2$. Let us denote this ensemble of trajectories by $\mathcal{R}$. However, among these trajectories, some may have already crossed or touched the $L_2$ boundary at an earlier time, and therefore should not contribute to $\mathcal{F}_1 \left( m|  x_0 = 0, \{ x_{j} \} \big|_{j=1}^{m-1} < L_2 \right)$. Therefore, one needs to systematically subtract or filter out these unwanted trajectories from the ensemble $\mathcal{R}$. The filtration method is a technique to carry out this subtraction recursively, using one-boundary first-passage probabilities. Eventually, the first-passage probability $\mathcal{F}_1 \left( m|  x_0 = 0, \{ x_{j} \} \big|_{j=1}^{m-1} < L_2 \right)$ with two absorbing boundaries is expressed completely in terms of the one-boundary first-passage probabilities \cite{Sakamoto2024}.}

\subsection{Normalization of $\mathcal{F}_1 \left( m|x_0 = 0, \{ x_{j} \} \big|_{j=1}^{m-1} < L_2 \right) $}
Due to the presence of an absorbing boundary at $x_{\rm abs} = L_2$, the probability $\mathcal{F}_1 \left( m|x_0 = 0, \{ x_{j} \} \big|_{j=1}^{m-1} < L_2 \right)$ is not normalised to unity. Rather in the first line of Eq.~\eqref{abkmao1a}, we saw that
\begin{align}
\sum _{m=0}^{\infty}   \mathcal{F}_1 \left( m|x_0 = 0, \{ x_{j} \} \big|_{j=1}^{m-1} < L_2 \right)   = \frac{L_2}{L}. \label{abo08564}
\end{align}
In the remaining part of this appendix, we will explicitly prove this normalisation. Following Eq.~\eqref{sok-surv-eq-2}, we can write
\begin{align}
\sum _{m=0}^{\infty}   \mathcal{F}_1 \left( m|x_0  \right. & \left. = 0, \{ x_{j} \} \big|_{j=1}^{m-1} < L_2 \right) =  \delta _{L_1,0}  \nonumber \\
 +& \left(1-\delta_{L_1,0} \right) \mathcal{J}(L,L_1),  \label{abo08563}
\end{align}
where the function $\mathcal{J}(L,L_1)$ is defined as
\begin{align}
\mathcal{J}(L,L_1) = \frac{1}{L}~ \sum _{\theta =1}^{L} \cot \left( \frac{\pi \theta}{2L} \right)~\sin \left( \frac{\pi \theta L_1}{L} \right).
\end{align}
One can rewrite this expression as
\begin{align}
\mathcal{J}(L,L_1) & =  \frac{1}{L}~ \sum _{\theta =1}^{2L-1} \cot \left( \frac{\pi \theta}{2L} \right)~\sin \left( \frac{\pi \theta L_1}{L} \right) \nonumber \\
& -\frac{1}{L}~ \sum _{\theta =L+1}^{2L-1} \cot \left( \frac{\pi \theta}{2L} \right)~\sin \left( \frac{\pi \theta L_1}{L} \right). \label{abo08561}
\end{align}
Replacing $\theta \to (2L-\theta)$ in the second summation, we observe that it becomes identical to $\mathcal{J}(L,L_1)$ and Eq.~\eqref{abo08561} reduces to
\begin{align}
\mathcal{J}(L,L_1) = \frac{1}{2L}~ \sum _{\theta =1}^{2L-1} \cot \left( \frac{\pi \theta}{2L} \right)~\sin \left( \frac{\pi \theta L_1}{L} \right). \label{abo08562}
\end{align}
Finally, we use the following summation formula \cite{Beck2010}
\begin{align}
\sum _{j =1}^{\alpha _1-1} \cot \left( \frac{\pi j }{\alpha _1} \right)\sin \left( \frac{2 \pi j \alpha _2}{\alpha _1} \right)  = \alpha _1-2 \alpha _2,\text{ for }\alpha _2 < \alpha _1.
\end{align}
and simplify Eq.~\eqref{abo08562} to
\begin{align}
\mathcal{J}(L,L_1) = \frac{L_2}{L}.
\end{align}
Inserting this form of $\mathcal{J}(L, L_1)$ in Eq.~\eqref{abo08563}, we recover the normalisation of $\mathcal{F}_1 \left( m|x_0 = 0, \{ x_{j} \} \big|_{j=1}^{m-1} < L_2 \right)$ quoted in Eq.~\eqref{abo08564}. Proceeding in the similar manner, one can find
\begin{align}
\sum _{m=0}^{\infty} \mathcal{F}_2 \left( m|x_0 = 0, \{ x_{j} \} \big|_{j=1}^{m-1} >- L_1 \right)  = \frac{L_1}{L}.
\end{align}

\section{First-passage probability for a 1d random walker with a reflecting boundary}
\label{appen-FPT-1d-2}
This section will present a detailed derivation of the first-passage probability $\mathcal{F}_2 \left( m|x_0 = -L_1 \right)  $ for reaching $L_2$ from an initial position $x_0 = -L_1$, in the presence of a reflecting boundary at $x_{\rm ref} = -L_1$. To this end, we first look at the propagator $\mathcal{P}_{\rm ra}(x,n|x_0)$ of finding the walker at some position $x$ (with $-L_1 \leq x \leq L_2$) after $n$-th time step. The subscript `$\rm ra$' denotes that the propagator obeys a reflecting boundary condition at $x = -L_1$ and an absorbing boundary condition at $x = L_2$. It satisfies the evolution equation
\begin{align}
\mathcal{P}_{\rm ra}(x,n+1|x_0) = \frac{1}{2} \mathcal{P}_{\rm ra}(x+1,n|x_0) + \frac{1}{2} \mathcal{P}_{\rm ra}(x-1,n|x_0), \label{1d-probra-eq-1}
\end{align}
with the boundary conditions
\begin{align}
\mathcal{P}_{\rm ra}\left( -L_1, n \right) = \mathcal{P}_{\rm ra}\left( -L_1-1, n \right),~\mathcal{P}_{\rm ra}\left( L_2, n \right) = 0. \label{1d-probra-eq-2}
\end{align}
Consistent with the absorbing boundary condition, we consider the solution 
\begin{align}
\mathcal{P}_{\rm ra}(x,n|x_0) = \sin \left(  \pi \omega (L_2-x) \right)  ~ \bar{\mathcal{P}}_{\rm ra}(\omega,n|x_0) . \label{1d-probra-eq-3}
\end{align}
The reflecting boundary condition then gives
\begin{align}
\omega = \frac{(2 \theta +1)  }{2 \left( L +\frac{1}{2}  \right)},~~~\text{with }\theta = 0,1,2,...L,\label{1d-probra-eq-4}
\end{align}	
where $L = (L_1+L_2)$. Plugging this in Eq.~\eqref{1d-probra-eq-4} yields
\begin{align}
\mathcal{P}_{\rm ra}(x,n|x_0) = \sum _{\theta =0}^{L} \sin \left[ \frac{ \pi (2 \theta +1) (L_2-x)}{2 \left( L +\frac{1}{2}  \right)  } \right]  ~ \bar{\mathcal{P}}_{\rm ra}(\theta,n|x_0). \label{1d-probra-eq-5}
\end{align}
The inverse transformation can be obtained using Eq.~\eqref{1d-probaa-eq-5}
\begin{align}
\bar{\mathcal{P}}_{\rm ra}(\theta,n|x_0) = & \frac{2}{\left( L +\frac{1}{2}  \right)} \sum _{x = -L_1}^{L_2}\sin \left[ \frac{ \pi (2 \theta +1) (L_2-x)}{2 \left( L +\frac{1}{2}  \right)  } \right] \nonumber \\
& ~~~~~~~~~~~~~~~~~~~~~\times \mathcal{P}_{\rm ra}(x,n|x_0). \label{1d-probra-eq-6}
\end{align}	
One can solve Eq.~\eqref{1d-probra-eq-1} to obtain
\begin{align}
\bar{\mathcal{P}}_{\rm ra}(\theta,n|x_0) = & \cos ^n\left[ \frac{\pi (2 \theta +1)}{ 2 \left( L +\frac{1}{2}  \right) }  \right] \bar{\mathcal{P}}_{\rm ra}(\theta,0|x_0).
\label{1d-probra-eq-7}
\end{align}	
To get the unknown $\bar{\mathcal{P}}_{\rm ra}(\theta,0|x_0)$, we utilize the initial condition $\mathcal{P}_{\rm ra}(x,0|x_0) = \delta _{x,x_0}$ and yield
\begin{align}
\bar{\mathcal{P}}_{\rm ra}(\theta,0|x_0) =  \frac{2}{\left( L +\frac{1}{2}  \right)}~\sin \left[ \frac{ \pi (2 \theta +1) (L_2-x_0)}{2 \left( L +\frac{1}{2}  \right)  } \right].
\end{align}
Substituting this in Eq.~\eqref{1d-probra-eq-5} gives the propagator
\begin{align}
& \mathcal{P}_{\rm ra}(x,n|x_0) = \frac{2}{\left( L +\frac{1}{2}  \right)} \sum _{\theta =0}^{L} \cos ^n\left[ \frac{\pi (2 \theta +1)}{ 2 \left( L +\frac{1}{2}  \right) }  \right] \nonumber \\
&\times \sin \left[ \frac{ \pi (2 \theta +1) (L_2-x)}{2 \left( L +\frac{1}{2}  \right)  } \right] ~\sin \left[ \frac{ \pi (2 \theta +1) (L_2-x_0)}{2 \left( L+\frac{1}{2}  \right)  } \right], \label{1d-probra-eq-8}
\end{align}	
which for $x_0=-L_1$ reduces to
\begin{align}
& \mathcal{P}_{\rm ra}(x,n|-L_1) = \frac{2}{\left( L +\frac{1}{2}  \right)} \sum _{\theta =0}^{L} \cos ^n\left[ \frac{\pi (2 \theta +1)}{ 2 \left( L +\frac{1}{2}  \right) }  \right] \nonumber \\
& \times \cos \left[ \frac{\pi (2 \theta +1)}{ 4  \left( L +\frac{1}{2}  \right) }  \right] ~\cos \left[ \frac{\pi  \left(  x+L_1+\frac{1}{2} \right) (2 \theta +1)}{ 2  \left( L +\frac{1}{2}  \right) }  \right] . 
\end{align}
With the propagator now at hand, we proceed to compute the first-passage probability as follows \cite{Rednerbook}
\begin{align}
\mathcal{F}_2 \left( m|x_0 = -L_1 \right)   = \frac{1}{2} \mathcal{P}_{\rm ra}(L_2-1,m-1|-L_1).
\end{align}
As explained in Appendix~\ref{appen-FPT-1d}, this relation is based on the fact that the walker can reach $L_2$ at the $m$-th step only if it was at $L_2 - 1$ at the $(m-1)$ time step and then made a successful jump to $x = L_2$ with probability $1/2$. Employing the expression of propagator derived above, we find
\begin{align}
& \mathcal{F}_2 \left( m|x_0 = -L_1 \right)   =  \sum _{\theta = 0 }^{L}      (-1)^{\theta} \cos^{m-1} \left( \frac{\pi (2 \theta +1)}{2 \left( L + \frac{1}{2} \right)  }  \right) \nonumber \\
& ~~~~~~~~~~\times \cos \left( \frac{\pi (2 \theta +1)}{4 \left( L + \frac{1}{2} \right)  }  \right) ~\sin \left( \frac{\pi (2 \theta +1)}{2 \left( L + \frac{1}{2} \right)  }  \right) \nonumber \\
&~~~~~~~~~~~~ \times  \frac{\left(1-\delta_{m,0} \right) \left(1-\delta_{L_1+L_2,0} \right)}{ \left( L + \frac{1}{2} \right)} .
\end{align}
By symmetry, we have $\mathcal{F}_2 \left( m|x_0 = -L_1 \right) = \mathcal{F}_1 \left( m|x_0 = L_2 \right)$ and the same expression holds for $\mathcal{F}_1 \left( m|x_0 = L_2 \right)$ as well. Moreover, it is derived under the assumption of the reflecting boundary condition. Consequently it should be correctly normalised to unity
\begin{align}
\sum _{m=0}^{\infty} \mathcal{F}_2 \left(m |x_0 = -L_1 \right) = \sum _{m=0}^{\infty} \mathcal{F}_1 \left(m |x_0 = L_2 \right) =1.
\end{align}

{\color{black}
\section{Filtration method for first-passage problem with two absorbing boundaries}
\label{app:filtration}
Here we use the filtration method of Sakamoto and Sakue \cite{Sakamoto2024} to derive the first-passage probability $\mathcal{F}_1 \left( m|  x_0 = 0, \{ x_{j} \} \big|_{j=1}^{m-1} < L_2 \right)$, namely the probability that the walker first reaches $-L_1$ at the $m$-th time step, conditioned on not having touched $L_2$ at earlier times. The expression obtained from the spectral decomposition of the absorbing propagator is given in Eq.~\eqref{ajkw861}. In this appendix, we show that the filtration method leads to the same result. We assume that $L_1 > 0$, $L_2 > 0$. 

To begin with, let us consider the first-passage problem with only one absorbing boundary at $-L_1$ (\emph{i.e.} take $L_2 \to \infty$). We then denote by $\mathcal{G}(m; x_0 \to -L_1)$ the first-passage probability from the initial position $x_0$ to $-L_1$ at the $m$-th time step.

We now have two first-passage problems: the original problem of $\mathcal{F}_1 \left( m|  x_0 = 0, \{ x_{j} \} \big|_{j=1}^{m-1} < L_2 \right)$ with two absorbing boundaries at $-L_1$ and $L_2$, and the auxiliary one-boundary problem described by $\mathcal{G}(m; x_0 \to -L_1)$. The main result of Sakamoto and Sakue is that the former can be constructed from the latter. To see this, let us consider $\mathcal{G}(m; x_0 \to -L_1)$. Physically, it involves all trajectories that reach $-L_1$ for the first time at $m$-th step. Among these trajectories, some may have already touched the other boundary at $L_2$ at an earlier time. Thus, the one-boundary first-passage probability contains trajectories that are not admissible in the two-boundary problem, leading to an overestimate of the desired probability. The filtration procedure removes precisely these inadmissible contributions and leaves only those paths that reach $-L_1$ before ever touching $L_2$.

Sakamoto and Sakue showed \cite{Sakamoto2024}
\begin{equation}
\scalebox{0.85}{$
\begin{split}
\mathcal{F}_1 &\left( m|  x_0 = 0, \{ x_{j} \} \big|_{j=1}^{m-1} < L_2 \right) = \mathcal{G}(m; 0 \to -L_1)   \\ 
& -\sum_{u=1}^{m} \mathcal{F}_2 \left( u|  x_0 = 0, \{ x_{j} \} \big|_{j=1}^{u-1} >- L_1 \right) \mathcal{G}(m-u; L_2 \to -L_1), 
\end{split}$} \label{appen-FM-eq-1}
\end{equation}
where $\mathcal{F}_2 \left( u|  x_0 = 0, \{ x_{j} \} \big|_{j=1}^{u-1} >- L_1 \right)$ is the first-passage probability that the walker first reaches the right boundary $L_2$ at the $u$-th time step, and have not touched the left boundary $-L_1$ at earlier times. Therefore, the second line in Eq.~\eqref{appen-FM-eq-1} accounts for those trajectories where the random walker reaches $L_2$ without reaching $-L_1$ at $u$-th time step, and subsequently reaches $-L_1$ for the first time in the remaining $(m-u)$ time steps. We then sum over all possible values of $u$.

Eq.~\eqref{appen-FM-eq-1} still has $\mathcal{F}_2 \left( u|  x_0 = 0, \{ x_{j} \} \big|_{j=1}^{u-1} >- L_1 \right)$ which involve two absorbing boundaries. However, it can also be expressed as
\begin{equation}
\scalebox{0.85}{$
\begin{split}
\mathcal{F}_2 &\left( m|  x_0 = 0, \{ x_{j} \} \big|_{j=1}^{m-1} >- L_1 \right) = \mathcal{G}(m; 0 \to L_2)  \\ 
& -\sum_{u=1}^{m} \mathcal{F}_1 \left( u|  x_0 = 0, \{ x_{j} \} \big|_{j=1}^{u-1} < L_2 \right) \mathcal{G}(m-u; -L_1 \to L_2).
\end{split}$} \label{appen-FM-eq-2}
\end{equation}
Combining Eqs.~(\ref{appen-FM-eq-1}-\ref{appen-FM-eq-2}), we yield
\begin{equation}
\scalebox{0.85}{$
\begin{split}
&\mathcal{F}_1 \left( m|  x_0 = 0, \{ x_{j} \} \big|_{j=1}^{m-1} < L_2 \right) = \mathcal{G}(m; 0 \to -L_1) -\sum_{u=1}^{m} \mathcal{G}(u; 0 \to L_2)  \\ 
&  \times \mathcal{G}(m-u; L_2 \to -L_1) +\sum_{u=1}^{m} \sum_{v=1}^{u}   \mathcal{F}_1 \left( v|  x_0 = 0, \{ x_{j} \} \big|_{j=1}^{v-1} < L_2 \right) , \\
&~~~~~\times \mathcal{G}(u-v; -L_1 \to L_2)  ~\mathcal{G}(m-u; L_2 \to -L_1)
\end{split}$} \label{appen-FM-eq-3}
\end{equation}
This relation expresses the two-boundary first-passage probability $\mathcal{F}_1 \left( m|  x_0 = 0, \{ x_{j} \} \big|_{j=1}^{m-1} < L_2 \right)$ completely in terms of the one-boundary first-passage probabilities $\mathcal{G}(m; x_0 \to x)$.

Looking at the convolution structure of Eq.~\eqref{appen-FM-eq-3}, we take its $Z$-transform as
\begin{align}
& \widetilde{F}_1(z) = \sum_{m=1}^{\infty} z^m \mathcal{F}_1 \left( m|  x_0 = 0, \{ x_{j} \} \big|_{j=1}^{m-1} < L_2 \right), \label{appen-FM-eq-4} \\
& \widetilde{\mathcal{G}}(z;x_0 \to x) = \sum_{m=1}^{\infty} z^m \mathcal{G}(m; x_0 \to x). \label{appen-FM-eq-5}
\end{align}
Eq.~\eqref{appen-FM-eq-3} now simplifies to
\begin{equation}
\scalebox{0.85}{$
\begin{split}
\widetilde{F}_1(z) = \frac{\widetilde{\mathcal{G}}(z;0 \to -L_1)-\widetilde{\mathcal{G}}(z;0 \to L_2)~ \widetilde{\mathcal{G}}(z;L_2 \to -L_1)}{1-\widetilde{\mathcal{G}}(z;-L_1 \to L_2)~\widetilde{\mathcal{G}}(z;L_2 \to -L_1)}.
\end{split}$} \label{appen-FM-eq-6}
\end{equation}
For the symmetric, nearest-neighbor random walk, we have \cite{Rednerbook}
\begin{equation}
\scalebox{0.85}{$
\begin{split}
\widetilde{\mathcal{G}}(z;0 \to x) = \zeta(z)^{|x|},~~\text{where }\zeta(z) = \frac{1-\sqrt{1-z^2}}{z}.
\end{split}$} \label{appen-FM-eq-7}
\end{equation}
Eq.~\eqref{appen-FM-eq-6} yields
\begin{align}
\widetilde{F}_1(z)  = \frac{\zeta(z)^{L_1}  - \zeta(z)^{2L-L_1}  }{1-\zeta(z)^{2L}}, \label{appen-FM-eq-8}
\end{align}
where recall $L=(L_1+L_2)$.
\subsection{Inversion of $\widetilde{F}_1(z)$}
The inverse-$Z$-transform is 
\begin{align}
\mathcal{F}_1 \left( n \right) = \frac{1}{2\pi i } \oint_C dz \frac{\widetilde{F}_1(z) }{z^{n+1}}, \label{appen-FM-eq-9}
\end{align}
where we use the short-hand notation $\mathcal{F}_1(n)$ instead of $\mathcal{F}_1 \left( n|  x_0 = 0, \{ x_{j} \} \big|_{j=1}^{n-1} < L_2 \right)$. Also $C$ is a counterclockwise contour enclosing the origin and is entirely within the radius of convergence of $\widetilde{F}_1(z)$.
%hence $C$ does not enclose the non-zero singularities of $\widetilde{F}_1(z)$.

The singularities of $\widetilde{F}_1(z)$ are simple poles, which are obtained from Eq.~\eqref{appen-FM-eq-8} as
\begin{align}
1-\zeta(z_{\rm pole})^{2L}=0, \implies z_{\rm pole} = \frac{1}{\cos\left( \frac{\pi \theta}{L} \right)}, \label{appen-FM-eq-10}
\end{align}
where $\theta=1,2, \ldots, (L-1)$. Note that $\theta=0$ is not included because $z=1$ is not a pole. Both numerator and denominator of $\widetilde{F}_1(z)$ in Eq.~\eqref{appen-FM-eq-8} vanish at $z=1$. Moreover all poles are non-zero and satisfy $|z_{\rm pole} | > 1$.

As mentioned before, the contour $C$ in Eq.~\eqref{appen-FM-eq-9} lies within the radius of convergence of $\widetilde{F}_1(z)$. Hence $C$ does not enclose any of the $z_{\rm pole}$. However, one can now enlarge the contour $C \to C_R$ and in doing so, some poles of $\widetilde{F}_1(z)$ are enclosed by $C_R$. By the residue theorem, the difference between the integrals over $C$ and $C_R$ is given by the sum of the residues of the enclosed poles. In particular, in the limit $|C_R|\to\infty$, the contour $C_R$ encloses all poles of $\widetilde{F}_1(z)$, and Eq.~\eqref{appen-FM-eq-9} can be expressed as
\begin{align}
\mathcal{F}_1 \left( n \right) & = \frac{1}{2\pi i } \oint_{|C_R|\to \infty} dz~ \frac{\widetilde{F}_1(z) }{z^{n+1}}  \nonumber \\
&- \sum_{\theta=1}^{L-1} \text{Residue}\left[ \frac{\widetilde{F}_1(z) }{2 \pi i z^{n+1}};z= \frac{1}{\cos\left( \frac{\pi \theta}{L} \right)}\right], \label{appen-FM-eq-11}
\end{align}
For $|C_R|\to \infty$, the integrand decays to zero and therefore the integration is zero. Consequently, we get 
\begin{equation}
\scalebox{0.85}{$
\begin{split}
\mathcal{F}_1 \left( n \right) & =- \sum_{\theta=1}^{L-1} \text{Residue}\left[ \frac{\widetilde{F}_1(z) }{2 \pi i z^{n+1}};z= \frac{1}{\cos\left( \frac{\pi \theta}{L} \right)}\right]
\end{split}$} \label{appen-FM-eq-12}
\end{equation}
Let us next calculate the residue and denote $\mathcal{R}_{\theta} = \text{Residue}\left[ \frac{\widetilde{F}_1(z) }{2 \pi i z^{n+1}};z= z_{\theta}\right] $ where $z_{\theta} = \frac{1}{\cos\left( \frac{\pi \theta}{L} \right)}$. Using Eq.~\eqref{appen-FM-eq-8}, we get
\begin{equation}
\scalebox{1.0}{$
\begin{split}
\mathcal{R}_{\theta} = \frac{\zeta \left(  z_{\theta} \right)^{L_1}- \zeta \left(  z_{\theta}  \right)^{2L-L_1} }{ \left[- \frac{d }{dz} \zeta(z)^{2L}  \right]_{z=z_{\theta}}} \times \left( \cos\left( \frac{\pi \theta}{L} \right) \right)^{n+1} 
\end{split}$} \label{appen-FM-eq-13}
\end{equation}
From Eq.~\eqref{appen-FM-eq-10}, we get $\zeta \left(  z_{\theta} \right) = \exp \left( \frac{i \pi \theta}{L}\right) $. Plugging this form in Eq.~\eqref{appen-FM-eq-13} yields
\begin{align}
\mathcal{R}_{\theta}  = - \frac{1}{L} \sin \left( \frac{\pi \theta L_1}{L} \right) ~ \sin \left( \frac{\pi \theta }{L} \right) \left( \cos\left( \frac{\pi \theta}{L} \right) \right)^{n-1}. 
\end{align} 
Substituting the form of residue in Eq.~\eqref{appen-FM-eq-12} and returning to the original notation for $\mathcal{F}_1(n)$ gives
\begin{equation}
\scalebox{0.85}{$
\begin{split}
\mathcal{F}_1 \left( n|  x_0 = 0, \{ x_{j} \} \big|_{j=1}^{n-1} < L_2 \right) =& \sum _{\theta =1}^{L-1} \frac{1}{L} \sin \left( \frac{\pi \theta L_1}{L} \right) ~ \sin \left( \frac{\pi \theta }{L} \right)  \\
& \times  \left( \cos\left( \frac{\pi \theta}{L} \right) \right)^{n-1}
\end{split}$} \label{appen-FM-eq-14}
\end{equation}
This expression matches with Eq.~\eqref{ajkw861} derived using the spectral decomposition of the absorbing propagator.}

\section{Survival probability $Q \left( n|L_1, L_2 \right)$ for large $n$}
\label{appen-surv-longn}
In the large-$n$ limit, we saw that the survival probability $Q \left( n|L_1, L_2 \right)$ for fixed $L_1$ and $L_2$ decays exponentially with time, see Eq.~\eqref{sok-surv-eq-18}. In order to prove this, we first note that for the walker to survive for a long duration, the interval $[-L_1, L_2]$ over which it moves also has to be large. In the asymptotic regime where both $n$ and $L$ are large while keeping the ratio $n/L^2$ finite, we find that the first-passage probabilities appearing in Eqs.~\eqref{sok-surv-eq-2}-\eqref{sok-surv-eq-4} satisfy the scaling relation
\begin{align}
&\mathcal{F}_1 \left( n|x_0 = 0, \{ x_{j} \} \big|_{j=1}^{m-1} < L_2  \right)  \simeq g_1 \left( n/L^2 \right), \nonumber \\
& \mathcal{F}_2 \left( n|x_0 = 0, \{ x_{j} \} \big|_{j=1}^{m-1} >- L_1  \right) \simeq g_2\left( n/L^2 \right), \nonumber \\
& \mathcal{F}_1 \left( n|x_0 = L_2  \right) \simeq g_3 \left( n/L^2 \right),  \label{sok-surv-eq-7} 
\end{align}
with the $g$-functions defined as
\begin{align}
g_1(z)  & = \frac{\pi}{L^2}~ \sum _{\theta =1}^{\infty} \theta ~\sin \left( \frac{\pi \theta L_1}{L}  \right)~\exp \left( -\frac{z \pi ^2 \theta ^2}{2 }\right),  \label{sok-surv-eq-8} \\
g_2(z)  & = \frac{\pi}{L^2}~ \sum _{\theta =1}^{\infty} \theta ~\sin \left( \frac{\pi \theta L_2}{L}  \right)~\exp \left( -\frac{z \pi ^2 \theta ^2}{2 }\right),   \label{sok-surv-eq-9} \\
g_3(z) & = \frac{\pi}{L^2}~\sum _{\theta =0}^{\infty} (-1)^{\theta} \frac{(2 \theta +1)}{2}~\exp \left( -\frac{z \pi ^2 (2\theta+1) ^2}{8}\right).  \label{sok-surv-eq-10}
\end{align} 
Plugging Eq.~\eqref{sok-surv-eq-7} in Eq.~\eqref{sok-surv-eq-1}, we obtain
\begin{align}
P \left( n_{\rm T} | L_1, L_2 \right) & \simeq \sum _{m=0}^{ n_{\rm T}} \left[ g_1 \left( m/L^2 \right)+ g_2 \left( m/L^2 \right) \right] \nonumber \\
&~~\times g_3 \left(  ( n_{\rm T}-m)/L^2   \right) \left(1-\delta_{L,0} \right).  \label{sok-surv-eq-11}
\end{align}
We perform the variable transformation $\kappa = m/L^2$ and for large $L$, we can treat $\kappa$ as a continuous variable to rewrite Eq.~\eqref{sok-surv-eq-11} as
\begin{align}
P \left( \omega \Bigm| L_1, L_2 \right) \simeq & \int _{0}^{\omega}d\kappa~ \left[ g_1(\kappa) + g_2(\kappa)  \right]~g_3(\omega -\kappa), \nonumber \\
& ~~~\times \left(1-\delta_{L,0} \right) \label{sok-surv-eq-12}
\end{align}
where $\omega = n_{\rm T}/ L^2$. Taking Laplace transformation with respect to $\omega \to s$ and using the convolution structure of the right-hand side, we obtain
\begin{align}
\bar{P} \left( s | L_1, L_2 \right)  \simeq   \left(1-\delta_{L,0} \right)  \left[ \bar{g}_1(s) +\bar{g}_2(s)  \right] \bar{g}_3(s), \label{sok-surv-eq-13}
\end{align}
where $\bar{P} \left( s | L_1, L_2 \right) $ stands for the Laplace transform of $P \left( \omega | L_1, L_2 \right) $ and similarly for the $g$-functions. Following Eqs.~\eqref{sok-surv-eq-8}-\eqref{sok-surv-eq-10}, we have
\begin{align}
\bar{g}_1(s) = & \frac{\sinh (\sqrt{2s} L_2)}{\sinh (\sqrt{2s} L)},~~\bar{g}_2(s) = \frac{\sinh (\sqrt{2s} L_1)}{\sinh (\sqrt{2s} L)}, \nonumber \\
&~~~~ \bar{g} _3(s) = \frac{1}{\cosh  ( \sqrt{2s} L)} , \label{sok-surv-eq-14}
\end{align}
and inserting them in Eq.~\eqref{sok-surv-eq-13} yields
\begin{align}
\bar{P} \left( s | L_1, L_2 \right) & \simeq   2  \left(1-\delta_{L,0} \right)  \csch \left( 2 \sqrt{2s} L  \right) \nonumber \\
& \times \left[ \sinh (\sqrt{2s} L_1)+\sinh (\sqrt{2s} L_2) \right]. \label{sok-surv-eq-15}
\end{align}
This gives us the trapping time probability in the Laplace domain when both $L_1$ and $L_2$ are large. One can use it to obtain various moments of $n_{\rm T}$. For example, the mean $\langle n_{\rm T} \rangle$ can be found to be
\begin{align}
\langle n_{\rm T} \mid L_1, L_2 \rangle = L_1^2 + L_2 ^2 + 3 L_1 L_2.
\end{align}
In fact, one can perform the inverse Laplace transform of Eq.~\eqref{sok-surv-eq-15} and get the full distribution
\begin{align}
P \left( n_{\rm T} \right. & \left. | L_1, L_2 \right)  = \frac{\pi}{2L^2}~\sum _{\theta}(-1)^{\theta -1} \theta \left[ \sin \left( \frac{\pi \theta L_1}{2L} \right)  \right. \nonumber \\
& \left.  + \sin \left( \frac{\pi \theta L_2}{2L} \right)  \right]~\exp \left(  -\frac{\theta^2 \pi ^2 n_{\rm T}}{8L^2} \right)   \left(1-\delta_{L,0} \right) , \label{sok-surv-eq-16}
\end{align}
where recall that $\omega = n_{\rm T}/ L^2$. From this, the survival probability that the particle has not been trapped follows to be
\begin{align}
Q \left(  \right. &  \left.   n|L_1,  L_2 \right)  \simeq  L^2 \int _{n/L^2}^{\infty} d\kappa~  P \left( n_{\rm T}=\kappa L^2 | L_1, L_2 \right) , \nonumber \\
& \simeq \frac{4}{\pi} ~\sum _{\theta =1}^{\infty}~ \frac{(-1)^{\theta-1}}{\theta}~ \left[ \sin \left( \frac{\pi \theta L_1}{2L} \right) + \sin \left( \frac{\pi \theta L_2}{2L} \right)  \right]   \nonumber \\
& ~~~~~~~~~~~~~~~~~~~\times \exp \left(  -\frac{\theta^2 \pi ^2 n}{8L^2} \right)   \left(1-\delta_{L,0} \right) . \label{sok-surv-eq-17}
\end{align}
Then, the leading order decay of the survival probability $n/L^2 \gg 1$ reads as
\begin{align}
Q \left( n|L_1, L_2 \right) \simeq & \frac{4}{\pi}~\left[ \sin \left( \frac{\pi  L_1}{2L} \right) + \sin \left( \frac{\pi  L_2}{2L} \right)  \right] \nonumber \\
& ~~~\times \exp \left(  -\frac{ \pi ^2 n}{8L^2} \right) ~  \left(1-\delta_{L,0} \right) .
\label{sok-surv-eq-18qjsh}
\end{align} 

\subsection{Survival probability for the Sokoban walker with $N_{\rm P}=1$}
\label{appen-surv-1d-sok}
We next compute the average of $Q \left( n|L_1, L_2 \right)$ for the Sokoban model using the joint probability of $q(L,L_i|N_{\rm P}=1) $ with $i \in \{1,2 \}$. As seen before in Eq.~\eqref{ahbaioi7153}, the joint probability for $N_{\rm P}=1$ can be calculated as
\begin{align}
q(L,L_i|N_{\rm P}=1) = & \rho^4 (L_i+1) \left(L-L_i+1 \right) \left( 1-\rho \right)^{L} \nonumber \\
&\times ~~\Theta(L-L_i).\label{eq-1d-sok-neiat-eq-4}
\end{align}
The survival probability is given
\begin{align}
S(n|N_{\rm P}=1)& \simeq \frac{4}{\pi} \sum _{L=1}^{\infty}  \sum_{i=1}^{2} \sum _{L_i=1}^{L}~q(L,L_i|N_{\rm P}=1) \nonumber \\
& \times \sin \left( \frac{\pi  L_i}{2L} \right) ~\exp\left( -\frac{ \pi ^2 n}{8L^2} \right), \nonumber \\
& \simeq \frac{8\rho ^4}{\pi} ~\sum _{L=1}^{\infty} \mathcal{H}(L) ~ (1-\rho)^{L}~e^{ -\frac{ \pi ^2 n}{8L^2}}.\label{sok-surv-eq-23}
\end{align}
with $\mathcal{H}(L)$ defined as
\begin{align}
 \mathcal{H}(L) & = \sum _{L_1=1}^{L}\left( L_1+1 \right) \left(L- L_1+1 \right)~\sin \left( \frac{\pi L_1}{2 L} \right) , \\
& = \frac{1}{4} \left[ 2+L+\cot \left(\frac{\pi}{4L} \right) \left\{ 3+2 L-L\cot \left(\frac{\pi}{4L} \right) \right. \right. \nonumber \\
&~~~~~~~~~~~~~~~~~\left.  \left. +\cot^2 \left(\frac{\pi}{4L} \right)   \right\}    \right], \\
& \simeq \frac{4(4-\pi) L^3}{\pi ^3} + \frac{2L^2}{\pi} + \frac{(24+5 \pi)L}{12 \pi} + \frac{(12-\pi)}{24}, \nonumber \\
&~~~~~~~~~~~~~~~~~~~~~~~~~~~~~~~~~~~~~~~~\text{for }L \gg 1. \label{sok-surv-eq-24}
\end{align}
In the last expression, we have also written down the large--$L$ behavior of $\mathcal{H}(L)$ for later use. Turning to Eq.~\eqref{sok-surv-eq-23}, we perform a change of variable $u = 2 \sqrt{2}L / \pi \sqrt{n}$ and change the summation into an integral for large $n$ 
\begin{align}
& S(n|N_{\rm P}=1) \simeq 2 \sqrt{2n} \rho ^4 \int _0^{\infty} du~\mathcal{H} \left( \frac{2 \alpha}{\lambda} u \right) ~e^{ -\frac{1}{u^2}-2 \alpha u }, \label{sok-surv-eq-25}\\
&~~~~~~~~~~ \text{where } \alpha = \frac{\lambda \pi \sqrt{n}}{4 \sqrt{2}},~~\lambda = \mid \ln(1-\rho)  \mid.  \label{sok-surv-eq-25aj1o}
\end{align}
For $n \gg 1$, one also has $\alpha \gg 1$ and this enables us to use the asymptotic expression for $\mathcal{H}$-function from Eq.~\eqref{sok-surv-eq-24}. The survival probability then turns out to be
\begin{align}
S(n|N_{\rm P}=1)\simeq &   2 \sqrt{2n} \rho ^4 \int _0^{\infty} du ~\left[  \frac{32 \alpha ^3 (4-\pi)}{\pi ^3 \lambda ^3} u^3  +  \frac{8 \alpha ^2}{\pi \lambda ^2} u^2 \right. \nonumber \\
& ~\left. \frac{\alpha(24+5 \pi)}{6 \pi \lambda} u + \frac{(12-\pi)}{24} \right]~e^{ -\frac{1}{u^2}-2 \alpha u }. \label{sok-surv-eq-27}
\end{align}
As illustrated in Eq.~\eqref{sok-surv-eq-28}, the integration over $u$ can be carried out in terms of Meijer-G function. After this, Eq.~\eqref{sok-surv-eq-27} becomes
\begin{equation}
\scalebox{0.85}{$
\begin{split}
& S(n|N_{\rm P}=1)\simeq  \left[   \frac{32  (4-\pi)}{\pi ^3 \lambda ^3}   \MeijerG{3}{0}{0}{3}{-}{0,2,\frac{5}{2}}{\alpha ^2}            +  \frac{8 }{\pi \lambda ^2}  \MeijerG{3}{0}{0}{3}{-}{0,\frac{3}{2},2}{\alpha ^2}  +  \right. \nonumber \\
&\left. \frac{(24+5 \pi)}{6 \pi \lambda}  \MeijerG{3}{0}{0}{3}{-}{0,1,\frac{3}{2}}{\alpha ^2} + \frac{(12-\pi)}{24} \MeijerG{3}{0}{0}{3}{-}{0,\frac{1}{2},1}{\alpha ^2}  \right]  \frac{8 \rho ^4}{\lambda \pi ^{\frac{3}{2}}}.
\end{split}$} \label{sok-surv-eq-29}
\end{equation}
By noting that $\alpha$ depends on $n$ through Eq.~\eqref{sok-surv-eq-25aj1o}, one can further simplify this expression by employing the asymptotic form of the G-functions using \textit{Mathematica}
\begin{align}
\MeijerG{3}{0}{0}{3}{-}{0,\frac{1+m}{2},\frac{2+m}{2}}{\alpha ^2} \simeq &~ e^{-3 \alpha^{\frac{2}{3}}} \left[ \mathbb{R}_m(\alpha) + \mathcal{O}\left( \frac{1}{\alpha ^{\frac{1}{3}}} \right)  \right], \label{sok-surv-eq-30}
\end{align}
where
\begin{align}
&\mathbb{R}_0(\alpha) = \frac{2 \pi}{\sqrt{3}} \alpha ^{\frac{1}{3}}, \label{sok-surv-eq-31} \\
& \mathbb{R}_1(\alpha) = \frac{2 \pi}{\sqrt{3}} \alpha +\frac{17 \pi}{18 \sqrt{3}} \alpha ^{\frac{1}{3}}, \label{sok-surv-eq-32} \\
& \mathbb{R}_2(\alpha) = \frac{2 \pi}{\sqrt{3}} \alpha^{\frac{5}{3}} +\frac{35 \pi}{18 \sqrt{3}} \alpha +\frac{1225 \pi}{1296 \sqrt{3}} \alpha ^{\frac{1}{3}}, \label{sok-surv-eq-33} \\
& \mathbb{R}_3(\alpha) = \frac{2 \pi}{\sqrt{3}} \alpha^{\frac{7}{3}} +\frac{59 \pi}{18 \sqrt{3}} \alpha^{\frac{5}{3}} +\frac{3745 \pi}{1296 \sqrt{3}} \alpha+\frac{199115 \pi}{139968 \sqrt{3}} \alpha^{\frac{1}{3}} . \label{sok-surv-eq-34}
\end{align}
With these asymptotic forms, the long-time behavior of the survival probability simplifies to
\begin{align}
S(n|N_{\rm P}=1) \simeq \mathcal{K}_{1}(n) \exp \left( -\frac{3 \pi ^{2/3}}{2^{5/3}} \lambda ^{2/3} n^{1/3} \right),  \label{sok-surv-eq-3ajhoiqbd5}
\end{align}
where the prefactor $\mathcal{K}_{1}(n)$ reads as
\begingroup
\small
\begin{align}
\mathcal{K}_{1}(n)  & = \frac{8 \rho ^4}{\lambda \pi ^{3/2}}  \left[   \frac{32  (4-\pi)}{\pi ^3 \lambda ^3}  \mathbb{R}_3 \left( \alpha  \right)           +  \frac{8 }{\pi \lambda ^2}  \mathbb{R}_2 \left( \alpha  \right)  \right. \nonumber \\
&~~~~~~~ \left.  + \frac{(24+5 \pi)}{6 \pi \lambda} \mathbb{R}_1 \left( \alpha  \right)  + \frac{(12-\pi)}{24}\mathbb{R}_0 \left( \alpha  \right)  \right], \label{sok-surv-eq-36} \\
& \simeq  \frac{2^{\frac{19}{6}} (4-\pi) \rho ^4}{\sqrt{3} \pi ^{\frac{7}{6}} \lambda ^{\frac{5}{3}}}~n^{\frac{7}{6}},~~\text{for } n \gg 1.
\end{align}
\endgroup
\subsection{Survival probability at intermediate times}
\label{appen-intermediate-time}
The survival probability $Q\left(n |L_1, L_2 \right)$ given in Eq.~\eqref{sok-surv-eq-18qjsh} works only for $n/L^2 \gg 1$. For small values of $n/L^2$, one needs to retain all terms in Eq.~\eqref{sok-surv-eq-17}. In this section, we show that it is possible to derive another expression for the survival probability which works well when the rescaled variable $n/L^2 \ll 1$. Our starting point is to rewrite the Laplace transform in Eq.~\eqref{sok-surv-eq-15} as
\begin{equation}
\scalebox{0.9}{$
\begin{split}
\bar{P} \left( s |  L_1, \right. & \left. L_2 \right)  \simeq  \sum _{m=0}^{\infty} \Big[  \exp \left( -\sqrt{2s} \left[ -(4m+1)L_1+(4m+2)L_2  \right] \right) \Big.  \\
& +\exp \left( -\sqrt{2s} \left[-(4m+2)L_1+(4m+1)L_2  \right] \right)  \\
& - \exp \left(-\sqrt{2s} \left[ -(4m+3)L_1+(4m+2)L_2  \right] \right)   \\
& \Big.-  \exp \left(-\sqrt{2s} \left[-(4m+2)L_1+(4m+3)L_2  \right] \right) \Big] 2   \left(1-\delta_{L,0} \right)  .
\end{split}$} \label{appen-intermediate-time-eq-1}
\end{equation}
Our goal is to analyze this expression for $s L^2 \gg 1 $ (equivalently $n/L^2 \ll 1$). In this limit, the leading order contribution to $\bar{P} \left( s | L_1, L_2 \right) $ will come from $m=0$ term in the first line of Eq.~\eqref{appen-intermediate-time-eq-1}
\begin{align}
\bar{P} \left( s | L_1, L_2 \right) = & 2 \left[  e^{-\sqrt{2s}(L_1+2 L_2)}+e^{-\sqrt{2s}(2L_1+ L_2)}   \right] \nonumber \\
& \times  \left(1-\delta_{L,0} \right) .\label{appen-intermediate-time-eq-2}
\end{align}
Performing the inverse Laplace transformation
\begin{align}
P \left(n_{\rm T}|L_1, L_2 \right) & \simeq  \sqrt{\frac{2}{\pi n_{\rm T}^3}} \left[  (L_1+2 L_2) ~e^{-\frac{(L_1+2 L_2)^2}{2n_{\rm T}}} \right. \nonumber \\
& \left. + (L_1+2 L_2) ~e^{-\frac{(2L_1+ L_2)^2}{2n_{\rm T}}}  \right] \left(1-\delta_{L,0} \right)  ,
\end{align}
which then gives the survival probability
\begin{align}
& Q(n|L_1, L_2) =1 - \int _0^{n} dn_{\rm T} P \left(n_{\rm T}|L_1, L_2 \right), \nonumber \\
& \simeq 1-2 \left[ \text{erfc} \left( \frac{L_1+2 L_2}{\sqrt{2n}}\right) +\text{erfc} \left( \frac{2L_1+ L_2}{\sqrt{2n}}\right)  \right]  \nonumber \\
& ~~~~~~~~~~~~~~~~\times \left(1-\delta_{L,0} \right) ,
\end{align}
for $n/L^2 \ll 1$. We now have to average over $L_1$ and $L_2$. For small densities
\begin{align}
q(L_i|N_{\rm P}=0) \simeq \rho e^{-\rho L_i},~q(L_i|N_{\rm P}=1) \simeq \rho^2 L_{i} e^{-\rho L_i},
\end{align}
where $ i \in \{ 1,2 \}$. With this at hand, we get
\begin{align}
S(n| N_{\rm P}=0)  \simeq & ~1- \rho ^2 \int _0 ^{n} d n_{\rm T}   \sum _{L_1, L_2=1}^{\infty}e^{-\rho(L_1+L_2)} \nonumber \\
& ~~~~~\times P \left(n_{\rm T}|L_1, L_2 \right).
\end{align}
This expression can be simplified to
\begin{align}
S(n| N_{\rm P}=0) \simeq ~ 1-& \frac{2 \sqrt{2}\rho ^2}{\sqrt{\pi}} \int _0^{n} d n_{\rm T}  ~2^{3/2}\int _0^{\infty}dw_1 
 \nonumber \\
 \times  \int _0^{\infty}dw_2 & ~(w_1+2 w_2) ~e^{-(w_1+2 w_2)^2}.
\end{align}
Upon carrying out the integrations over $w_1$ and $w_2$, we obtain
\begin{align}
S(n| N_{\rm P}=0)  \simeq 1-n\rho ^2.
\end{align}
Proceeding in a similar manner for $N_{\rm P}=1$ yields
\begin{align}
S(n| N_{\rm P}=1) \simeq 1-\frac{n^2\rho ^4}{8}.
\end{align}

\begin{figure}[t]
	\centering
	\includegraphics[scale=0.19]{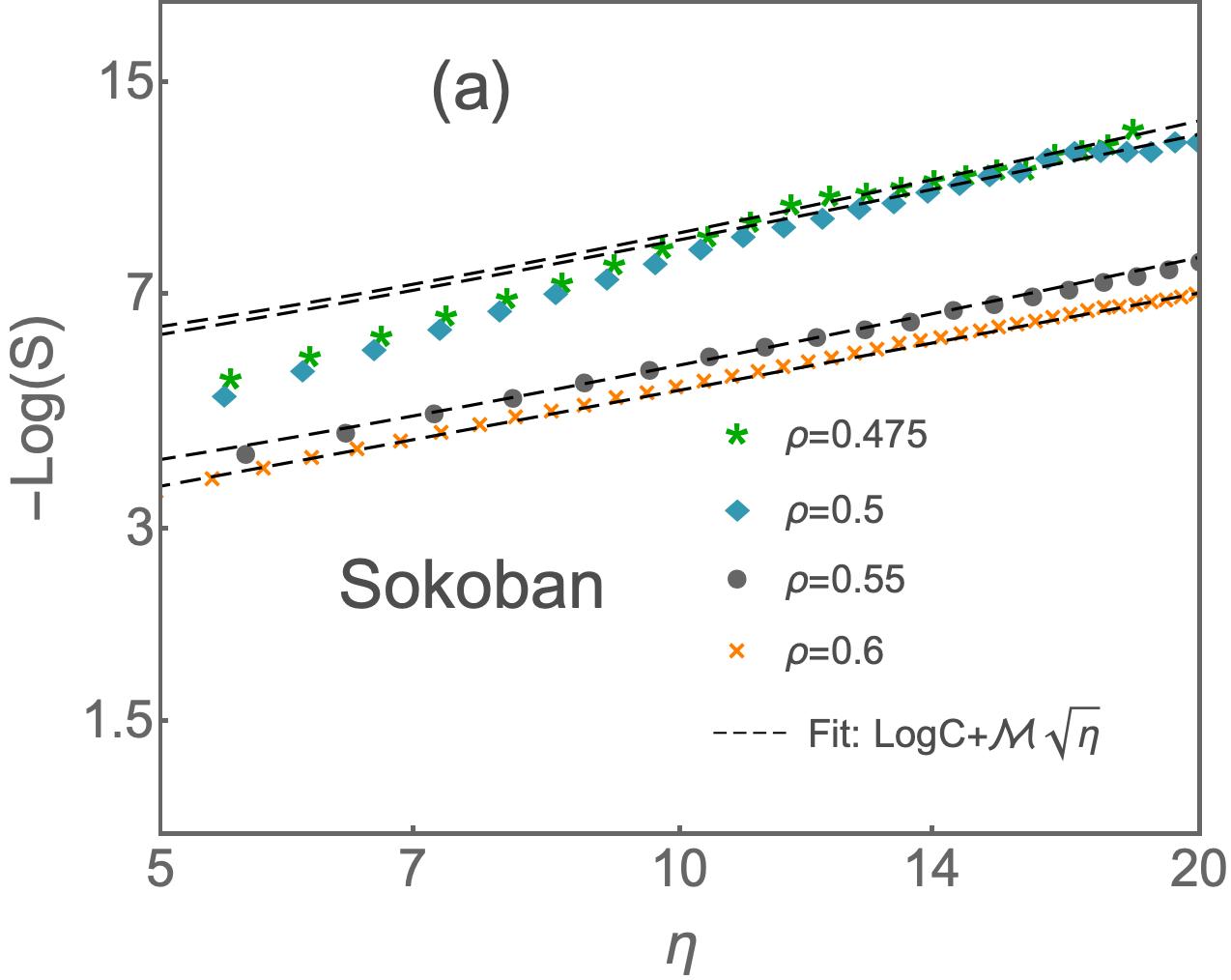}
	\includegraphics[scale=0.19]{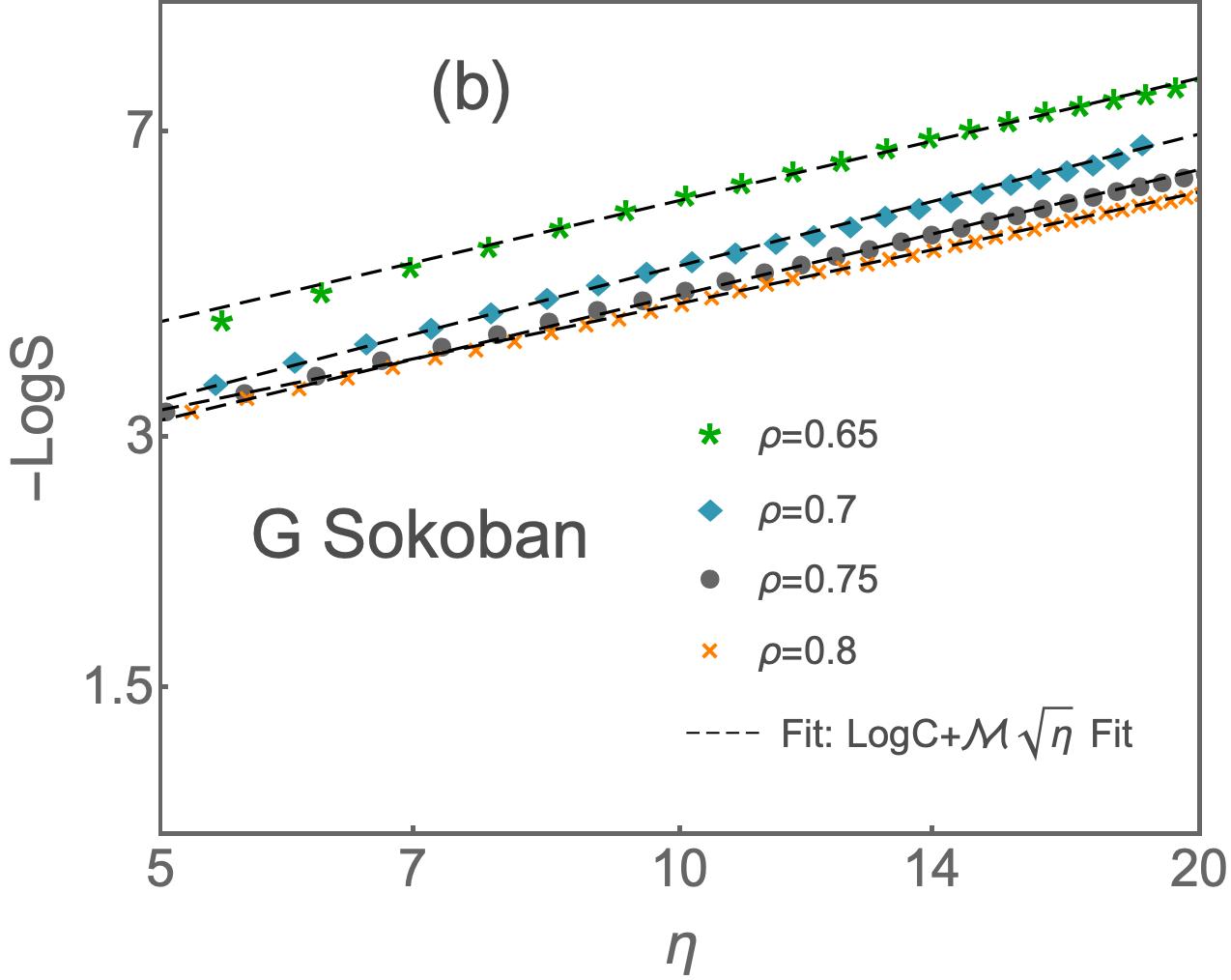}
	\caption{Plot of $\log S(\eta)$ versus $\eta$ for the two-dimensional Sokoban and G-Sokoban models in the log-log scale. Each panel shows simulation data for different densities (shown by symbols). The dashed lines represent fits to the data based on the stretched-exponential form in Eq.~\eqref{appen-fit-2d-eq-2}. The linear trend at large $\eta$ in the log-log scale reflects the $\sqrt{\eta}$ behavior of $\text{log}S(\eta)$. The slope of each fit yields the decay coefficient $\mathcal{M}_{\#}(\rho)$, while the vertical offset provides an estimate of the prefactor $C_{\#}(\eta)$. The estimated values of $\mathcal{M}_{\#}(\rho)$ are given in Eq.~\eqref{2d-st-surv-eq-3}, while that of $C_{\#}(\eta)$ are presented  in  Table~\ref{tab:prefactor-values}.
	}
\label{fig-fit-2dd}
\end{figure}

\begin{table}[htbp]
\begin{tabular}{|l|c|c|}
\hline
\textbf{Model} & \textbf{Density $\rho$} & \textbf{Prefactor $C_{\#}(\eta)$} \\
\hline
%\multirow{4}{*}{AIL} 
 % & 0.45 & 1.0 \\
 % & 0.50 & 1.0 \\
 % & 0.55 & 1.221 \\
 % & 0.60 & 1.8 \\
%\hline
\multirow{4}{*}{Sokoban} 
  & 0.475 & $\eta^{2.8}$ \\
  & 0.50  & $\eta^{1.55}$ \\
  & 0.55  & $\eta^{1.25}$ \\
  & 0.60  & $\eta^{0.1}$ \\
\hline
\multirow{4}{*}{G-Sokoban} 
  & 0.65 & 0.8604 \\
  & 0.70 & 1.349 \\
  & 0.75 & 1.0 \\
  & 0.80 & 0.576 \\
\hline
\end{tabular}
\centering
\caption{Numerical estimates of the prefactor $C_{\#}(\eta)$ for different values of the density $\rho$ in each model.}
\label{tab:prefactor-values}
\end{table}
\section{Numerical estimation of $C_{\#}(\eta)$ and $\mathcal{M}_{\#}(\rho)$}
\label{appen-fit-2d}
In this appendix, we will explain how we estimate the quantities $C_{\#}(\eta)$ and $\mathcal{M}_{\#}(\rho)$ appearing in Eq.~\eqref{2d-st-surv-eq-2}, where $\#$ stands for the sok or Gsok depending on the model. These quantities are not known analytically in two dimensions and must therefore be extracted through a fitting procedure. In Fig.~\ref{fig-fit-2dd}, we plot $\text{log}S(\eta)$ versus $\eta$ for both models. Each panel corresponds to one of the two models studied, and within each panel, we show multiple curves corresponding to different values of the density $\rho$. These plots provide a direct visualization of how the survival probability varies across densities, and form the basis of our fitting procedure.

For a fixed density, the large-$\eta$ regime exhibits a clear linear trend in the $\log S(\eta)$ versus $\eta^{1/2}$ plot, consistent with the expected stretched-exponential decay
\begin{align}
S(\eta) \simeq C_{\#}(\eta)\exp\left(-\mathcal{M}_{\#}(\rho)\, \eta^{1/2}\right). \label{appen-fit-2d-eq-1}
\end{align}
Taking the logarithm yields
\begin{align}
\log S(\eta) \simeq \log C_{\#}(\eta) - \mathcal{M}_{\#}(\rho)\, \eta^{1/2}, \label{appen-fit-2d-eq-2}
\end{align}
so that $\log S(\eta)$ becomes approximately linear in $\eta^{1/2}$ for sufficiently large $\eta$. From the slope of this linear region, we extract $\mathcal{M}_{\#}(\rho)$, which quantifies the rate of decay and depends on the density. The vertical offset of the fit gives the prefactor $C_{\#}(\eta)$, which is sub-exponential and varies more slowly with $\eta$. By repeating this procedure for different densities, we obtain the decay coefficient $\mathcal{M}_{\#}(\rho)$, as outlined in Eq.~\eqref{2d-st-surv-eq-3}. On the other hand, while we do not have an expression for the prefactor $C_{\#}(\eta)$, its values for considered densities in Fig.~\ref{fig-fit-2dd} are given in 
Table~\ref{tab:prefactor-values}.\\

%Although we do not determine the precise form of $C_{*}(\eta)$, the numerical fits suggest that it grows polynomially with $\eta$, consistent with the behavior of prefactors in other stretched-exponential processes. This procedure allows us to obtain consistent estimates of both $\mathcal{M}_{*}(\rho)$ and $C_{*}(\eta)$ across all three models and a broad range of densities.

\bibliography{Bib_new}

\end{document}